\documentclass{elsart4-1}

\usepackage{graphicx}
\usepackage{epsfig}

\usepackage{amssymb}

\usepackage[english]{babel}
\newtheorem{e-proposition}[theorem]{Proposition}

\newtheorem{e-definition}[theorem]{Definition\rm}

\renewcommand{\theequation}{\arabic{equation}}
\def\og{\leavevmode\raise.3ex\hbox{$\scriptscriptstyle\langle\!\langle$~}}
\def\fg{\leavevmode\raise.3ex\hbox{~$\!\scriptscriptstyle\,\rangle\!\rangle$}}

\begin{document}
% Select a primary header Physics or Astrophysics
% You can place after the header (classification), if you know it.

%\centerline{Active galactic nuclei at $\gamma$-ray energies}
\begin{frontmatter}


% \title{The Ankle}%\thanksref{label1}}
% \author{Name\thanksref{label2}}

\selectlanguage{english}

\vskip-1.3in
\title{Active galactic nuclei at $\gamma$-ray energies}

% If all authors are at the same address, the [label1] can be suppressed

\selectlanguage{english}
%\author{~}
%\author{Charles Dermer$^a$ and Berrie Giebels$^b$}
%\ead{deligny@ipno.in2p3.fr}
%\author[authorlabel2]{Author Name2}
%\ead{author.name2@email.address2}

%\address{IPN Orsay, 15 Rue Clemenceau\\
%91406 Orsay CEDEX, France}
%\address[authorlabel2]{Address2}

% If your know the dates of reception, and acceptation you can put them now;
%    idem the name of the person presenting your article
% ===================================================================
%       IN CASE OF TWO AUTHORS WITH DIFFERENT ADDRESSES,
%             PLEASE USE THE FOLLOWING EXAMPLE
\author[a]{Charles Dennison Dermer} and
\author[b]{Berrie Giebels}
\ead{CharlesDermer@outlook.com{\rm (Charles Dermer)}}
\ead{Berrie.Giebels@CNRS.fr}
\address[a]{Code 7653, Naval Research Laboratory\\
Washington, DC, United States}
\address[b]{LLR Ecole polytechnique, CNRS/IN2P3, Universit\'e Paris-Saclay\\
91128 Palaiseau,France}
% ===================================================================
%\medskip
\begin{center}
{\small Received *****; accepted after revision +++++}
\end{center}

\begin{abstract}

  Active Galactic Nuclei can be copious extragalactic emitters of MeV-GeV-TeV $\gamma$ rays, a phenomenon
  linked to the presence of relativistic jets powered by a super-massive black hole in the center of the host galaxy. Most of
  $\gamma$-ray emitting active galactic nuclei, with more than 1500 known at GeV energies, and more than 60 at
  TeV energies, are called ``blazars''. The standard blazar paradigm features a jet of relativistic magnetized plasma
  ejected from the neighborhood of a spinning and accreting super-massive black hole, close to the observer direction.
  Two classes of blazars
  are distinguished from observations: the flat-spectrum radio-quasar class (FSRQ) is characterized by strong
  external radiation fields, emission of broad optical lines, and dust tori. The BL Lac class (from the name of
  one of its members, BL Lacertae) corresponds to weaker advection-dominated
  flows with $\gamma$-ray spectra dominated by the inverse Compton effect on synchrotron photons.
  This paradigm has been very successful for modeling the broadband spectral energy distributions of blazars.
  However, many fundamental issues remain, including the role of hadronic processes and the rapid variability of those
  BL Lac objects whose synchrotron spectrum peaks at UV or X-ray frequencies.
  A class of $\gamma$-ray--emitting radio galaxies, which are thought
  to be the misaligned counterparts of blazars, has emerged from the results of the {\it Fermi}-Large Area Telescope and
  of ground-based Cherenkov telescopes. Soft $\gamma$-ray emission has been detected from a few nearby Seyfert
  galaxies,
%e.g.,
%NGC 6814 and
%Circinus,
though it is not clear whether those $\gamma$ rays originate from the nucleus.  Blazars and their misaligned
counterparts make up most of the $\gtrsim 100$ MeV extragalactic $\gamma$-ray background (EGB), and are
suspected of being the sources of ultra-high energy cosmic rays.
%An estimate of the contribution of radio galaxies
%to the EGB is presented, and is relevant to unification studies, which are briefly
%reviewed here.
The future ``Cherenkov Telescope Array'', in synergy with the {\it Fermi}-Large Area Telescope and a wide range of telescopes in space and on
the ground, will write the next chapter of blazar physics.
%% {\it To cite this article: O. Deligny, C. R. Physique XX (2014).}

\vskip 0.5\baselineskip

%\selectlanguage{french}
\noindent{\bf R\'esum\'e} Les noyaux actifs de galaxie peuvent \^etre de puissants \'emetteurs dans tout le domaine
$\gamma$ du MeV au TeV, un ph\'enom\`ene d\^u \`a la pr\'esence de jets relativistes, en liaison avec un trou noir super-massif au
centre de la galaxie h\^ote. La classe d'\'emetteurs de rayons $\gamma$ la plus abondante parmi les
noyaux actifs de galaxie, avec plus de 1500 sources \'etablies aux \'energies du GeV, et plus de 60
aux \'energies du TeV, sont les ``blazars''. Le paradigme actuel du blazar met en jeu un jet de plasma magn\'etis\'e, orient\'e
\`a faible angle de la ligne de vis\'ee, et \'eject\'e depuis le voisinage d'un trou noir accr\'etant et super-massif en rotation.
Les observations permettent de distinguer deux types de blazars~: les quasars radio \`a spectre plat
(ou FSRQ) comprennent des champs de rayonnement externes puissants, des zones avec des raies d'\'emission optiques larges,
et des tores de poussi\`eres. La classe des BL Lac (du nom d'un de ses membres, BL Lacertae)
poss\`ede des flots d'accr\'etion plus faibles, domin\'es par l'advection, et dans lequel l'\'emission
des rayons $\gamma$ vient essentiellement de l'effet Compton inverse sur les photons synchrotron. Ce paradigme permet de mod\'eliser
l'\'emission des blazars sur tout le spectre \'electromagn\'etique. Cependant, beaucoup de
probl\`emes fondamentaux restent sans r\'eponse, notamment le r\^ole des processus hadroniques, et la variabilit\'e
tr\`es rapide de l'\'emission de certains objets BL Lac, ceux dont le spectre synchrotron \'emet le maximum de puissance
dans les domaines UV et X.
Les observations du satellite {\it Fermi}-LAT et celles des observatoires Tcherenkov au sol ont \'egalement mis en \'evidence
une nouvelle classe de radio-galaxies \'emettrices de rayons $\gamma$, consid\'er\'ees comme les
contreparties non-align\'ees des blazars. On a aussi d\'etect\'e l'\'emission de rayons $\gamma$ de basse \'energie
provenant de galaxies de type Seyfert,
mais il n'est pas encore s\^ur que cette \'emission vienne du noyau.
Les blazars avec leurs contreparties non-align\'ees sont \`a l'origine de la
plus grande partie de l'\'emission gamma extragalactique diffuse au-dessus de $100\,{\rm MeV}$, et sont
soup\c{c}onn\'es d'\^etre les sources des rayons cosmiques d'ultra-haute \'energie. Le futur r\'eseau ``Cherenkov Telescope
Array'' (CTA), en synergie avec le  t\'elescope spatial {\it Fermi} et une grande vari\'et\'e de t\'elescopes dans l'espace et
au sol, \'ecriront le prochain chapitre de la physique des blazars.

%\vskip 0.5\baselineskip
%\noindent

\vskip 0.5\baselineskip
\noindent{\small{\it Keywords~:} Active Galactic Nuclei; Gamma rays; Supermassive Black Holes \vskip 0.5\baselineskip
\noindent{\small{\it Mots-cl\'es~:} Noyaux Actifs de Galaxie~; Rayons Gamma~; Trous Noirs Supermassifs~}}

\end{abstract}
\end{frontmatter}

% now the Version française abrégée, if it exists
%\selectlanguage{french}
%\section*{Version fran\c{c}aise abr\'eg\'ee}
% Text of your Version française abrégée here

\selectlanguage{english}

%%%%%%%%%%%%%%%%%%%%%%%%%%%%%%%%%%%%%%%%%%%%%%%%%%%%%%%%%
\section{Introduction: AGN detected at $\gamma$-ray energies}
\label{sec:intro}
%%%%%%%%%%%%%%%%%%%%%%%%%%%%%%%%%%%%%%%%%%%%%%%%%%%%%%%%%

%%   Besides black-hole powered AGN, the other main class of HE and VHE $\gamma$-ray
%% galaxies are the star-forming galaxies whose high-energy activity is ultimately powered by the kinetic energy
%% of exploding stars.  Only a handful are known.
%EGRET detected $\gamma$ rays from The Large Magellanic Cloud already as early as 1992 \cite{sre92},
%and $\gamma$-ray detection of
%its sibling, the Small Magellanic Cloud, was .

Among the astrophysical $\gamma$-ray emitters located well beyond the Milky Way are the $\gamma$-ray galaxies. From the
recent advances in the field we can classify, simply, two types of $\gamma$-ray galaxies. First are the {\it black-hole
  galaxies}, %%% which divide into the generally distant and powerful blazars with low spatial object densities,
%%% and the radio galaxies with weak $\gamma$-ray emissions and correspondingly larger spatial object densities
%%% (in comparison with blazars), relating to blazar/radio-galaxy unification discussed below.
which are powered by infalling gas onto a massive black hole in their center, and will be thouroughly
discussed here. Second are the generally much weaker and likely more numerous {\it cosmic-ray galaxies} with their
$\gamma$-ray emission powered by stellar explosions rather than black holes, which make shocks that accelerate
cosmic rays \cite{ohm}. Indeed, galaxies hardly have to harbor a radio-luminous black hole to be $\gamma$-ray
luminous, as confirmed by a quick glance at the {\it Fermi}-LAT all-sky image, which shows the Milky Way lit
up by cosmic rays colliding with diffuse gas and dust.  Besides most $\gamma$ rays produced by nuclear
collisions of cosmic ray to make pions, cosmic-ray galaxies are also illuminated in $\gamma$ rays by
pulsars and pulsar-wind nebulae.  Analysis shows that black-hole galaxies %%% ,
%%% i.e., blazars and radio galaxies,
make the bulk of the high-energy (HE; $\gtrsim 100$ MeV) and Very High Energy (VHE; $\gtrsim 100$ GeV)
extragalactic $\gamma$-ray background (EGB, \cite{ack15}). Cosmic-ray galaxies, because they
essentially partake in hadronic processes, should have comparable neutrino and $\gamma$-ray luminosities. Note how
different these two types of $\gamma$-ray galaxies are in comparison with the two types of %% radio-quiet and
%% radio-loud
$\gamma$-ray emitting black-hole galaxies identified in the CGRO days \cite{dg95}.

The black-hole $\gamma$-ray galaxies are Active Galactic Nuclei (AGN), which are among the most powerful known
astrophysical sources of non-thermal radiation and most luminous known electromagnetic emitters, with
luminosities in the range $10^{35} - 10^{41}\,{\rm W}$.

\subsection{The active galaxy zoo}
\begin{figure}[!b]
  \centering
  \includegraphics[width=13.0cm]{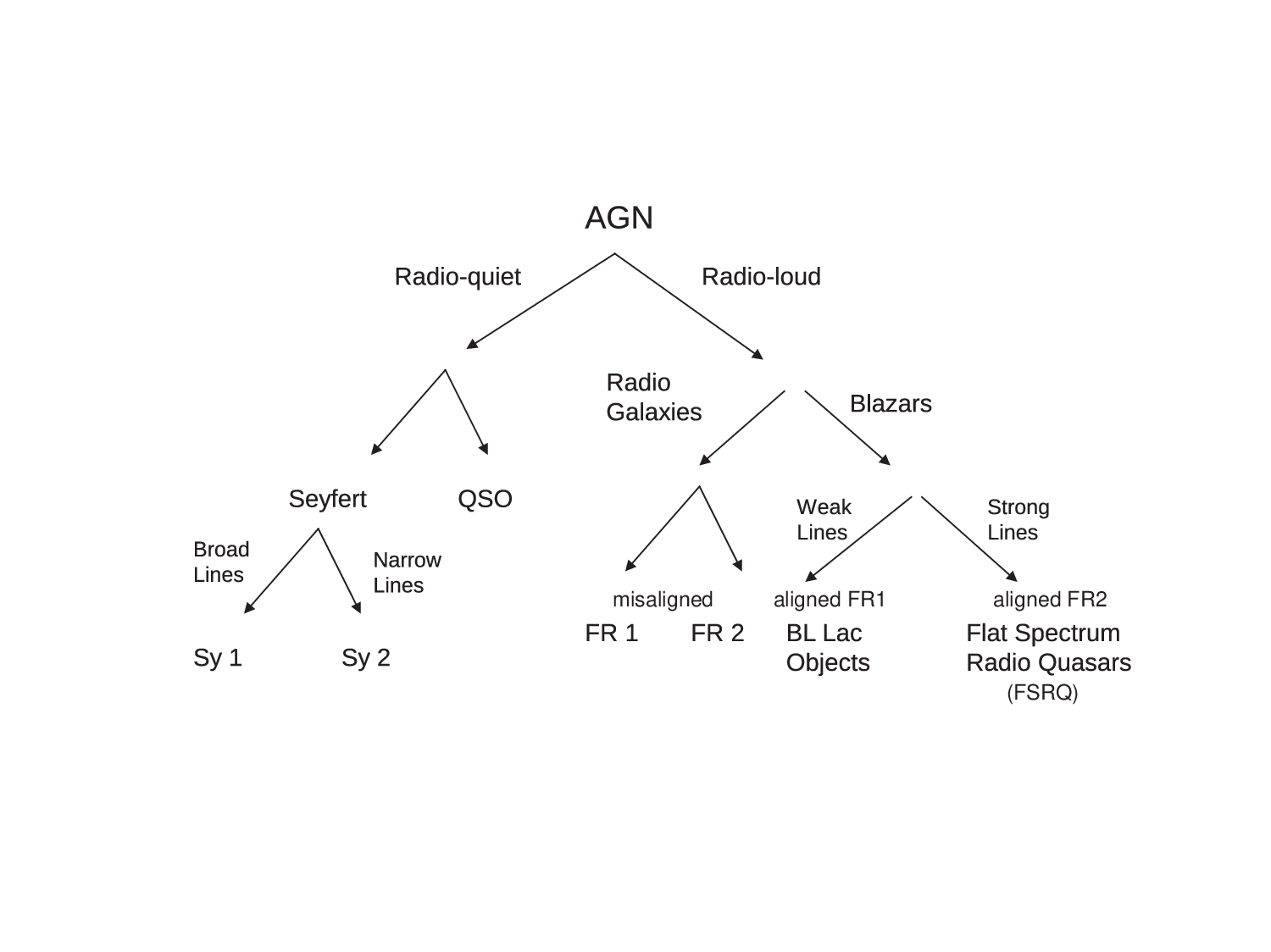}
  \caption{{\bf Observational classification of active galaxies.} AGN are subdivided into classes depending on
  observational aspects, such as their radio loudness or the presence of optical lines in their spectra. QSO = quasi-stellar objects; Sy1 and Sy2 = Seyfert 1 and 2; FR1 and FR2 = Fanaroff-Riley 1 and 2.}
  \label{fig:class}
\end{figure}

The observational classification of AGN, dominated by
the dichotomy between radio-quiet and radio-loud classes, with the latter constituting 10\% of the population,
is represented in the chart of Fig. \ref{fig:class}. The less numerous radio-loud AGN are about 3~orders
of magnitude brighter in the radio band than their radio-quiet counterparts.

\begin{figure}[!t]
  \centering
  \includegraphics[width=15.0cm]{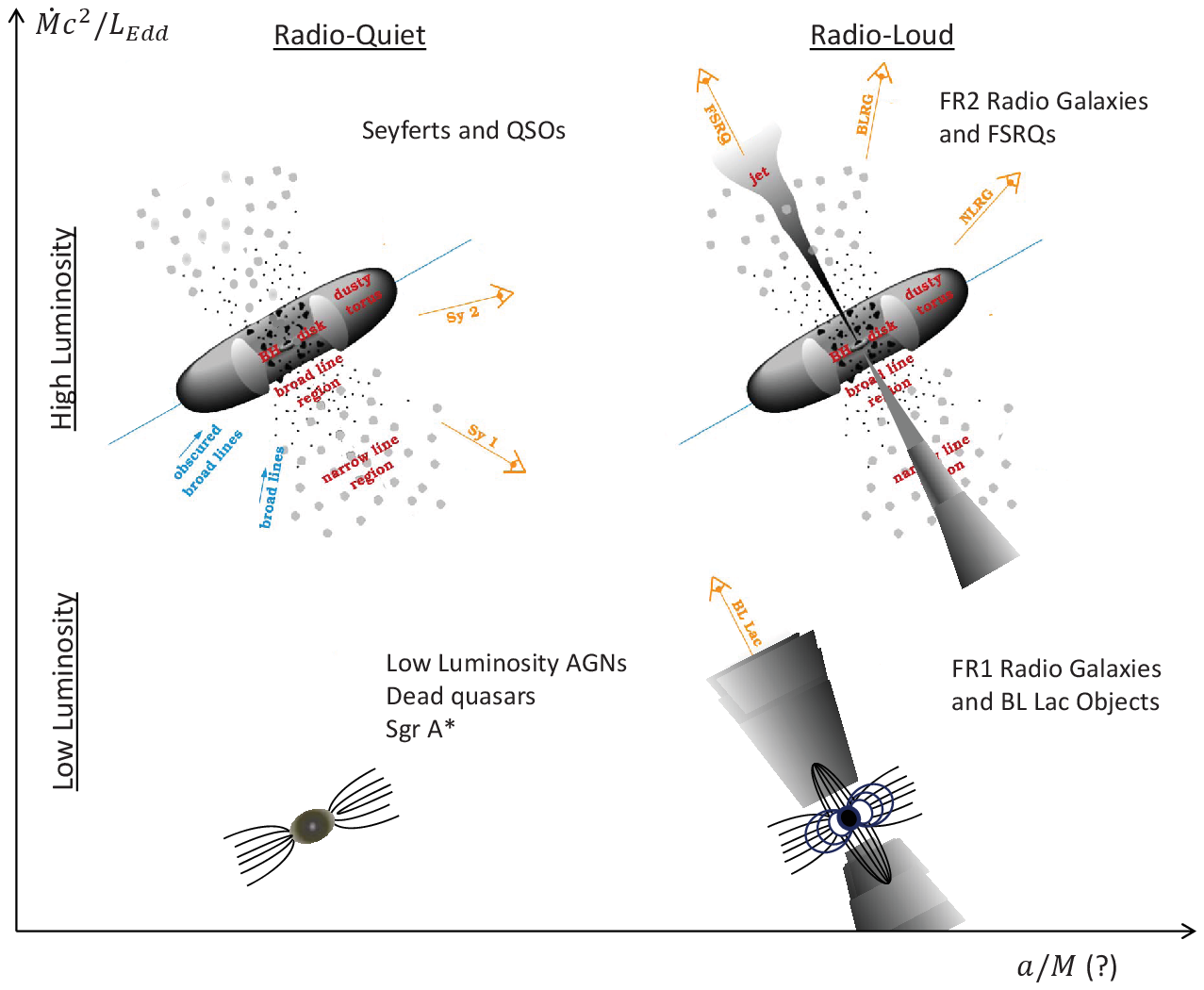}

  \caption{{\bf Types of active galaxies.} Cartoon illustration of AGN taxonomy, following the unification
    scheme for radio-quiet and radio-loud galaxies, {\it cf}.\ \cite{Biteau2013} as adapted from
    \cite{Urry1995}.  In this two-parameter model, including orientation, any given AGN is either radio-quiet
    or radio-loud, here speculated to depend on black-hole rotation $a/M$ (where $M$ is the black hole mass,
    and $a$ its angular momentum), or low power or high power, as
    determined by the mass-accretion rate. The misaligned AGN sources are the radio galaxies, including the
    low luminosity FR1 counterparts to BL Lac objects, and high luminosity FR2 radio galaxies, which divide
    into narrow- and broad-line radio galaxies, depending on orientation.}
  \label{fig:unification}
\end{figure}

The AGN unification scheme is based on the sketch by Urry and Padovani \cite{Urry1995} (see Fig.\ \ref{fig:unification}) which
displays the composite AGN phenomenon (black hole, disk, torus, clouds and jet), and illustrates how
orientation effects, different accretion powers, and different spin parameters\footnote{The spin parameter
is the ratio of the angular momentum $a$ of the black hole to its mass $M$.} $a/M$ of the black hole could
account for the wide range of AGN types. According to Fig.\ \ref{fig:unification}, the appearance of an AGN
depends crucially on the orientation of the observer with respect to the symmetry axis of the accretion
disk. In this picture, the difference between radio-loud and radio-quiet AGN depends on the presence or
absence of radio-emitting jets powered by the central nucleus, which in turn may be induced by the rotation of the
black hole. In Fig.\ \ref{fig:unification}, at high accretion rates (relative to the Eddington limit) and
large luminosities, both radio-loud and radio-quiet AGN have dusty tori, broad-line regions (BLRs),
narrow-line regions, and strong big blue/UV bump emissions from an optically thick accretion disk.
BLR clouds illuminated by the accretion-disk radiation are obscured in Seyfert 2 AGN when viewing through the
dusty torus, and so only narrow lines from high-altitude\footnote{The altitude is the distance to the central nucleus
projected along the rotation axis of the accretion-disk/torus system.} material far from the black hole are seen, resulting in a narrow-line AGN. By
comparison, broad lines are seen from Seyfert 1 AGN when viewing at higher latitudes\footnote{Similarly, the latitude
is defined as the angle with respect to the accretion disk considered as an equatorial plane.} so that the central
nucleus and BLR are visible.

When the jet is directed close to the line of sight (``aligned'' jet), two AGN subclasses are distinguished from
observations~:
\begin{itemize}
\item BL Lac objects correspond to the aligned jets of low-luminosity Fanaroff-Riley 1 (FR1)\footnote{According to the
Fanaroff-Riley classification \cite{fr74}, FR1 have radio jets that are brighter in the center, while FR2 object jets are fainter
in the center but feature brighter radio spots towards the end of their jets.}
radio galaxies (but this classification could miss the type of BL Lac objects that arises from the beamed
radiation overwhelming the lines and disk radiation).
\item Flat Spectrum Radio Quasars (FSRQ, with a radio spectral index $\alpha \sim 0$ at a few GHz) correspond to the
aligned jets of higher luminosity Fanaroff-Riley 2 (FR2) radio galaxies.
\end{itemize}
As illustrated in Fig. \ref{fig:unification}, BL Lac objects occur in AGN with no
significant accretion disk, broad lines, or dusty torus \cite{gio12}. In low-luminosity AGN, advective effects
in accretion likely play an important role in accounting for the relationship between escaping photon power
and accretion power.

Narrow-line and broad-line radio galaxies likewise depend on the direction of the observer's line of sight
with respect to the angle of the disk-jet system, and would more likely be associated with FR2 radio
galaxies. Blazars are those sources for which we happen to viewing at an angle $\theta \lesssim 1/{\Gamma}$,
that is, within the Doppler beaming cone of the relativistically outflowing plasma moving with bulk Lorentz factor
$\Gamma$.  The powerful FSRQs also have strong broad optical lines, indicating the presence of accretion-disk
radiation and dense broad-line region material.
%% Note that the detection of BLR radiation is more feasible
%% whereas the beamed radiation in ISP objects can lead to small equivalent width, BL Lac-type objects. With HSP
%% blazars, where the BLR lines would be expected to re-emerge, the accretion-disk radiation seems to be mostly
%% gone.
The low-luminosity counterparts of the radio-loud AGN are the BL Lac objects and their misaligned
counterparts, the FR1 radio galaxies. Since blazars are relativistically beamed and aligned objects, their misaligned counterparts should be more
numerous, but the interpretation of the measured source counts of the two classes must take into
account beaming corrections.\\
%%  Fig.\ \ref{fig:unification} is
%% based on the sketch by \cite{Urry1995} which displays the composite AGN phenomenon (black hole, disk,
%% torus, clouds and jet), and illustrates how orientation effects, different accretion powers,
%% and different spin parameter $a/M$ of the black hole can account for the wide
%% range of AGN types. Here the BL Lac objects are the aligned jets of
%% low-luminosity FR1 radio galaxies, but this classification
%% could miss the type of BL Lac objects that
%% arise from the beamed radiation overwhelming
%% the lines and disk radiation. As illustrated here,
%% BL Lac objects occur in AGN with no significant accretion disk, broad lines,
%% or dusty torus \cite{gio12}. In  low-luminosity AGN, advective
%% effects in accretion likely play an important role in accounting
%% for the relationship between escaping photon power and accretion power.

\subsection{History of active galactic research at gamma-ray energies}

The growth and development of space-based $\gamma$-ray astronomy (see \cite{intro} in the first volume of this review)
owes much to the pioneering missions OSO-3
(1967-1968), SAS-2 (1972-1973) and COS-B (1975-1982) \cite{Swanenburg1978}, but the $\gamma$-ray astronomy of
active galactic nuclei (AGN) cannot be said to have begun until the launch of the {\it Compton} Gamma Ray
Observatory (CGRO) in 1991. Prior to the EGRET (Energetic Gamma ray Experiment Telescope) on CGRO \cite{djt},
five $\gamma$-ray emitting AGN were known, but what a mixed bag! As reviewed by Bassani and Dean \cite{bd83},
these included the Seyfert galaxies NGC 4151 and MCG 6-11-11, the radio galaxy Centaurus A, the ``peculiar
galaxy" NGC 1275, and the quasar 3C 273. Only 3C 273 was detected at $\gtrsim 35$ MeV energies---by
COS-B---whereas the others had emission extending only to a few hundred keV.

\begin{figure}[!t]
  \centering
  \includegraphics[width=7.0cm]{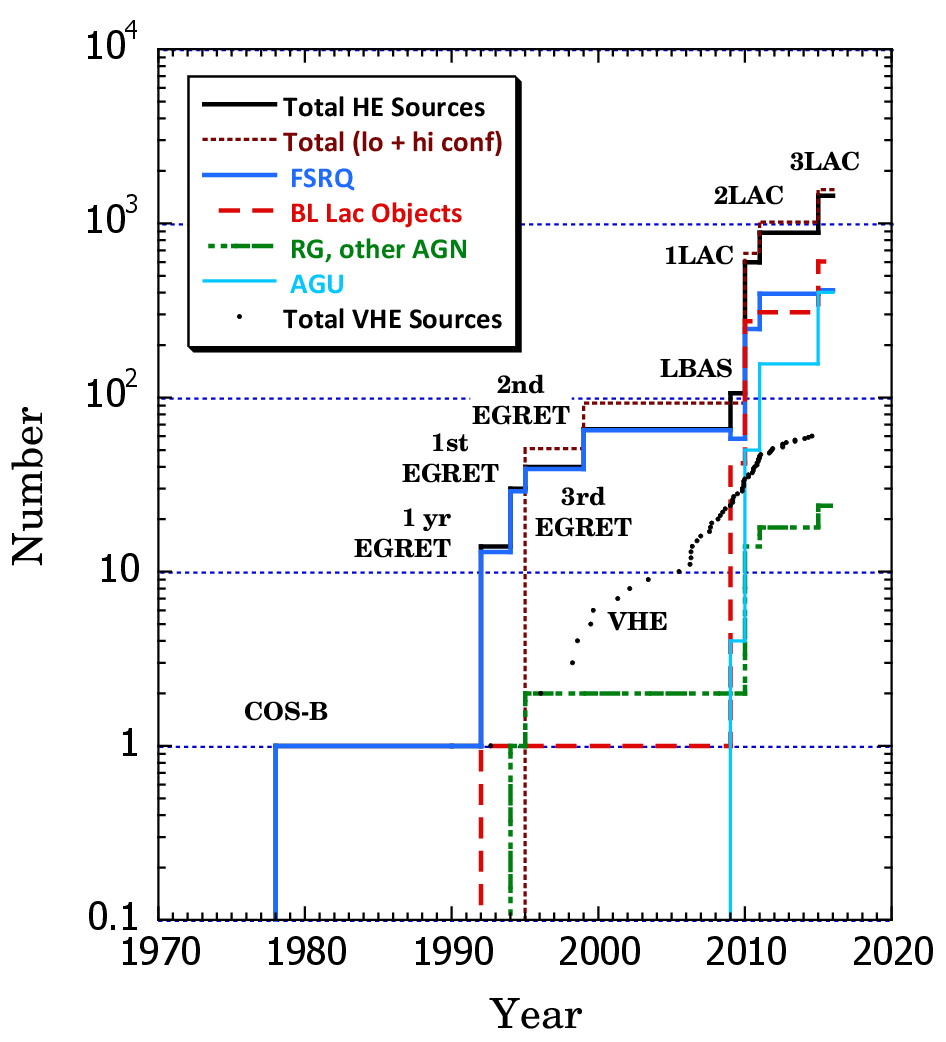}\hskip0.5in
    \includegraphics[width=7.0cm]{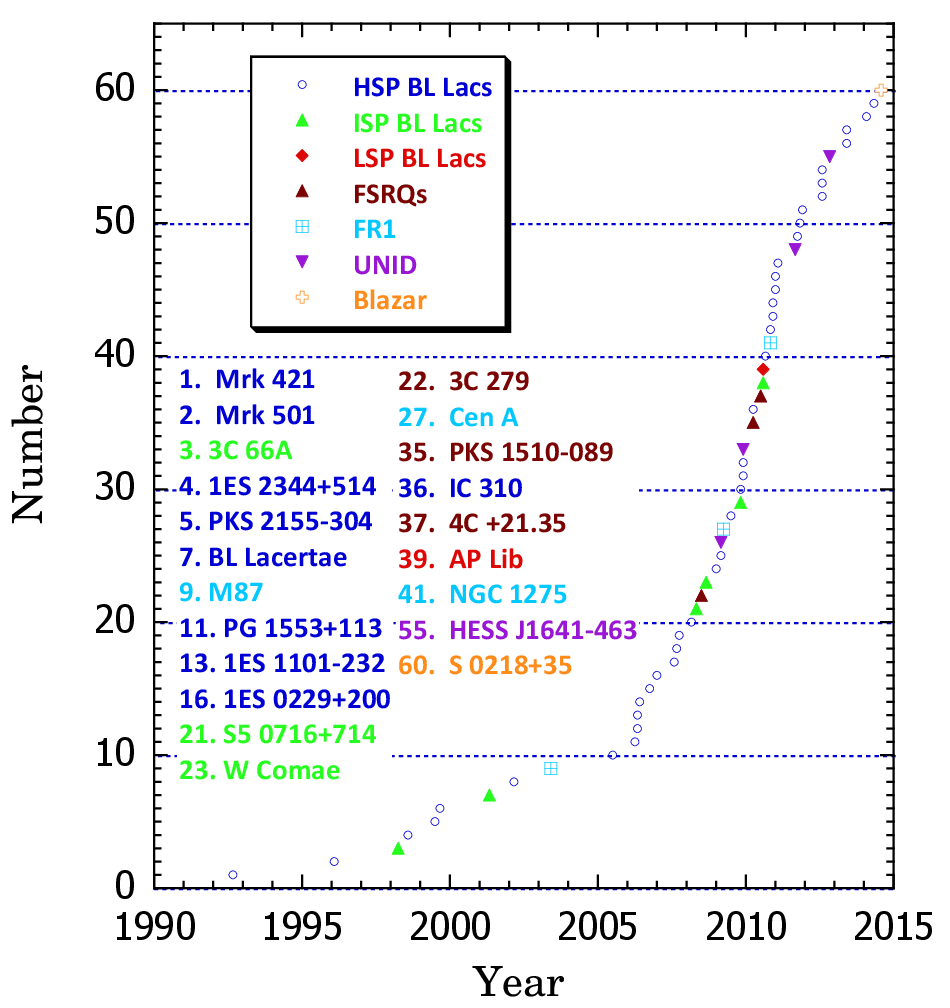}
  \caption{{\bf The growth rate of $\gamma$-ray AGN at GeV and TeV energies.} {\it (a; left):} Growth of
    sources at high-energy (HE), $\gtrsim 100$ MeV, $\gamma$ rays. The number of sources reported in the
    different catalogs and lists, as noted in the figure, are from the 1st EGRET Catalog \cite{fic94}; 2nd
    EGRET Catalog \cite{tho95}; 3rd EGRET catalog \cite{Hartman1999}; LAT Bright AGN Sample (LBAS)
    \cite{LBAS}, First LAT AGN Catalog (1LAC) \cite{1LAC}, 2LAC \cite{2LAC}, and the 3LAC \cite{3LAC}.  {\it
      (b; right):} Number of VHE AGN detected by Atmospheric Cherenkov Detectors (ACTs), obtained using the TeVCat.
    While the brightest and
    nearest AGN have been detected by the previous generation of ACTs such as the Whipple observatory, HEGRA, and CAT, a clear
    break in the discovery rate occurs around 2004 when the new generation of instruments start
    observing. While the bulk of extragalactic VHE sources are BL Lac type blazars (such as Mrk421, the first
    discovered of that type), a few radio-galaxies (M87, Cen A, and NGC 1275) and FSRQ-type blazars (3C279,
    PKS 1510-89, and 4C +21.35) are now also significantly detected. See section 2.1 for the definition of
    other acronyms used in these figures.}
  \label{fig:tevcat}
\end{figure}

The first EGRET pointing towards 3C 273 revealed a bright flaring source, but at the position
of the quasar 3C 279. By the end of the first year of the mission, more than 14 $\gamma$-ray emitting
AGN between  $\sim 100$ MeV and  5 GeV were found.
Most detections were prominent radio-loud quasars, including PKS 0528+134, 3C 454.3, and CTA 102,
but also included the BL Lac object Mrk 421. The strong
connection with apparently superluminal\footnote{When the radiative zone is moving at relativistic velocity
along a direction close to the line of sight, its apparent velocity as measured on the basis of the observer's
proper time may be greater than $c$.} radio sources implied that the $\gamma$ rays come from a nearly aligned
relativistic jet of a black hole  \cite{dsm92}, and the $\gamma$-ray blazar class emerged. By the end
of the CGRO mission, 66 high-confidence and 27 low-confidence detections of blazars had been made, including the radio
galaxy Cen A \cite{Hartman1999}; see Fig.\ \ref{fig:tevcat} left.

During the same time, a major advance in ground-based $\gamma$-ray astronomy took place when the on-off
approach was superseded by the imaging Atmospheric Cherenkov Technique (ACT), \cite{nama}, leading to the significant
detection of the Crab nebula at Very High Energies (VHE; $\gtrsim 100$ GeV) with the pioneering Whipple array
\cite{wee89}. Soon after the recognition that blazars are EGRET sources, Mrk 421 was found to be a VHE source
\cite{pun92}. The VHE discovery of Mrk 501 was reported in 1995, and the pace of discovery has since
quickened, particularly with the introduction of new detectors and arrays.  The largest class of VHE AGN
sources consists of BL Lac objects, which all show a characteristic double-humped spectral energy distribution
(SED) in $\nu F_\nu$ representation\footnote{For the definition and the interest of the SED, see \cite{intro};
$\nu$ is the frequency and $F_\nu$ is the power received per unit area and frequency.}; this structure will be further
described and interpreted in the following
sections. For most of them, the low-energy peak is typically located in the UV and X-ray ranges ($\nu_s
> 10^{15}$ Hz). Soon however radio galaxies (e.g., M87), low-peaked BL Lac objects like AP Lib and BL Lac
itself, and even FSRQs have been detected at VHE; see Fig.\ \ref{fig:tevcat} right.

Extragalactic $\gamma$-ray emitters now constitute more than half of the high-energy (HE; $E>100\,{\rm MeV}$)
emitters identified by the Large Area Telescope (LAT) on the {\it Fermi} Gamma ray Space Telescope \cite{djt}, and are
the second largest population, after pulsar-wind nebulae, in the ground-based VHE regime.\footnote{According
  to TeVCat; see http://tevcat.uchicago.edu} Although radio-loud AGN represent only $\sim 10$\% of all AGN,
the overwhelming majority of known extragalactic $\gamma$-ray sources are radio-loud AGN, where some physical
process ``turns on" the supermassive black hole to make a radio jet. Much speculation has focused on tapping
the energy of rotation through processes occurring in the spinning black hole's magnetosphere.  Plasma
processes in the jet activate a particle acceleration mechanism whose details still elude us, though shock
acceleration and magnetic reconnection are plausible mechanisms.  The interaction of the energetic particles
(whether leptonic or hadronic) with ambient radiation or magnetic fields then generate, through well-known
relativistic radiation physics \cite{dm09}, the copious amounts of $\gamma$ rays that are often observed from
$\gamma$-ray blazars, including extreme states with rapid flux variations, large apparent luminosities, and HE
and VHE $\gamma$-ray photons escaping from very compact volumes.

%% \begin{figure}[!t]
%%   \centering
%%   \includegraphics[width=9.0cm]{myssc.eps}
%%   \caption{{\bf Observational classification of active galaxies.} AGN are subdivided into classes depending on
%%   observational aspects, such as their radio loudness or the presence of atomic lines in their spectra.}
%%   \label{fig:ssc}
%% \end{figure}

%%%%%%%%%%%%%%%%%%%%%%%%%%%%%%%%%%%%%%%%%%%%%%%%%%%%%%%%%
\section{Classification of AGN detected in $\gamma$ rays}
\label{sec:classes}
%%%%%%%%%%%%%%%%%%%%%%%%%%%%%%%%%%%%%%%%%%%%%%%%%%%%%%%%%

Back in 1983, when extragalactic $\gamma$-ray astronomy was in its infancy with a handful of sources, early
classification attempts were made by e.g. Bassani \& Dean \cite{bd83}. The listed sources were divided into a
class of hard X-ray/radio-quiet Seyferts with emission spilling over into the soft $\gamma$-ray band, a class
of MeV radio galaxies, and 3C 273, at redshift $z = 0.158$. This last object is an extraordinary
ultra-luminous infrared galaxy and an AGN with a prominent blue bump, strong radio emissions and episodes of
superluminal motion of radio emitting blobs observed with high-resolution VLBI\footnote{VLBI: Very Long Baseline Interferometry.}
milli-arc-second radio imaging
\cite{1985ApJ...289..109U}. It was this source that heralded the $\gamma$-ray blazar class, whose existence
was perhaps most clearly predicted by K\"onigl \cite{kon81} on the basis of its associated synchrotron and
inverse Compton $\gamma$-ray emission.\footnote{The importance of $\gamma$ rays in extragalactic jet astronomy
  otherwise received little attention at that time \cite{bbr84}.}

Fast-forwarding to 2015, the high-confidence clean sample of the Third Large Area Telescope
Catalog of Active Galactic Nuclei (3LAC) \cite{3LAC} using the first four years of the {\it Fermi}-LAT data
lists 1444 $\gamma$-ray AGN, divided into:
\begin{itemize}
\item $\approx 30$\% FSRQs (404 sources);
\item $\approx 40$\% BL Lac objects (604 sources);
\item $\approx 30$\% blazars of unknown type (402 sources), which are {\it
  Fermi}-LAT high-latitude $\gamma$-ray excesses over the background associated with counterpart radio sources
having inadequate optical follow-up to determine whether the source is a weak-lined BL Lac or strong-lined
FSRQ;
\item $<2$\% non-blazar AGN (a mere 24 sources), which are mainly radio galaxies, radio-loud narrow line Seyfert galaxies, and
  candidate Seyfert AGN.
\end{itemize}

Besides black-hole powered AGN, the other main class of HE and VHE $\gamma$-ray galaxies are the star-forming
galaxies whose high-energy activity is ultimately powered by the kinetic energy of exploding stars. In
addition to the Milky Way, they include the Large Magellanic Cloud, detected with EGRET in 1992 \cite{sre92},
and the Small Magellanic Cloud reported by the {\it Fermi}-LAT Collaboration in 2010 \cite{abd10SMC}.  The
detection of star-forming galaxies outside the Milky Way and its satellites commenced with the joint
announcement of {\it Fermi}-LAT \cite{abd10sb} and VHE detections of NGC 253 \cite{ace09} and M82
\cite{VER09}.  {\it Fermi}-LAT has since announced the detection of Andromeda, but no ultraluminous infrared
galaxy (not counting 3C 273) has been detected at HE or VHE.
%The search for $\gamma$-ray emission
%from Arp 220 finds significant enhancement in the vicinity of
%Arp 220, but not coincident.

The growth rates of the different primary source classes are plotted in Fig. \ref{fig:tevcat} left.
Combined efforts of the VERITAS, MAGIC, and H.E.S.S.\footnote{H.E.S.S.: High Energy Stereoscopic System, in
  Namibia; MAGIC: Major Atmospheric Imaging Cherenkov, in the Canary Islands; VERITAS: Very Energetic
  Radiation Imaging Telescope Array System, in Arizona} ground-based arrays has led to the explosion of new
AGN sources at VHE (Fig. \ref{fig:tevcat} right), importantly assisted by the all-sky capability of {\it
  Fermi}-LAT.

%% From the HE-VHE advances, we can classify, simply, two types of $\gamma$-ray galaxies. First are the {\it
%%   black-hole galaxies}, which divide into the generally distant and powerful blazars with low spatial object densities,
%% and the radio galaxies with weak $\gamma$-ray emissions and correspondingly larger spatial object densities (in
%% comparison with blazars), relating to blazar/radio-galaxy unification discussed below. Second are the
%% generally much weaker and more numerous cosmic-ray galaxies with their $\gamma$-ray emission powered by
%% stellar explosions rather than black holes, which make shocks that accelerate cosmic rays.  Besides most
%% $\gamma$ rays produced by nuclear collisions of cosmic ray to make pions, {\it cosmic-ray galaxies} are also
%% illuminated in $\gamma$ rays by pulsars and pulsar-wind nebulae.  Analysis shows that black-hole galaxies,
%% i.e., blazars and radio galaxies, make the bulk of the HE and VHE EGB, though cosmic-ray galaxies, because
%% they essentially partake in hadronic processes, have comparable neutrino and $\gamma$-ray luminosities. Note
%% how different these two types of $\gamma$-ray galaxies are in comparison with the two types of radio-quiet and
%% radio-loud $\gamma$-ray emitting black-hole galaxies identified in the CGRO days \cite{dg95}.

%%LMC and SMC in VHE?
Excepting the local group galaxies and the starburst galaxies NGC 253 and M82 ,
all other extragalactic VHE sources are AGN, most of them being radio-loud, but a few nearby radio-quiet
Seyfert galaxies.\footnote{As of 2015 May, no GRB has been detected with a ground-based VHE instrument.}
A complex background due to the Fermi bubbles \cite{suel} and possible background blazars
has hampered the search for {\it Fermi}-LAT detection of the famous ultra-luminous infrared galaxy Arp 220.
%%[Berrie--more info on VHE from star-forming galaxies? Arp 220. What's new with HESS?]
%%[no other starburst galaxies to date - recently the superbubble in the LMC, not sure it should be mentioned
%%here since it's in the local group, as the sentence states.]

BL Lac objects with peak synchrotron frequency of the $\nu F_\nu$ spectral energy distribution (SED) at UV and
X-ray energies dominate the VHE sample of AGN.  A smaller fraction of sources, including radio galaxies, have
lower peak synchrotron frequencies. Almost all of the VHE AGN are also detected with the {\it Fermi}-LAT
\cite{3LAC}, allowing complementary studies of the SED over $\approx 6$ orders of magnitude in energy, limited on the
lower end, $\ll 100$ MeV, by poor sensitivity of MeV-regime telescopes, and on the higher end,
$\gtrsim 10$ TeV, by $\gamma$-ray pair production attenuation with optical/IR photons of the extragalactic
background light (EBL).  Because BL Lac objects have weak lines, either due to an intrinsically weak accretion
disk and BLR, or due to the BLR radiation being ``washed out" by beamed nonthermal
radiation \cite{gio12}, a large fraction of BL Lac objects, $\approx 50$\%, do not have precisely measured
redshifts. Attenuation of $\gamma$ rays with the EBL provides a technique to determine redshift \cite{san15},
which can be used with a range of other techniques to constrain redshift \cite{sha13,Ajello2014}.  The
redshift incompleteness problem is a major issue in statistical studies of blazars.

\subsection{Radio-Loud Blazars and Model Outlines}

The prime AGN science for HE and VHE $\gamma$-ray astronomy is the science of radio-loud blazars.  How
radio-loud are these objects? A blazar's {\it apparent isotropic radio luminosity} $L_r = 4\pi d_L^2 \Phi_r$,
where $d_L$ is the luminosity distance\footnote{The luminosity distance $d_L$ is conventionally defined
in such a way that the ratio of the absolute to the apparent luminosity (assuming isotropic emission) be equal to $4\pi d_L^2$.}
and $\Phi_r$ is the energy flux at radio frequencies, can reach
$\approx 10^{39}$ W in powerful blazars such as 3C 454.3 and PKS 1510-089.  Isotropic {\it bolometric} blazar
luminosities dominated by HE $\gamma$ rays can reach apparent values $ > 10^{43}$ W, as in the case of 3C
454.3 during a flaring state in 2010 November \cite{abd11}.  The blazar engine is required to generate large
amounts of power irregularly over remarkably short timescales $\ll R_{\rm S}/c$ (the Schwarzschild radius
light crossing time), collimate the relativistic jet outflow, and amplify the received flux via Doppler
boosting.  Beaming corrections even as small as 0.1\% make the energetics of anything other than an $\approx
10^9 M_\odot$ supermassive black hole infeasible. Note that the radio (i.e., $\lesssim$ 100 GHz) luminosity $L_{r}$ of the Milky
Way is $ \cong 2\times 10^{31}$ W, $\approx 3$~orders of magnitude less than the radio luminosity of the
nearby FR-1 radio galaxy Cen A.
%The bolometric luminosity of BL Lac objects, on the other
%hand, rarely exceeds $L_\gamma\cong 10^{46}$ erg s$^{-1}$.

The two main classes of $\gamma$-ray blazars, namely FSRQs and BL Lacs,
are usually defined according to conventional criteria
according to which a blazar is a BL Lac object if the
equivalent width of the strongest optical emission line is $<5$\AA,
and the optical spectrum shows a Ca II H/K break ratio  $< 0.4$ in order to ensure that the radiation is predominantly nonthermal (the Ca II break arises from old stars in elliptical galaxies).
The {\it Fermi}-LAT collaboration also
introduced \cite{Abdo2010} a new blazar classification, alluded to earlier,
 based on the frequency  $\nu^{\rm
  s}_{\rm peak}$ of the
$\nu F_\nu$ synchrotron SED, with low-synchrotron-peaked (LSP) blazars having $\nu^{\rm s}_{\rm
  peak}<10^{14}\,{\rm Hz}$, intermediate-synchrotron-peaked (ISP) blazars having $10^{14}\,{\rm Hz}<\nu^{\rm
  s}_{\rm peak}<10^{15}\,{\rm Hz}$, and high-synchrotron-peaked (HSP) blazars for sources with $\nu^{\rm
  s}_{\rm peak}>10^{15}\,{\rm Hz}$.\footnote{When redshift is unknown, the measured $\nu F_\nu$ peak synchrotron
frequency is an uncertain factor $(1+z)$ smaller than the rest frame $\nu L_\nu$ peak synchrotron frequency.} The FSRQ blazars are found to have mostly soft $\gamma$-ray
spectra, with $\Gamma \geq 2.2$, and also to be in the LSP blazar class,
indicating that their rapidly falling spectra are unlikely to provide
significant fluxes at TeV energies. On the other hand, the type of BL Lac objects that are strong VHE sources, indicating that
 the peak energy of the $\gamma$-ray component is $\gtrsim 100$ GeV, are primarily HSP blazars.

Understanding the dynamic, broadband multiwavelength and multimessenger
data from blazars depends on an underlying model.
In the Marscher-Gear \cite{mg85} model, sometimes referred to as a ``shock-in-jet" model,
a background plasma ejected along the jet axis, possibly moving outward at relativistic speeds,
supports a shock that deposits nonthermal power in the form of electrons with a power-law spectrum.
The flow magnetic field as a function of radius is parameterized, as is the
electron injection power with radius. From this basis, the total
radio emission or radio intensity as a function of time can be calculated for comparison
with multi-band radio data \cite{fuh14} and interferometric images.

Surely a model like this must apply to outflowing plasma shocked by irregularities in the flow. Nevertheless,
in contrast to the shock-in-jet model, a simpler ``one-zone model" has been widely adopted as the standard
paradigm for modeling the multiwavelength SEDs of blazars. In its simplest form, a spherical ball (in the
fluid frame) of magnetoactive plasma entrains a nonthermal electron distribution. The magnetic field is
assumed to be randomly oriented, and the pitch-angle distribution of particles is isotropic.  Instabilities or
shocks in these plasma blobs are thought to accelerate very energetic particles, which radiate photons whose
flux is amplified in the jet axis along which the magnetoactive plasma travels at highly relativistic
speeds.  The highly Doppler-boosted nonthermal electromagnetic emission is characterized by a double-humped
SED in $\nu F_\nu$ representations, revealing much interesting blazar/black-hole astrophysics.

%These objects are the bulk of the extragalactic $\gamma-$ray emitters, and their numbers have
%grown from the single 3C273 detected by the COS B experiment~\cite{Swanenburg1978}, 66 in the third catalog of
%EGRET sources~\cite{Hartman1999}, to now nearly thousand (and counting) with the currently operating {\it
%  Fermi-}LAT~\cite{Ackermann2011,Acero2015}. While blazars can be further subdivided into different subclasses, the
%current picture seems to indicate that the accretion rate of gas falling onto the black hole might well be at
%the origin of luminosity and peak emission properties~\cite{Ajello2014}.

%The abundance of
%strong atomic emission lines from the broad-line regions in FSRQs also
%make redshift determination possible, whereas a large fraction of
%BL Lac objects lack redshift, amounting to

\subsection{Radio Galaxies}

%%% \begin{figure}[!t]
%%%   \centering
%%%   \includegraphics[width=15.0cm]{f2.eps}
%%%   \caption{{\bf Types of active galaxies.} Cartoon illustration of  AGN taxonomy,
%%% following the unification scheme for radio-quiet and radio-loud galaxies,
%%% {\it cf}.\  \cite{Biteau2013} as adapted from \cite{Urry1995}.
%%% In this two-parameter model, including orientation,
%%% any given AGN is either radio-quiet or radio-loud, here speculated to depend on black-hole
%%% rotation $a/M$, or low power or high power, as determined by the
%%% mass-accretion rate. The misaligned AGN sources are the radio galaxies, including
%%% the low luminosity FR1 counterparts to BL Lac objects,
%%% and high luminosity FR2 radio galaxies, which divide into narrow- and broad-line radio
%%% galaxies, depending on orientation.}
%%%   \label{fig:unification}
%%% \end{figure}

According to the unification scenario \cite{Urry1995}, as sketched in Fig.\ \ref{fig:unification}, radio
galaxies are the misaligned counterparts of blazars. Conversely, blazars are radio galaxies in the
circumstance when we happen to be looking down a black hole's jets. Therefore, radio galaxies and blazars are
the same objects viewed from different directions. To complicate things further, the low-luminosity {\it
Fanaroff and Riley} \cite{fr74} FR1 radio galaxies showing twin-jet morphology are believed to be the
misaligned counterparts of BL Lac objects, whereas the high radio-luminosity FR2 class of radio galaxies
showing lobes and hot spots are thought to be the counterparts to the FSRQs, with a dividing line at a luminosity
$L_{rad}\sim 10^{35}$ W. Establishing the underlying relationships between the aligned blazar and misaligned
radio-galaxy classes is difficult because Doppler boosting of the jetted blazar emission means that blazars do
not follow Euclidean behavior\footnote{If all sources with the same intrinsic isotropic luminosity were distributed uniformly
in Euclidean space, the number $N$ of sources with an apparent luminosity greater than $S$ would vary like $S^{-3/2}$; for
blazars, the slope of the cumulative distribution of $\log N$ versus $\log S$ does
not have an index of -1.5.} either at low, moderate, or high redshifts. In the local universe, blazars are
rare, with (an apparent) space density of BL Lac objects no greater than $\sim (10^2$ -- 10$^3$) Gpc$^{-3}$,
whereas the density of the misaligned counterparts can be $\sim 10^3$ times greater, depending on the Doppler beaming
factor
\footnote{When the angle $\theta$ between the velocity {\boldmath${\beta}$}$c$ of the radiative zone and the
  line of sight $\mathbf{n}$ is small enough, the frequency of the jet emission, which relativistic aberration confines
  within a half-angle of $\sim 1/\Gamma$ radians from {\boldmath${\beta}$}
  (with the Lorentz factor $\Gamma = (1-\boldmath{\beta}^2)^{-1/2}$),
  is seen by the observer multiplied by a
  Doppler beaming factor $\delta=[ \Gamma (1-\mathbf{\beta} \cos \theta)]^{-1}$. This enhances significantly the
  observed flux by a factor $\sim \delta^4$ compared to what is expected from isotropic $1/r^2$
  flux-dependency.}.
%  line of sight $\mathbf{n}$ is small enough, the jet emission, which relativistic aberration confines
% within a half-angle of $\sim 1/\Gamma$ radians from (\boldmath${\beta}$)
 %(with $\Gamma = (1-\boldmath{\beta}^2)^{-1/2}$),
%  is seen by the observer with a
%  Doppler beaming factor $\delta=[ \Gamma (1-\boldmath{\beta} \cdot \mathbf{n} )]^{-1}$ which enhances significantly the
%  observed flux by a factor $\sim \delta^4$ compared to what is expected from isotropic $1/r^2$
%  flux-dependency.}.

The first {\it Fermi}-LAT article \cite{MAGN} on  misaligned AGN (MAGN)
examined sources from the first LAT AGN Catalog (1LAC) \cite{1LAC} associated with steep
(flux density $F_\nu \propto \nu^{-\alpha_r}$, with $\alpha_r > 0.5$) 178 MHz radio-spectrum
objects in the Third Cambridge (3CRR) and Molonglo radio  catalogs that show extended radio structures in radio maps.
Because of the close relation between core dominance and $\gamma$-ray spectral properties,
core dominance can be used to infer the alignment of radio-emitting AGN  \cite{lh05}.
The original {\it Fermi}-LAT MAGN population comprises 11 sources, including
 7 FR1 radio galaxies, namely, 3C 78 (NGC 1218), 3C 84 (NGC 1275), 3C 120, M87,
Cen A, NGC 6251, and PKS 0625-354, two FR2 radio galaxies (3C 111 and PKS 0943-76),
and two steep spectrum radio quasars (3C~207, 3C~380) that are thought to be slightly misaligned FSRQs.

\begin{figure}[!t]
  \centering
  \includegraphics[width=7.5cm]{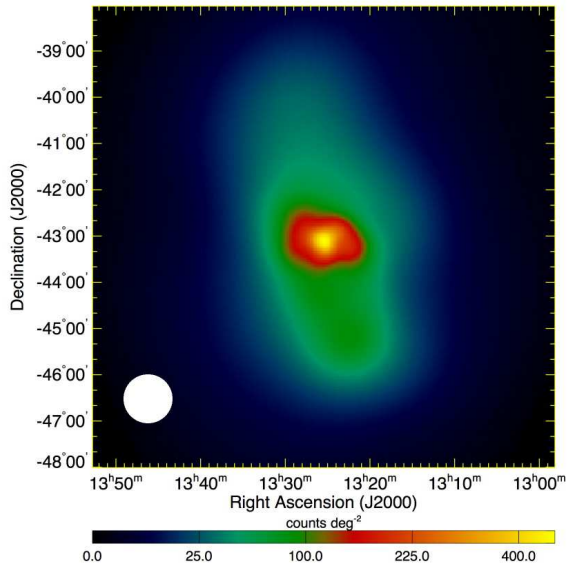}
  \caption{{\bf The first $\gamma$-ray map of a radio galaxy.} {\it Fermi}-LAT $> 200$ MeV map of the radio galaxy Cen A, after subtracting
point sources and the Milky Way's Galactic diffuse emission  \cite{abd10cena}.  The angular extent of the $\gamma$-ray emission is $\approx 10^\circ$.
At a distance of $\approx 3.5$ Mpc, these structures span $\approx 600$ kpc.  }
  \label{fig:cena}
\end{figure}

\begin{figure}[!b]
\vskip0.2in
  \centering
  \includegraphics[width=3.4cm]{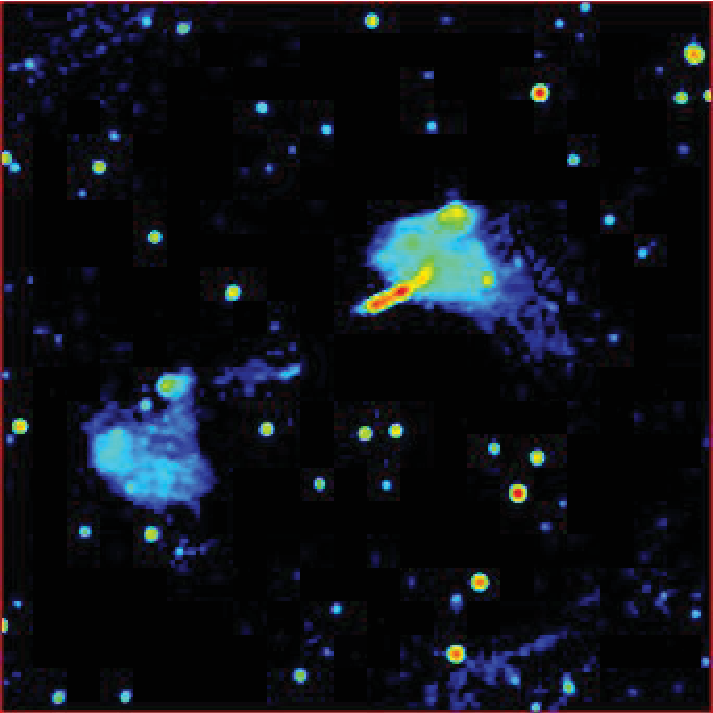}\hskip0.4in
  \includegraphics[width=3.7cm]{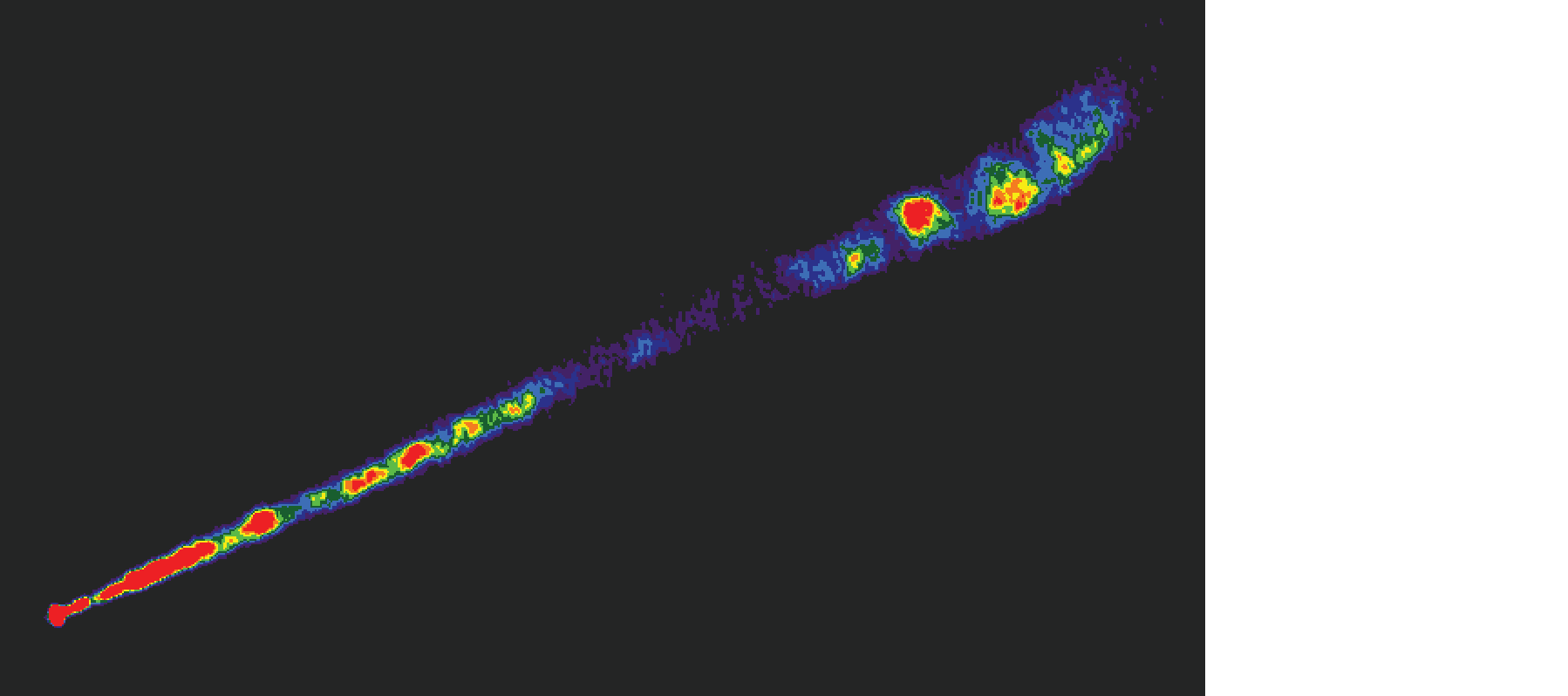}\hskip1.2in
  \includegraphics[width=4.6cm]{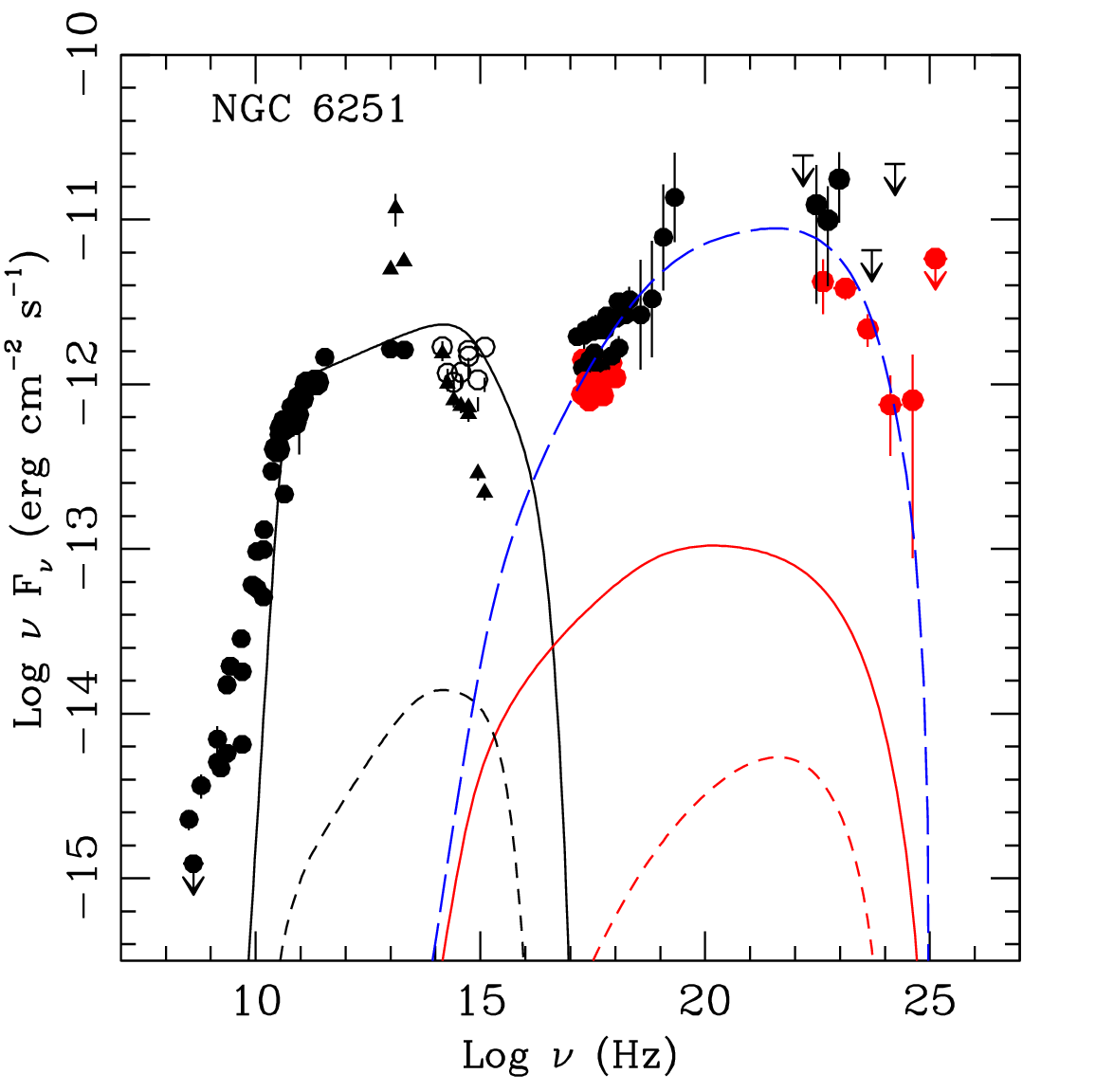}
  \caption{{\bf Three views of the same object.} {\it Left}: Radio image of NGC 6251 \cite{2001ASPC..250..294W}. The image size spans 72.5$^\prime$, which corresponds to a projected distance of
$\approx 2.0$ Mpc for an angular diameter distance of $\approx 98$  Mpc. {\it Center}: High resolution VLA image of the inner jet of NGC 6251, spanning a projected distance of $\sim 4$ kpc \cite{werner}. {\it Right}: Multiwavelength spectrum of NGC 6251, with synchrotron self-Compton model fit \cite{mig11}.
}
  \label{fig:ngc6251}
\end{figure}

The MAGN sources in the 3CRR catalog have large core dominance parameters compared to
the general 3CRR source population, implying that the beamed component makes an
appreciable contribution to the $\gamma$-ray flux. This is furthermore supported by
the fact that four of these sources---3C 78, 3C 111, PKS 0943-76, and 3C 120---do not appear in 2LAC \cite{2LAC},
evidently due to variability. By the time
of the 3LAC, the population of MAGN has nearly doubled. One new source is
IC 310, a head-tail radio galaxy or slightly off-axis FR1 radio galaxy \cite{2012A&A...538L...1K}
in the Perseus cluster with HSP BL Lac properties,
that was detected by  MAGIC \cite{ale10} after being alerted of a {\it Fermi} high-energy excess \cite{ner10}.
Three others are Fornax A, an FR1 radio galaxy, and Pictor A and 3C 303, both  FR2 radio galaxies.  Fornax A and Pictor A were widely suspected of being sources of high-energy $\gamma$ rays (e.g., \cite{geo08}).
Cen A is the first mapped radio galaxy at GeV energies, as shown in Fig.\ \ref{fig:cena} \cite{abd10cena}. The second
extended radio galaxy at HE could be NGC 6251 \cite{mig11}, given its extended radio structure (at z = 0.0247 and a luminosity distance $d_l \cong 103$ Mpc, the angular extent of its radio emission is $\approx 0.8^\circ$; see Fig.\ \ref{fig:ngc6251}).
%NGC 4945 is thought to be a starburst powered.

Radio-loud Narrow Line Seyfert 1 (RL-NLS1) galaxies have also been  established as a $\gamma$-ray source class \cite{abd09c}. These objects show narrow H$\beta$ lines with FWHM line widths $\lesssim 2000$ km s$^{-1}$, weak forbidden lines ($[OIII]/H\beta < 3$)
and a strong Fe II bump, and are therefore classified as narrow-line type I Seyferts \cite{pog00}. By comparison with the $\sim 10^9 M_\odot$ black holes in blazars, the host galaxies of RL-NLS1s are spirals that host nuclear black holes with relatively small ($\sim 10^6$ -- $10^8 M_\odot$) mass that accrete at a high Eddington rate. The detection of these objects challenges scenarios where radio-loud AGN are hosted by elliptical galaxies that form as a consequence of galaxy mergers \cite{Bottcher2002}.

The misaligned AGN and radio galaxies now found to be HE and VHE emitters
 make it clear that there are particle acceleration and $\gamma$-ray emission
sites far beyond the inner radio jet.
 The nearby giant radio galaxy M87 was the first non-blazar extragalactic VHE
$\gamma$-ray emitter~\cite{Aharonian2003}, followed later by Centaurus A and IC 310.  While only
Cen A was detected
by EGRET, {\it Fermi}-LAT has found HE counterparts to all VHE radio galaxies and, in addition,
discovered other HE-emitting radio galaxies~\cite{Abdo2010,Acero2015}. The large radio structures in Cen A
exceeds the field of view of IACTs, challenging conventional VHE detection that searches for point
sources of radiation. The CTA observatory however, with an order of magnitude improvement in sensitivity and
better angular resolution, should greatly improve our knowledge of radio structures with extended VHE emission.

While Cen A does appear to have rather stable VHE emission, M87 and IC 310 have exhibited large and rapid
flux variations. All current major IACT observatories are able to observe M87 quasi-simultaneously, and a joint
effort was led in 2010 to search for variability on timescales shorter than a day, which is of the
order of the light crossing time of the Schwarzschild radius for the supermassive central black
hole in M87. Multiwavelength signatures appear to be different from previous occurrences of VHE flux enhancement, but
an interesting pattern emerged, namely  a very significant X-ray flux enhancement of the core during two
such episodes \cite{Abramowski2012}. Even faster flux variability was found in the radio galaxy IC 310, on
time scales of a few minutes \cite{Aleksic2014}, shorter than the light crossing time across the event
horizon. This has previously only
been seen in HSP BL Lac objects.
%Large Doppler factors of the emission zone, often invoked
%for blazars, seem to be difficult to apply in this case.
%%WHY?

%%[Berrie: Please expand discussion below of radio galaxies with VHE; inc. M87, incl. X-ray/VHE correlated
%%flaring; Cen A; NGC 1275; others? (already mentioned IC 310)]
%%[Any VHE mapping of radio galaxies expected with HESS or CTA?]

\subsection{Other Galaxies}

A keen interest is taken in both the HE and VHE regimes to uncover evidence of $\gamma$-ray emissions from
classes of AGN other than those in the jet set (see Fig.\ 2). These ``jet-deprived" AGN are not decorated with
enormous radio displays. By lacking the jets, however, the less complex environment may allow a deeper view
into the interior regions using optical and X-ray probes. Classes of X-ray missions, from the currently operating
Chandra to the future Athena and ASTRO-H, provide deep insights into the inner accretion-disk geometry and
behavior.

Out of the 24 ``Other AGN" in the {\it Fermi}-LAT 3LAC clean sample \cite{3LAC}, 13 are radio galaxies,
discussed above, 5 are RL-NLS1 galaxies, 3 are steep spectrum radio quasars, and one (4C $+39.23$B) is a
compact steep spectrum radio source (CSS), a category for which $\gamma$-ray emission has been predicted
\cite{2008ApJ...680..911S}. The low-latitude radio/$\gamma$-ray source PMN J1603-4904 may be another
young, $\gamma$-ray emitting compact symmetric radio source \cite{2014A&A...562A...4M}.
%and besides CSS in
%the 3LAC, leaving 2 other ``AGN" in the 3LAC clean sample, and 6 other ``AGN" in the entire 3LAC sample of
%1591 sources.
The remaining two "AGN" out of the 24 are GB 1310+487 and PMN J1118-0413. Circinus, a Seyfert
2 galaxy, is located at $b= 3^\circ$, so is not technically part of the LAT AGN Catalog samples because of the
restriction to high ($|b|> 10^\circ$) Galactic latitudes, though it would be another example of a radio-quiet
galaxy possibly detected at $\gamma$-ray energies.  Earlier {\it Fermi}-LAT searches for emissions from
radio-quiet Seyferts \cite{ack12}, based on 120 bright hard X-ray radio-quiet Seyfert galaxies, obtained
marginal detections of NGC 6814 and ESO 323-G077, neither of which is reported in the 2LAC and
3LAC. Incidentally, PKS 0943-76 from the MAGN catalog remains out of the 3LAC source list.  The $\gamma$-ray
emission from other candidate radio-quiet Seyfert galaxies or other AGN might be attributed to cosmic-ray
interactions rather than radiations related to the central black-hole, but variability studies must contend
with weak fluxes.

%%%%%%%%%%%%%%%%%%%%%%%%%%%%%%%%%%%%%%%%%%%%%%%%%%%%%%%%%
\section{AGN observed at High Energies ($>0.1\,{\rm GeV}$) and Very High Energies ($>100\,{\rm GeV})$}
\label{sec:vhe}
%%%%%%%%%%%%%%%%%%%%%%%%%%%%%%%%%%%%%%%%%%%%%%%%%%%%%%%%%

At high- and very-high energies, $\gamma$-ray emission is thought to be a by-product of
black-hole nuclear activity that expels  oppositely directed
relativistic jets of plasma along the rotational axis of a spinning
black hole. Most AGN activity is fueled by accretion, but
the black hole's spin, which would be a consequence of the formation history
of the galaxy and its central pc, could explain why only a small fraction,
$\sim 3$\%, of AGN have observable jets \cite{bla90,dm09}.
Morphological studies of the host galaxies of blazars
are difficult because of the brightness of the nuclear light
and the generally large redshifts, $z\gg 0.1$, of the
host galaxies, particularly for FSRQs.

\subsection{Multiwavelength SEDs of $\gamma$-ray blazars}

\label{sec:multiwavelength}

Confining our attention to nonthermal jet radiation powered by black-hole activity, the most widely remarked feature of
the broadband SEDs of $\gamma$-ray sources powered by black-hole activity is their two-humped shape.  The
characteristic double-humped SEDs of blazars extend over 15 or more decades in energy, from radio to TeV
$\gamma$ rays. Since blazars are intrinsically variable, observations are most valuable when they occur as
nearly simultaneously as possible, in order to give an accurate picture of the AGN's emission across all
wavelengths. Establishing an accurate SED is important to determine the relative importance of the different
components in the continua, while the correlation of the variability in different wavebands gives insights
into the jet physics. Multiwavelength observations are also sometimes the only available means to infer the
nature of unidentified $\gamma$-ray sources, either through the characterization of the SED or, when a
 counterpart is found at different wavelengths, to correlate variability with the $\gamma$-ray flux variability.
The high quality long baseline {\it Fermi}-LAT light curves for scores of blazars is a tribute to its large field of view and
scanning strategy.

The most constrained instruments for multiwavelength (MWL) observations are usually the
ground-based optical and VHE telescopes, which provide critical information on the radiative particles since
the optical emission is often near the peak emission of the first hump in the SED, while the second probes
directly the most energetic emission. The nonthermal optical emission is however often contaminated with
optical emission from the host galaxy for objects with $z\lesssim 0.3$, while the $\gamma$-ray emission is
attenuated through interactions with the cosmological optical and infrared backgrounds. The flux attenuation
increases with distance and energy \cite{Horns2015}.

\begin{figure}
  \centering
  \includegraphics[width=7.3cm]{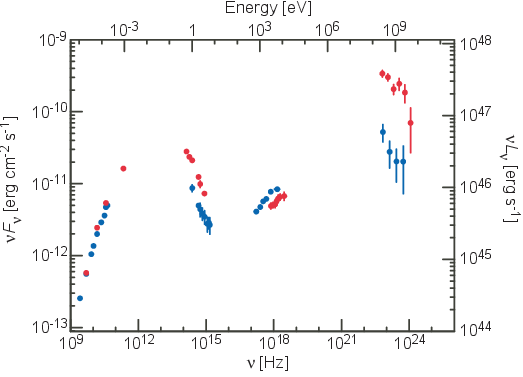}\hskip0.3in
  \includegraphics[width=7.5cm]{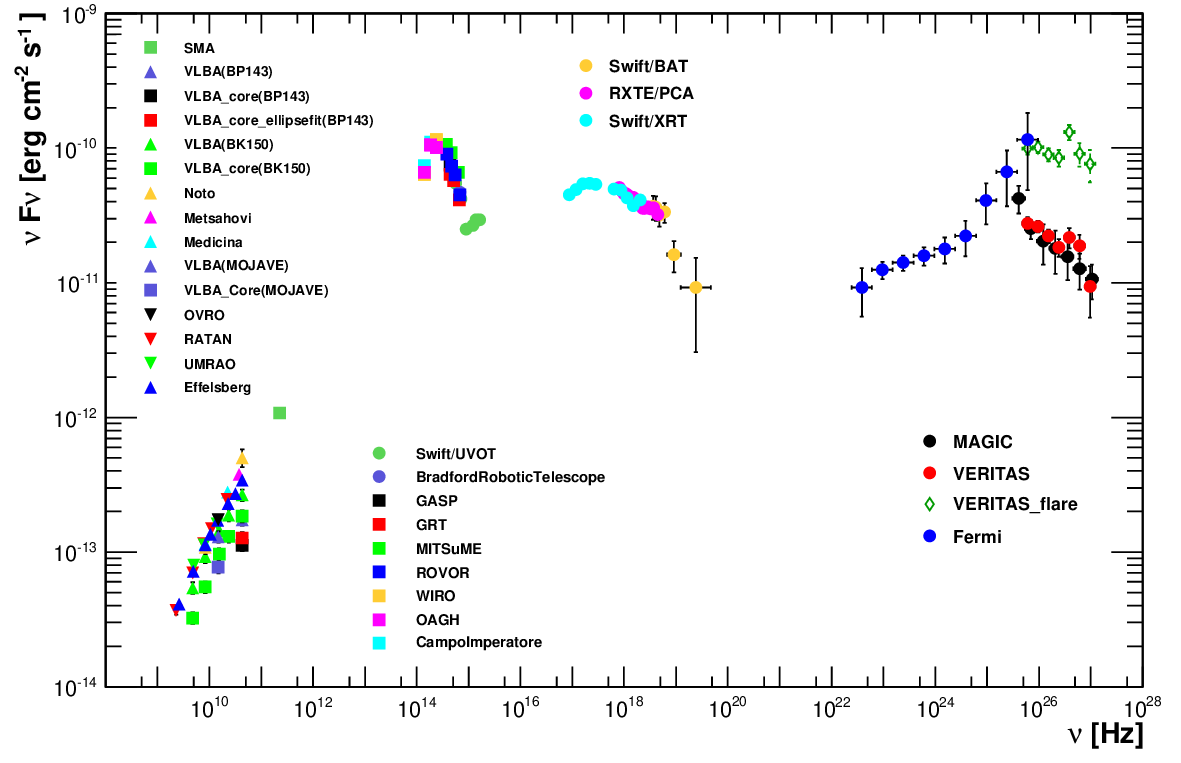}
  \caption{{\bf Multiwavelength blazar SEDs.}   {\it Left}: Data shows two
epochs of 3C 279 \cite{abd103C279}.  The red points
were measured between 54880 and 54885 MJD, when
a $\gamma$-ray flare was accompanied by a rapid
change in optical polarization.
{\it Right}: SED for Mrk 501 averaged over all observations taken during the
multifrequency campaign performed between 2009 March 15 (MJD 54905) and 2009
August 1 (MJD 55044) \cite{abd11mrk501}. The high-state flux detected by VERITAS
from MJD 54952.9 -– 54955.9 is shown by the green points, and this epoch is subtracted
from the average SED. The legend reports the correspondence
between the instruments and the measured fluxes. Galactic emission has not been
removed. }
  \label{fig:blazarsed}
\end{figure}

Fig.\ \ref{fig:blazarsed} shows the SEDs of two well known blazars, the FSRQ 3C 279 on the left and the HSP BL
Lac object Mrk 501 on the right. It is also convenient to convert $\nu F_\nu$ into its equivalent in absolute luminosity: $\nu L_\nu
= 4\pi d_L^2 \times \nu F_\nu$, in which $d_L$ is the luminosity distance.
The distance modulus $D_l = \log (4\pi d_L^2)$ provides a simple conversion from  $\nu F_\nu$ to
$\nu L_\nu$ in a log-log representation\footnote{Since c.g.s. units are used in most AGN SED's in the literature, $d_L$ is here expressed in cm.}.
The redshift of 3C 279 is $z = 0.538$, and during this epoch of observation
the peak synchrotron frequency (in the source frame) is $\nu_s \cong 10^{13}$ Hz. The value of $\nu L_\nu$ at
$\nu = \nu_s$ is $\nu_s L_{\nu_s} \cong 5\times10^{39}\,{\rm W} = 5\times 10^{46}$ erg s$^{-1}$, implying a bolometric isotropic synchrotron
luminosity (supposing only a $1/d^2$ flux dependency) a factor $\approx 3$ higher.  The redshift of Mrk 501 is
$z = 0.03366$, implying a distance of $\approx 149$~Mpc and a distance modulus $D_l=54.42$.  Its peak synchrotron frequency $\nu_s \cong 10^{17}$ Hz and, from Fig.\ \ref{fig:blazarsed} and
$D_l$, $\nu_s L_{\nu_s}^{pk} \cong 10^{37}\,{\rm W}$ ($10^{44}$ erg s$^{-1}$), so that the bolometric synchrotron flux is a factor
$\approx 5\times$ larger (the width of the synchrotron SED in Mrk 501 is wider than that of 3C 279).
%The redshift of Mrk 421 is $z = 0.30$, implying a distance of $\approx 130$ Mpc and distance modulus $D_l =
%\log (4\pi d_L^2) = 54.3$.  Its peak synchrotron frequency $\nu_s \cong 10^{17}$ Hz and, from
%Fig.\ \ref{fig:blazarsed} and $D_l$, $\nu_s L_{\nu_s}^{pk} \cong 10^{44}$ erg s$^{-1}$, so that the bolometric
%synchrotron flux is a factor $\approx 5$ larger.
Although the ratio of the synchrotron powers for the two blazars is $\sim 50$, the ratio
of the $\gamma$-ray powers can be $\gtrsim 10^3$ because of the large Compton dominance of
the FSRQ's SED in $\gamma$ rays.

Examination of the SEDs in Fig.\ \ref{fig:blazarsed} shows that several issues
need clarification. Considering only AGN emission requires that any residual galactic
continua be removed, as seen in the Mrk 501 SED that contains a strong IR feature from old stellar populations
of the host galaxy. The examination of the correlated activity of the nucleus requires, in addition
to simultaneity, either imaging or variability to identify these emissions themselves.
Where imaging is not possible, timing steps in, with the presumption that the most luminous
and highly variable radiations can only be produced by the nuclear black hole.
For SED correlation studies measured over specified time windows, the most
important observables are the measured
frequencies $\nu_{pk}^s$ at the peak of the $\nu F_\nu$ synchrotron component,
and $\nu_{pk}^{\rm C}$ at the peak of the $\nu F_\nu$ inverse Compton component.
In the source frame for a known redshift $z$, the peak frequencies are
$\nu_s=(1+z) \nu_{pk}^s$ and $\nu_{\rm C}=(1+z) \nu_{pk}^{\rm C}$.
The corresponding values $\nu L_\nu^s$ (respectively $\nu L_\nu^{\rm C}$)
at peak synchrotron (respectively Compton) frequency are then deduced.
The variability time $t_{var}$ is crucial to spectral modeling, but is often the most
elusive quantity to measure, because variable components may be hidden beneath a slowly
varying continuum, and the variability at different energy ranges can be very different.
Other important observables are the width and structure of the various components of the SED.
As can be seen from the 3C 279 SED, the extrapolation of its X-ray emission to the $\gamma$-ray
band is discontinuous, implying multiple emission components. In the simplest case, the same electrons would be responsible for
the synchrotron emission and for the inverse Compton effect by upscattering their own synchrotron radiation
(synchrotron-self Compton or SSC process). However, accurate modeling of
detailed SEDs of FSRQs require both synchrotron self-Compton and one or more external Compton components.

Even in the absence of a blazar SED model, studies of the statistics of many blazars can be
used to correlate the observables noted above. These correlations test
blazar SED models and suggest relationships between different blazar classes.
Significant blazar correlations have been established
from statistical studies, as described in Section~6.

\subsection{Multiwavelength SEDs of $\gamma$-ray radio galaxies}

Most radio galaxies are of the FR1 class, including the radio galaxies detected at $\gamma$-ray energies, in
which case their aligned counterparts are BL Lac objects, according to the unification scenario
\cite{Urry1995}.  By comparison with FR2 and FSRQ galaxies, these are the simplest ``one-zone" configurations,
consisting of a ball of magnetized plasma entraining nonthermal electrons making synchrotron radiation together with
SSC X-rays and $\gamma$ rays. Indeed, early studies of radio quasars concluded that the lack of
self-Compton X-rays was inconsistent with the assumption of a stationary emitting region or, more radically,
either the nonthermal synchrotron interpretation for extragalactic radio sources was incorrect or redshifts
were not cosmological. Emission regions moving at relativistic speeds preserved the cosmological
interpretation of quasars. These plasma jets require black-hole engines, and furthermore provide the energy to
form, over cosmic timescales, the extended radio lobes of radio galaxies.

\begin{figure}[!t]
%\vskip0.1in
  \centering
  \includegraphics[width=7.3cm]{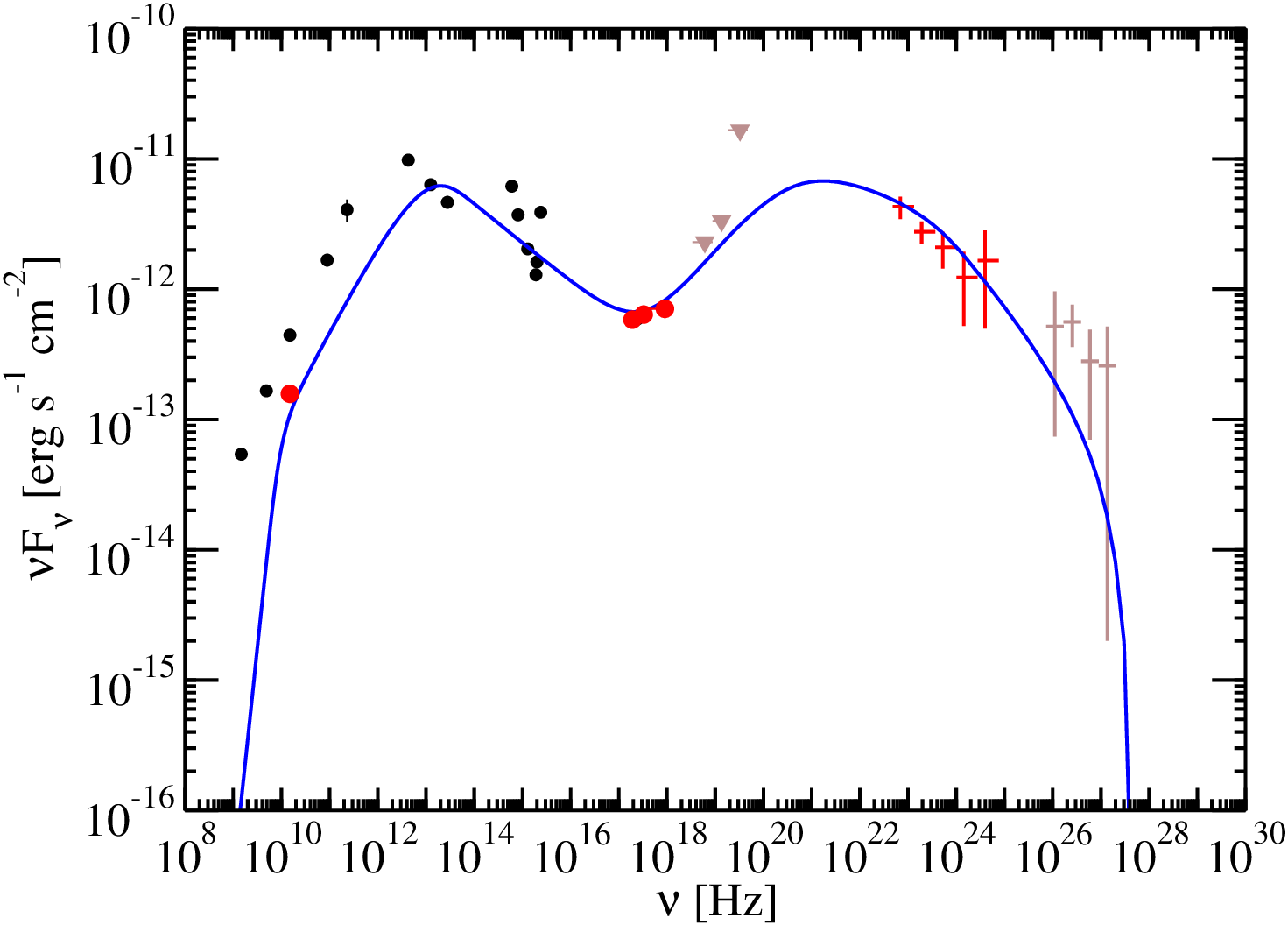}\hskip0.4in
  \includegraphics[width=7.0cm]{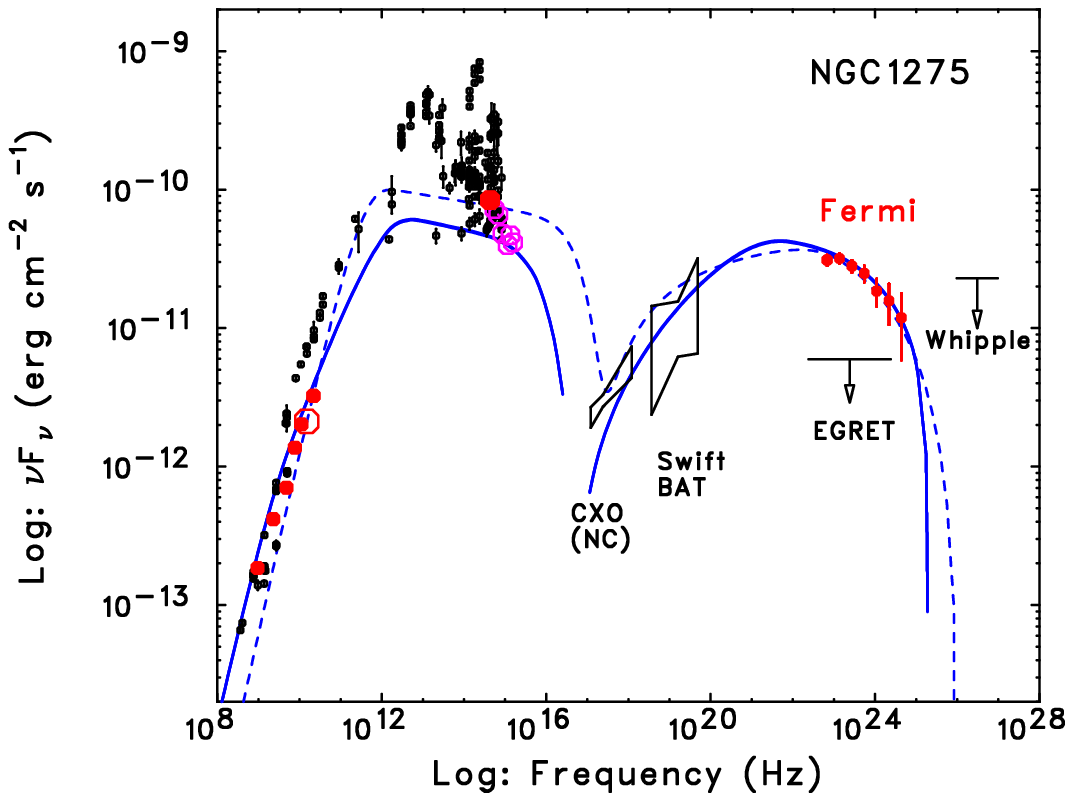}
  \caption{{\bf Multiwavelength radio-galaxy SEDs.}   {\it Left}: SED data of  M87 (where the TeV data from
    the HESS experiment are not contemporaneous with the {\it Fermi-}LAT data)  and one-zone model fit
 \cite{abd09M87}.
{\it Right}: SED data for NGC 1275, including one-zone model fits \cite{abd09NGC1275}.    }
  \label{fig:radiogalaxysed}
\end{figure}

Fig.\ \ref{fig:radiogalaxysed} shows data for M87 \cite{abd09M87} and NGC 1275 \cite{abd09NGC1275}, along with
spectral models of these radio galaxies. Fig.~\ref{fig:ngc6251} also shows one-zone modeling for NGC 6251
\cite{mig11}.  An interesting result of SED modeling of radio galaxies is that their
radio--through--$\gamma$-ray SEDs are generally well fit with large physical-dimension, slowly varying radio-emitting
plasma moving at mildly relativistic speeds, with bulk Lorentz factor $\Gamma \sim 1.5$ -- 2. Such small beaming requirements
are at odds with the larger values of $\Gamma \sim$ 10 -- 30, more typical of those derived for the radiating relativistic jet
plasma through BL Lac blazar spectral modeling, and suggest that emission regions with different velocities exist.
Because of the powerful beaming $\propto \delta^{6}$ for external Compton
radiations\footnote{$\delta$ is the Doppler beaming factor mentioned above.}, the central rapidly moving jet emission is hardly seen at large
angles from the jets. Instead, the $\gamma$-ray emissions from radio galaxies might be made
in structured layers
of the jet as in \cite{henri06} and in the spine-sheath model \cite{gtc05}, or could result from an extended emission region with
varying Doppler factor as a result of jet deceleration \cite{gk03}. See Ref.\ \cite{mig11} for
spine-sheath modeling of NGC 6251.

%%%%%%%%%%%%%%%%%%%%%%%%%%%%%%%%%%%%%%%%%%%%%%%%%%%%%%%%%
%%% \section{The Blazar/Radio Galaxy Paradigm}
%%% \label{sec:parad}
%%%%%%%%%%%%%%%%%%%%%%%%%%%%%%%%%%%%%%%%%%%%%%%%%%%%%%%%%

%\subsection{Unification of radio galaxies and blazars}

%%%%%%%%%%%%%%%%%%%%%%%%%%%%%%%%%%%%%%%%%%%%%%%%%%%%%%%%%
\section{Variability Properties of $\gamma$-ray Blazars}
\label{sec:var}
%%%%%%%%%%%%%%%%%%%%%%%%%%%%%%%%%%%%%%%%%%%%%%%%%%%%%%%%%

The dimension of time introduces all sorts of complexities
into blazar physics. Indeed, it was the rapidly
variable radio fluxes, on time-scales of months,
 that introduced  the blazar puzzle \cite{1966ApJ...146..634P,1973ApJ...186..791J}.
Blazars, as the
name implies, are highly variable sources, yet their variability properties are
 frequency- and class-dependent, and highly resistant to
 the identification of any simple underlying patterns or behaviors.
First is the question of what constitutes variability. The likelihood of
fluctuations away from an average value
is something that can be determined quantitatively from
a given data set.  Auto-correlation analyses can identify
preferred time scales. Cross-correlation analyses can constrain
jet models and identify consistent particle acceleration
and radiation processes. Predictions about correlated emissions
or changes in fluxes at different frequencies test jet models.

\subsection{$\gamma$-ray variability}

Day-scale $\gamma$-ray variability of blazars was known from the EGRET era,
most notably from the 1991 August flares of 3C 279 \cite{1993ApJ...411..133K}.
{\it Fermi}-LAT, with its better sensitivity, can probe variability on much
shorter time scales.
But for only a few  blazars does {\it Fermi}-LAT have sensitivity to probe to a few hour time
scale during major outbursts, namely  3C 454.3 \cite{2011ApJ...733L..26A},
PKS 1222+216 \cite{2011ApJ...733...19T},
PKS 1510-089 \cite{2013ApJ...766L..11S,2013MNRAS.431..824B},
and 3C 273 \cite{2011A&A...530A..77F,2013MNRAS.430.1324N}. These few
hours are in the ball park of simple expectations for minimum
variability time scales for supermassive black holes.
Only recently has evidence for variability at sub-ks timescales
at GeV energies been found in flares
of 3C 279 and PKS~1510-089~\cite{Lott14}.

Most simply, the natural scale of the blazar engine is the Schwarzschild radius $R_{\rm S}$ of a black hole with mass $M_{bh}$,
corresponding to a light-crossing time of:
\begin{equation}
{R_{\rm S}/ c} \,\cong\, 10^4 \; {\rm s}\; \times M_9 \: \mbox{where} \: M_9 = M_{bh}/10^9M_\odot\;.
\label{RSc}
\end{equation}
Supermassive black holes with masses $\approx 10^9 M_\odot$ are supposed to power most blazars; black-hole
masses in blazars are obtained, e.g., by the bulge/black-hole mass relationship \cite{2000ApJ...539L...9F}.  BL Lac objects, for
instance, are almost entirely hosted in otherwise normal elliptical galaxies \cite{urr00}, so that the bulge
mass is essentially equal to the host galaxy's normal matter mass, implying $\sim 10^9M_\odot$ black holes at
their centers.  It seems reasonable to expect a loss of power at smaller timescales than $\cong 3$ hrs, as
flaring on time scales shorter than $R_{\rm S}/c$ should reduce the size of the implied emission volume and
with it, the radiant luminosity.  Yet powerful flares varying on timescales $\sim 100\times$ shorter than
implied by eq.\ (\ref{RSc}) have been detected in VHE; mainly from HSP BL Lacs.

Hints of ultra-rapid variability in blazars were evident in early Whipple data
of Mrk 421 \cite{1995ApJ...449L..99M}, and in data from successor VHE telescopes preceding the
era of $\gamma$-ray Cherenkov  arrays. But it was with the operations
of the H.E.S.S., VERITAS and MAGIC VHE $\gamma$-ray observatories that measurements of
extremely rapid variability of BL Lac objects began to threaten simple
kinematic understanding of black-hole physics, presenting a puzzle that still
remains today. Variability on time scales as short as $\approx 5$ min in PKS 2155-304  \cite{Aharonian2007b},
and a few minutes in  Mrk 501 \cite{2007ApJ...669..862A} and Mrk 421 \cite{2012AIPC.1505..514F},
has now been reported by VHE telescopes.

\begin{figure}[!t]
  \centering
  \includegraphics[width=8.5cm]{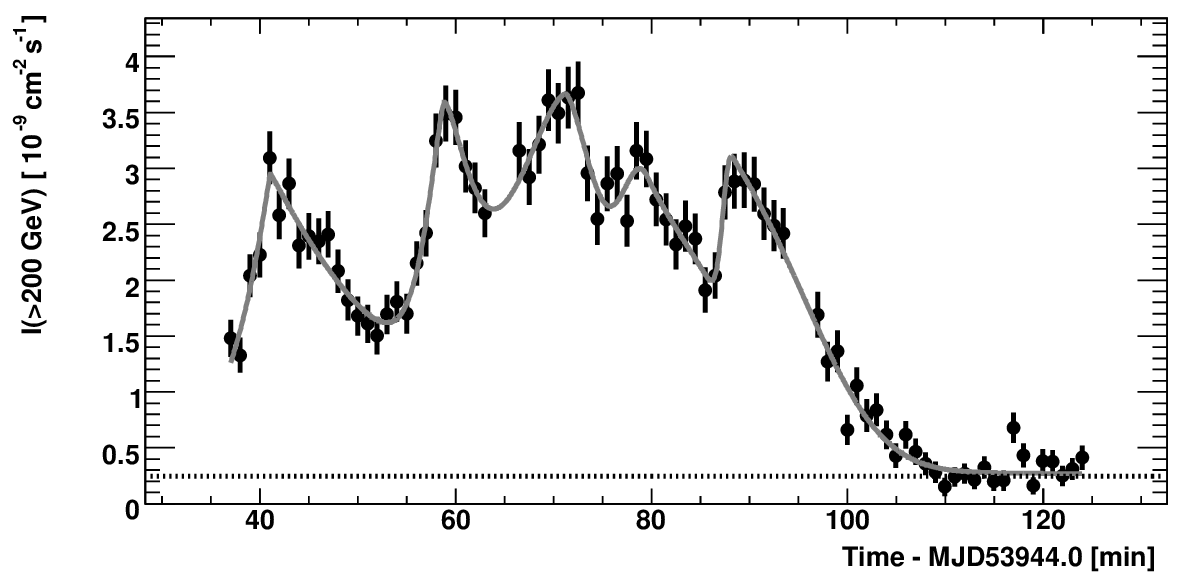}
  \includegraphics[width=6.3cm,height=6.3cm]{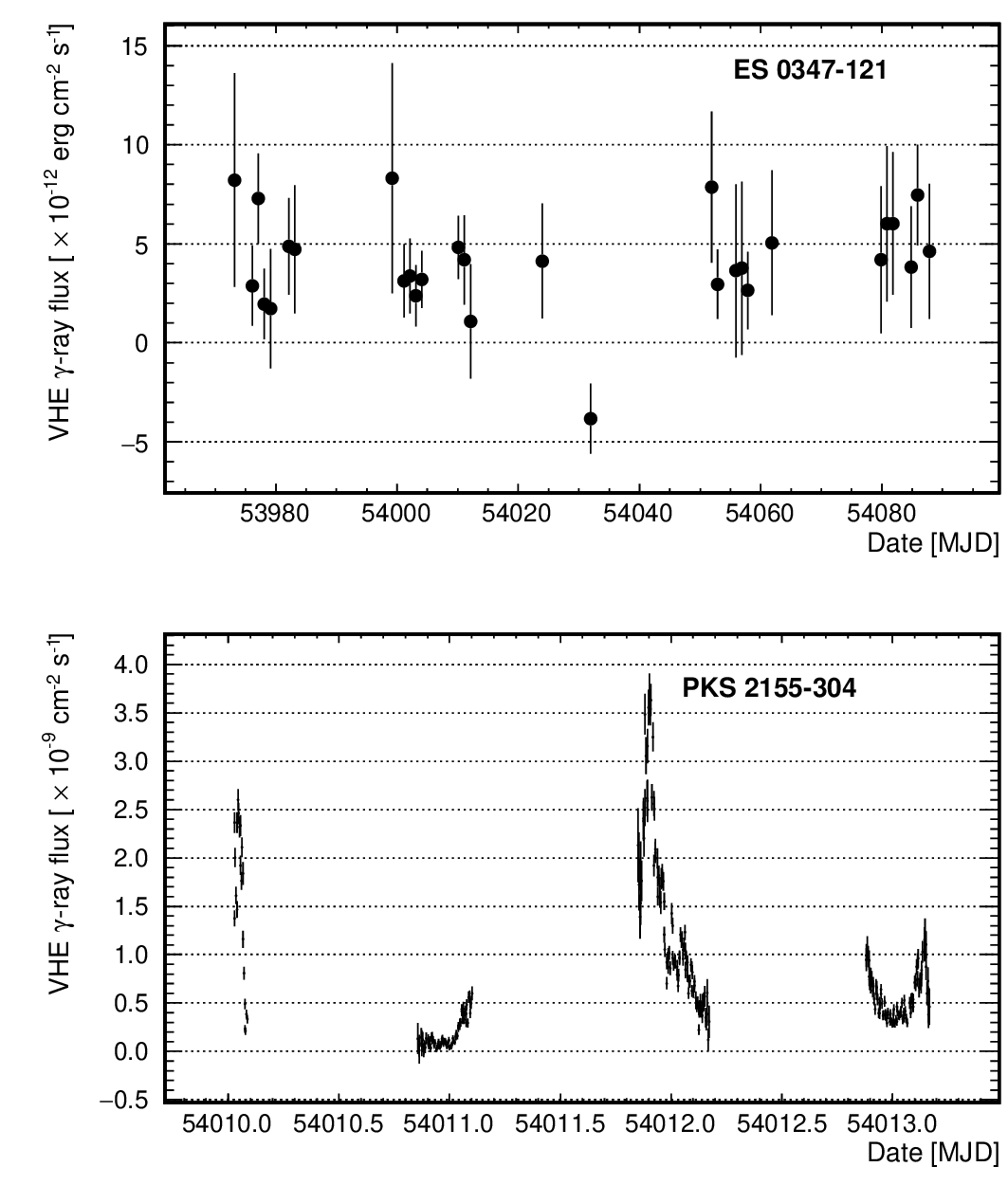}
  \caption{{\bf BL Lac light curves.}
Light curves of the HSP BL Lac object PKS 2155-304 ($z = 0.116$) in 2006
July 28 (left) \cite{Aharonian2007b} and during the period of the 2006 flares (lower right)
\cite{2012A&A...539A.149H}. H.E.S.S.\ VHE light curve of the HSP BL Lac object 1ES 0347-121
($z = 0.815$; upper right) over a period of $\sim 120$ days \cite{Aharonian2007}.
}
\label{fig:HESSlcs}
\end{figure}

Fig.\ \ref{fig:HESSlcs}, left and lower right, shows the extreme flaring behavior of PKS 2155-304 \cite{Aharonian2007b},
measured with H.E.S.S.\ on and around 2006 July 28. At its most intense phase, factor-of-two flux variations are
seen on times as short as 5 minutes. With apparent luminosities of the $\gamma$-ray component exceeding
$\sim 10^{39}$ W and $\gamma$ rays with energies as large as $\sim 3$ TeV, $\gamma\gamma$ opacity arguments
require bulk Lorentz factors $\Gamma \gtrsim 50$ \cite{2008MNRAS.384L..19B}. Though most HSP BL Lacs display
erratic and flaring outbursts, a class of HSP blazars has been established that show weakly or non-detectable
VHE fluxes. Fig.\ \ref{fig:HESSlcs}, upper right, shows the H.E.S.S.\ light curve of 1ES 0347-121
\cite{Aharonian2007}, which, in this figure, shows no significant evidence of variability. VERITAS results on
the HSP blazar
1ES 0229+200 $(z = 0.1396$) also shows extremely weak variability \cite{2013arXiv1307.8091C}.

Rapid variability does not appear to be peculiar to the TeV BL Lac objects.
The FSRQ 4C +21.35 has varied at 70 GeV -- 400 GeV energies on timescales
as short as 30 minutes \cite{2011ApJ...730L...8A}. By comparison, significant MAGIC VHE detections of 3C 279 took place on two successive
days, 22 and 23 February 2006 \cite{2008Sci...320.1752M},
after which it was also detected during  a flare on 16 Jan 2007 \cite{2011A&A...530A...4A}.
The VHE fluxes for the two days in 2006 were each significant, and
 differed by $>2\sigma$, indicating day-scale variability for VHE emission from 3C 279.

Just preceding the launch of the {\it Fermi} Large Area Telescope, the satellite experiment AGILE,\footnote{Astro-Rivelatore Gamma a Immagini L'Eggero, an Italian
  Space Agency project launched in April 2007.} with HE sensitivity comparable to EGRET, found that the FSRQ
3C 454.3 ($z = 0.859$) was ``ringing off the hook" \cite{2008ApJ...676L..13V}.  When {\it Fermi} LAT started
monitoring the sky, with science operations starting in August of 2008, 3C 454.3 provided a wealth of
information from $\gamma$ rays alone on its variability properties. For example,
\begin{enumerate}
  \item A broken power-law function to model the {\it Fermi}-LAT spectrum of 3C 454.3 shows a spectral break at $E_{br}\cong 1.5$ -- 3 GeV (observer's frame) \cite{2009ApJ...699..817A}, similar in energy  in the source frames for the spectral breaks of other LSP and ISP blazars \cite{2010ApJ...710.1271A};
  \item $E_{br}$ is only weakly dependent (showing slight evidence for a hardening) on increasing GeV flux \cite{2010ApJ...721.1383A};
  \item There seems to be a plateau phase preceding flares and a tendency for highest energy photons to come later in the flare \cite{2010ApJ...721.1383A};
  \item The largest flare reached apparent isotropic $\gamma$-ray luminosities $L_\gamma \approx 2\times 10^{43}$ W \cite{2011ApJ...733L..26A}, eclipsing the EGRET record for PKS 1622-297 \cite{1997ApJ...476..692M}.
  \end{enumerate}
A {\it Fermi}-LAT study of blazar variability \cite{2010ApJ...722..520A} shows that FSRQs are more variable
than BL Lac objects in the GeV range, even taking into account flux differences. The power density spectrum of
light curves of bright blazars are generally well fit with power-law power density spectra.

\begin{figure}[!t]
  \centering
  \includegraphics[width=8.0cm,height=7.8cm]{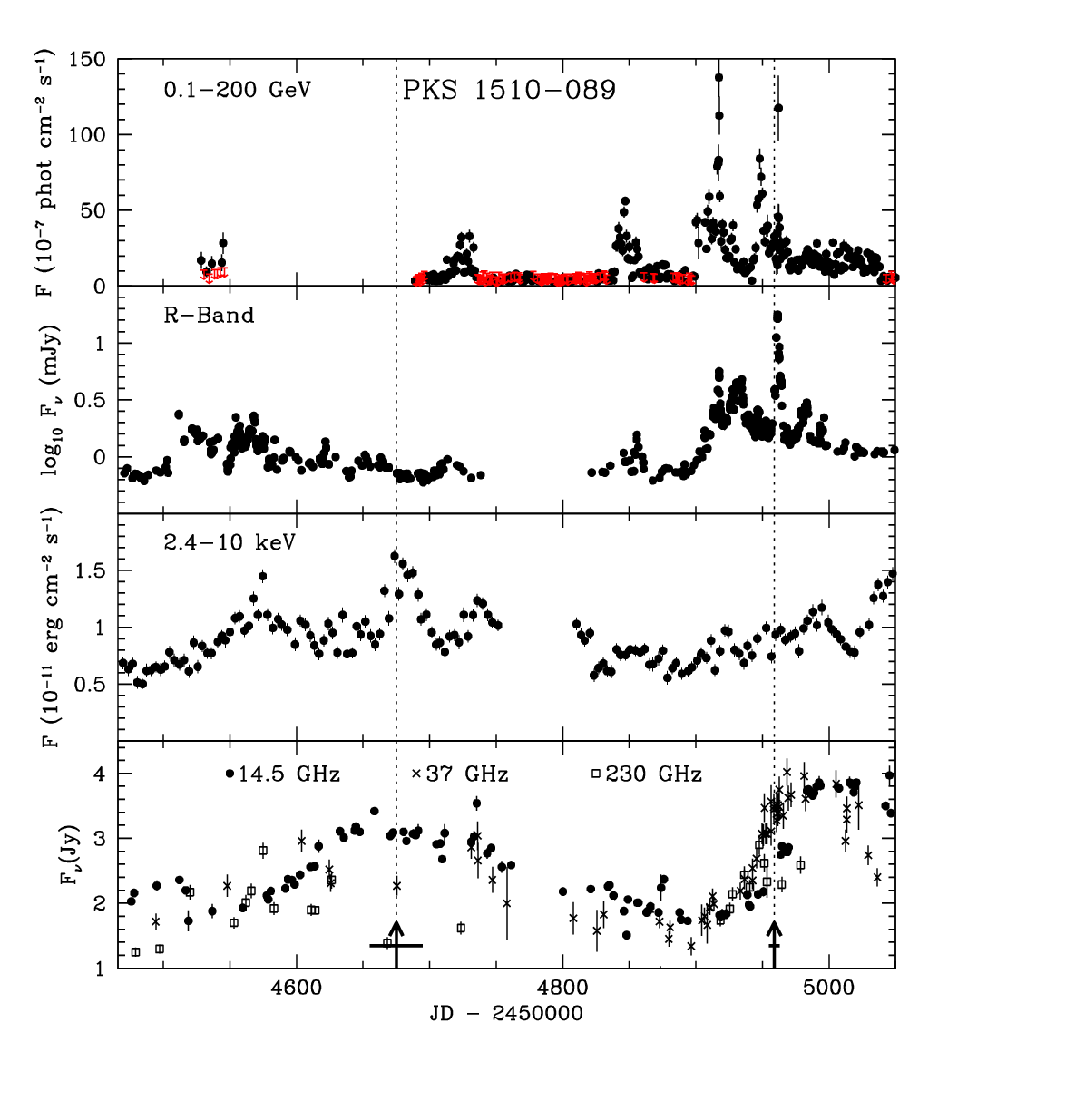}
  \includegraphics[angle=-90,origin=c,width=7.8cm,height=7.8cm]{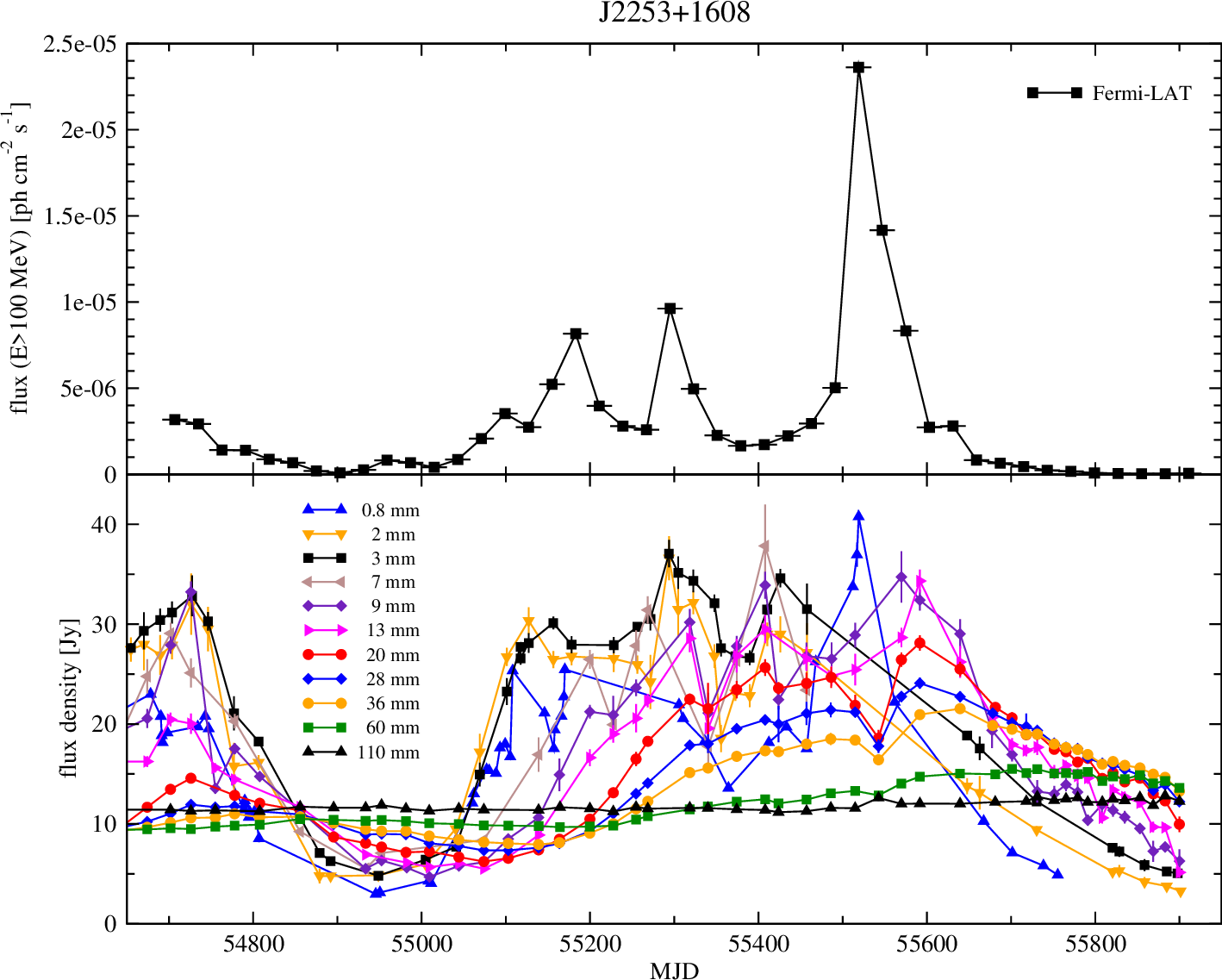}
%\vskip-0.3in
  \caption{{\bf Multiwavelength FSRQ light curves.}
Multiwavelength flaring behavior of PKS 1510-089 ({\it left}; $z = 0.36$)
	\cite{2010ApJ...710L.126M} and 3C 454.3 ({\it right}; $z = 0.86$) \cite{fuh14}. }
  \label{fig:radiogamma}
\end{figure}

\subsection{Multiwavelength variability}

Multiwavelength
campaigns provide valuable information to answer questions concerning the location and environment of the
blazar $\gamma$-ray emission site, which are frustratingly uncertain.  The precise location can, of course,
depend on the particular blazar and whether it is in a flaring or quiescent state.  Nevertheless, depending on
the blazar model, some answers to the question about the location of the usual $\gamma$-ray emission site in FSRQs
are
\begin{itemize}
  \item the deep inner jet, where abundant highly ionizing radiation is found \cite{2010ApJ...717L.118P,2011MNRAS.417L..11S};
  \item the sub-pc scale of the BLR \cite{Cerruti2013,der14};
  \item the pc-scale where the IR radiation from the dusty torus dominates \cite{2000ApJ...545..107B};
  \item the $\gtrsim 10$ pc scale where external radiation fields are weak;
\end{itemize}

    Radio/$\gamma$-ray correlations can be used to infer how far the $\gamma$-ray emission site is from the
    central black hole, and often infer that the $\gamma$ rays are produced at many pc removed from the
    black-hole engine \cite{2008Natur.452..966M}.

Two of many possible examples of multiwavelength blazar studies are shown in Fig.\ \ref{fig:radiogamma}.
On the left, light curves at radio, X-ray, optical, and $\gamma$-ray
frequencies for the FSRQ PKS 1510-089 are shown  \cite{2010ApJ...710L.126M}. Besides being one of only 3 VHE FSRQs
now known, and showing GeV variability on timescales $< 3$ hrs \cite{2013ApJ...766L..11S,Lott14},
optical polarization and electric vector polarization angle can be monitored.
In PKS 1510-089, a large swing of the electric vector polarization angle accompanies the optical and $\gamma$-ray flaring,
with the polarization fraction reaching 30\%  \cite{2010ApJ...710L.126M}. Clearly there is an ordered magnetic field
disrupted by the flaring activity. Marscher et al. \cite{2010ApJ...710L.126M} argue that this points to events taking place
past the optically thick radio core, implying a distant, $\gtrsim 10$ pc
location of the $\gamma$-ray emission site in sources like PKS 1510-089,
and the LSP BL Lac objects OJ 287 ($z = 0.305$) \cite{2011ApJ...726L..13A}
and AO 0235+164 ($z = 0.94$) \cite{2013arXiv1303.2039A} (neither of which
has been detected at VHE). Impressive optical polarization angle
swings correlated with $\gamma$-ray flaring activity in 3C 279 have
 been reported by the {\it Fermi}-LAT collaboration \cite{abd103C279}, leading
to much speculation about the jet's magnetic-field geometry ranging from
turbulent ejection to helical motions of the emitting plasma to bent jets.

The radio/$\gamma$-ray connection is of particular interest because there
is wide agreement that these radiations are fully nonthermal, as compared with
IR, optical, and X-ray, which can include quasi-thermal dust and accretion-disk
emissions. Fig.\ \ref{fig:radiogamma}, right, shows radio and $\gamma$-ray light
curves for 3C 454.3, also known as 2253+1508 \cite{fuh14}. Cross-correlation of
the cm to sub-mm light curves
with the $\gamma$-ray light curves of 3C 454.3 and several other blazars
reveals a frequency-dependent
lag of the radio with respect to the $\gamma$ rays, with longer lags at longer
wavelengths. The cm wavelength emission lags the radio by $\approx 2$ months,
in agreement with earlier claims \cite{2010ApJ...722L...7P}.
Detailed studies of the radio-$\gamma$-ray connections in
blazars are conducted at
the Owens Valley Radio Observatory \cite{2011ApJS..194...29R}.

\subsection{Extreme Flaring States}

The long-term average SEDs of FSRQs and low-peaked BL Lac objects
generally display a spectral softening above a few GeV \cite{2010ApJ...710.1271A}.
Some of the models to explain this softening
involve combined accretion-disk/BLR scattering by jet electrons \cite{2010ApJ...714L.303F},
Klein-Nishina effects at a few GeV when jet electrons scatter Ly
$\alpha$ radiation \cite{Cerruti2013},
or $\gamma\gamma$ attenuation of $\gamma$ rays made
deep in the BLR
by He II Ly $\alpha$ and {}
recombination radiation \cite{2010ApJ...717L.118P,2011MNRAS.417L..11S}.
Episodes of extreme blazar flaring, where  hard emission extends to $\gtrsim 10$ GeV
without a break, challenge FSRQ SED models. Such extreme
flaring states have been observed in the FSRQs
PKS 0805-08 ($z =1.84$), 3C 454.3 ($z = 0.86)$  \cite{2014ApJ...790...45P},
and 3C 279 ($z = 0.538$) \cite{2015arXiv150204699H}.
The rapid 30-min flaring time scale of VHE emission
observed from PKS 1222+216 (4C +21.35) by MAGIC and
{\it Fermi}-LAT \cite{2011ApJ...730L...8A,2011ApJ...733...19T} also challenge
scattering models \cite{2012MNRAS.425.2519N}.
To reach VHE energies requires scattering of IR photons,
as the scattering of BLR radiation faces both the
decline of the cross section and the limit to
scattered energy for Compton scattering
in the Klein-Nishina regime.

Monitoring the full sky for GeV $\gamma$-ray transients requires a large field-of-view instrument, like {\it Fermi}-LAT,
in its standard scanning mode. HAWC provides a weaker version of the same capability at TeV
energies. {\it Fermi} All-Sky Variability Analysis of 47 months of {\it Fermi}-LAT data reveal $\sim 27$
sources at low Galactic latitude that are likely to be blazars \cite{2013ApJ...771...57A}.  Triggers for
ground-based telescopes sensitive to the VHE part of the spectrum, and
real-time alerts for blazar $\gamma$-ray hot spots still rely, unfortunately, on luck and the phase of the
moon, though technologies replacing photomultipliers with photodiodes are being developed to mitigate the
latter problem \cite{2013JInst...8P6008A}, at least for the brightest sources.

%%%%%%%%%%%%%%%%%%%%%%%%%%%%%%%%%%%%%%%%%%%%%%%%%%%%%%%%%
\section{Statistical Properties of $\gamma$-ray Blazars}
\label{sec:stat}
%%%%%%%%%%%%%%%%%%%%%%%%%%%%%%%%%%%%%%%%%%%%%%%%%%%%%%%%%

Fundamental progress can be made by identifying a significant empirical correlation that can then be understood
theoretically.  This is the basis of the Hertzsprung-Russell diagram in stellar astronomy relating
stellar luminosity and temperature.  An analogous relationship in blazar physics would relate synchrotron
or Compton luminosity $L_{s({\rm C})}$ to ``temperature," of which the synchrotron ($\nu_s = \nu^{\rm s}_{\rm
  peak}$) and Compton ($\nu_{\rm C}$) peak
frequencies  are analogs in the nonthermal universe. This is the basis of the blazar sequence,
described next, followed by other robust correlations in blazar $\gamma$-ray astronomy.

\subsection{ Blazar sequence}

\begin{figure}
  \centering
  \includegraphics[width=5.3cm]{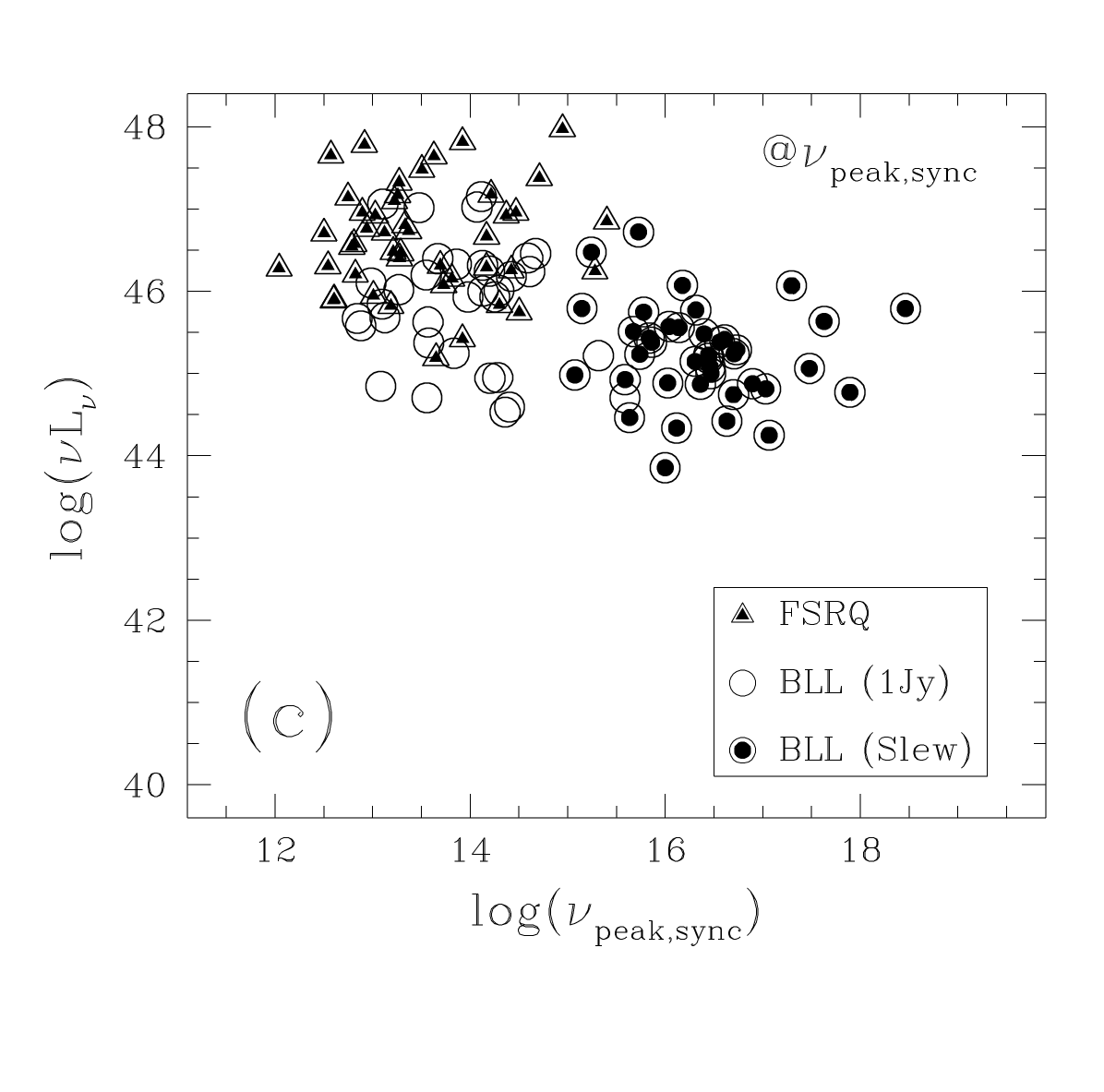}
 \includegraphics[width=4.85cm]{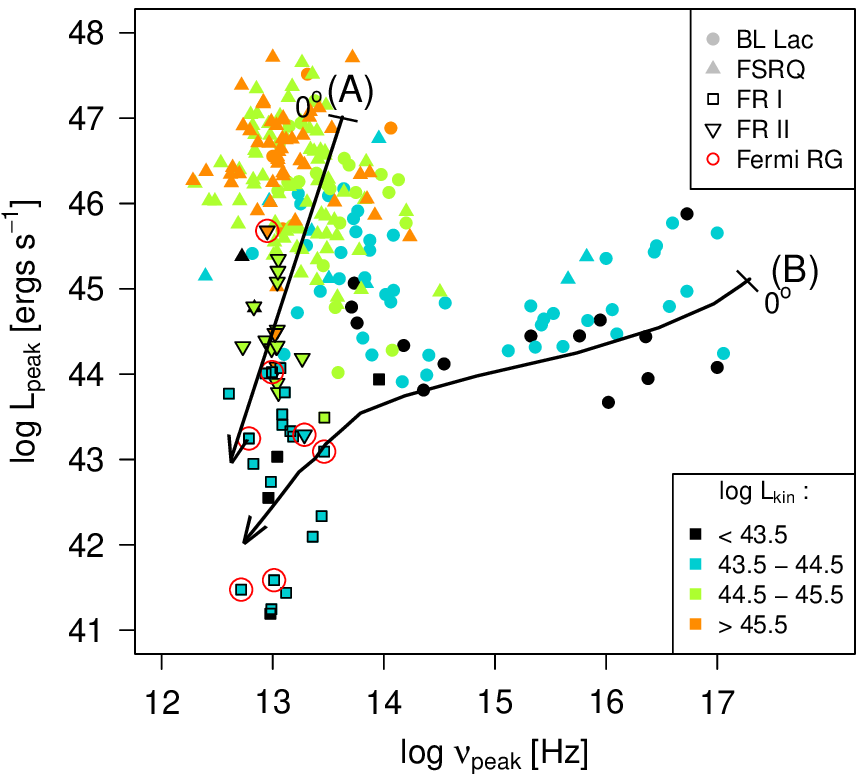}
\includegraphics[width=5.3cm]{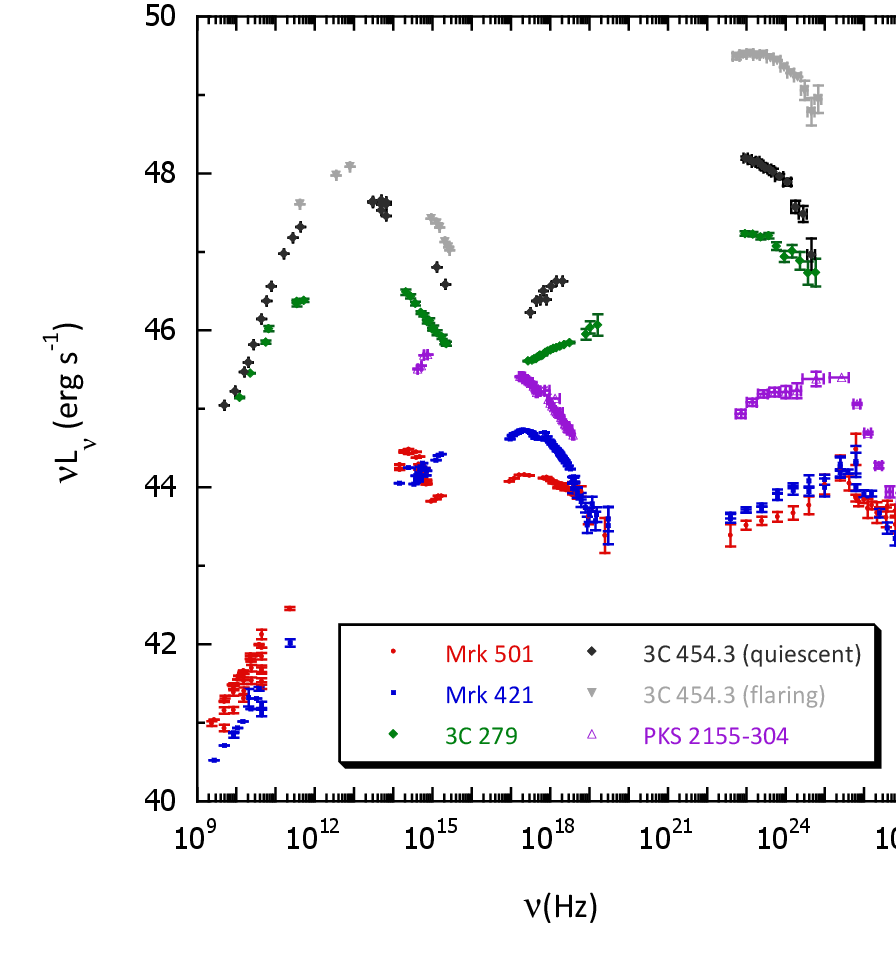}
  \caption{{\bf Three versions of the blazar sequence.}
Left figure is the original blazar sequence diagram presented by Fossati et al.\ (1998) \cite{Fossati1998}.
Center figure is the more recent version presented by Meyer et al.\ (2011) \cite{mey11}, including radio galaxies.
Right figure shows quasi-simultaneous multiwavelength blazar $\nu L_\nu$ SED for Mrk 421, Mrk 501, PKS 2155-304 \cite{Aharonian2009}, 3C
    279 (Epoch C from \cite{hay12}) and two epochs of 3C 454.3 \cite{Cerruti2013}.
    The blazar sequence showing the anti-correlation of bolometric synchrotron and Compton luminosities
    with peak synchrotron and Compton frequencies, respectively, can be deduced from the plot. }
  \label{fig:sequence}
\end{figure}

The luminosity of the lower-energy synchrotron hump in the SED of blazars appears to anti-correlate with the
peak synchrotron frequency $\nu_s$, as shown in the left panel in Fig.\ \ref{fig:sequence} \cite{Fossati1998}.
The sample used in Ref.\ \cite{Fossati1998} to construct this diagram is generally composed of powerful, LSP
FSRQs and weaker, HSP BL Lac objects.  A recent analysis of the blazar sequence \cite{mey11} notes that
a set of blazars simulated by standard one-zone models (i. e. a single Lorentz factor and a convex electron spectrum)
at a range of observation angles yields a corresponding diagram
in strong contradiction with the sequence of the left diagram of Fig.\ \ref{fig:sequence}.
Such one-zone blazar models would predict $L_{syn} \propto \nu_s^4$
(and $L_{syn} \propto \nu_s^3$ for a standing shock \cite{Lind1985}). Nevertheless, a low-luminosity, low-frequency extension must
be found in the $L_{syn}$ vs.\ $\nu_s$ plane from the off-axis emissions of blazars.  Yet the overall behavior
of the blazar sequence is orthogonal to this.

As updated by \cite{mey11}, the blazar sequence forms an ``L" or
even a ``y," as shown in the center panel of Fig.\ \ref{fig:sequence}.  This suggests two separate
populations, which are further distinguished by the jet kinetic powers inferred from the luminosities of the
extended radio emissions around radio galaxies and the host galaxies of blazars.  The highest luminosity
objects are invariably LSP FSRQs, and the lower-luminosity objects are mostly HSP BL Lac objects.
The trends which appear for these objects are compatible with what is expected from rapid angular debeaming
in a strong and homogeneous jet for FSRQs (track A in figure \ref{fig:sequence}) and less rapid debeaming
in the case of BL Lacs having weaker jets and velocity gradients (track B in figure \ref{fig:sequence}).
The radio galaxies occupy the positions expected from the off-angle tracks of FSRQs and low-synchrotron peaked BL Lac
objects.

This trend can be inferred from the associated blazar SED sequence
illustrated in the right panel of Fig.\ \ref{fig:sequence}. Note that a mixture of
quiescent and flaring states are shown there, and that the blazar sequence
is generally plotted from blazar SEDs averaged over long times, though it would
be interesting to compare with flaring states.
The transition from LSP to HSP objects was interpreted \cite{Ghisellini1998} as
a result of decreasing Compton cooling associated with the decreasing energy density of the external radiation
fields. This in turn allows an increase of the maximum electron energy, shifting the synchrotron and
Compton peak emissions blueward. Smaller external radiation energy densities are attributed \cite{Bottcher2002} to
smaller accretion powers, which occurs when the circumnuclear environment of supermassive black holes is
gradually depleted and hence less accretion-disk radiation is reprocessed. This trend is also
reproduced by \cite{Finke2013}, who used as parameters the magnetic field of the blazar jet emitting region, the external
radiation field energy density, and the jet angle to the line of sight.

The HSP population from low to high values of $\nu_s$ displays a contrary trend where
$L_{syn}$ is positively correlated with $\nu_s$. This is argued \cite{mey11} to be a consequence
of a decelerating flow that explains the apparent contradiction between the
large Doppler and $\Gamma$ factors inferred from spectral modeling and $\gamma\gamma$ opacity
arguments applied to BL Lac objects compared to the smaller values
 found from spectral modeling of radio galaxies and from VLBI observations of
BL Lac objects at the milli-arc-second scale \cite{mey11}.
%Rather than a decelerating flow from a force directed against the jet,
Intermittent black-hole ejecta with a variety of $\Gamma$ values, followed by colliding shells,
could make a decelerating jet in a colliding shell-type scenario.
Radiative processes would, however, seem to provide
a  feedback mechanism in objects whose photon power is dominated by external Compton processes.
The external Compton processes provide radiative braking,
unlike synchrotron and SSC emissions in BL Lac objects, where the radiation is
emitted isotropically in the jet frame.
%The $L_{syn}$ vs.\ $\nu_s$ blazar sequence  is also compatible with a spine-sheath scenario.

\subsection{ Blazar divide}

\begin{figure}[!t]
  \centering
  \includegraphics[width=10.0cm]{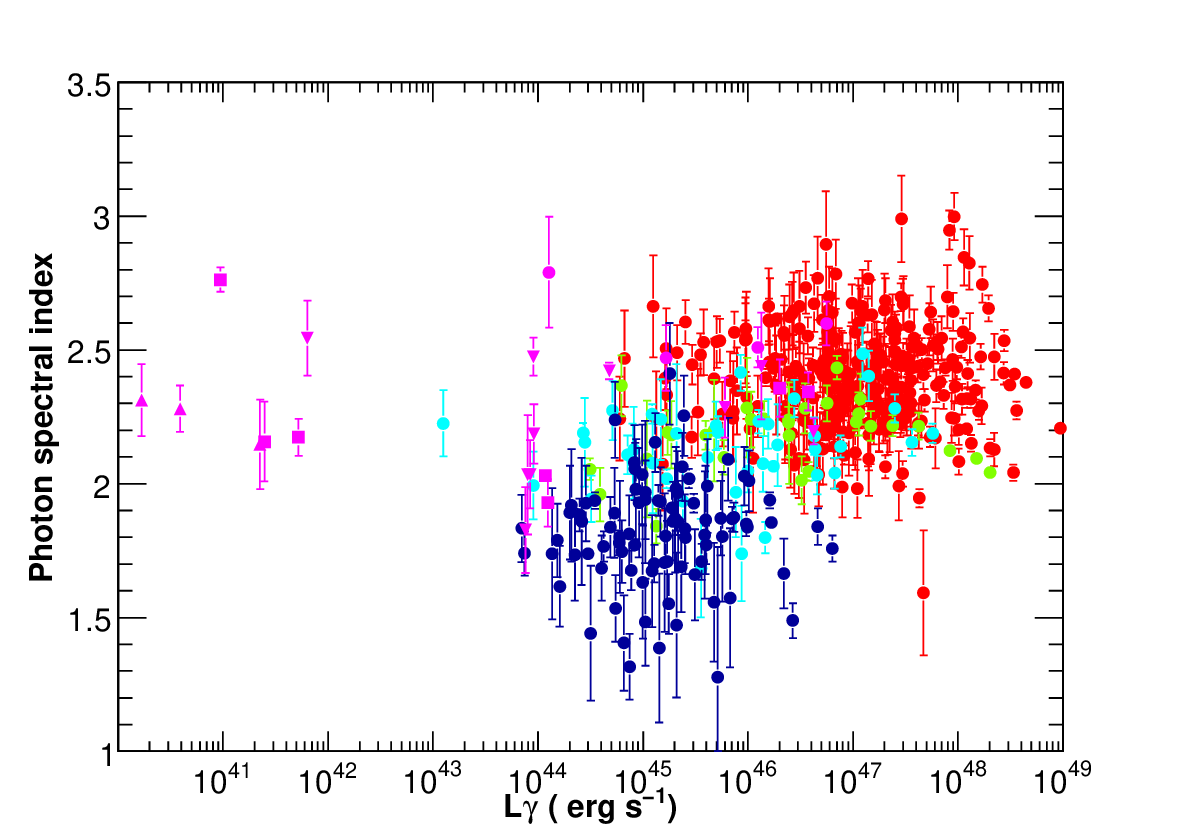}
  \caption{{\bf The blazar divide.}
The photon index $\Gamma_\gamma$ is plotted against $\gamma$-ray luminosity at
$\gamma$-ray energies 0.1 -- 100 GeV  \cite{2LAC}.
Red: FSRQs, green: LSP-BL Lac
objects, light blue: ISP-BL Lac objects, dark blue: HSP-BL Lac objects,
magenta: non-blazar AGN (circles: NLS1s, squares: misaligned AGN, up
triangles: starbursts, down triangles: other AGN). The blazar divide refers
to the abrupt change in spectral index between FSRQs and HSP BL Lacs
at $L_\gamma\approx 10^{46}$ erg s$^{-1}$ ($10^{39}\,{\rm W}$).}
  \label{fig:blazardivide}
\end{figure}

Another blazar correlation discovered in the {\it Fermi}-LAT data is a divide separating BL Lac and FSRQ-type
blazars, when the $\gamma$-ray number spectral index $\Gamma_\gamma$, or the energy spectral index $\alpha_\gamma$, measured in
the $0.1-100\,{\rm GeV}$ energy range, are plotted as a function of $\gamma$-ray luminosity $L_\gamma$
\cite{Ghisellini2009,1LAC}. The version shown in Fig.\ \ref{fig:blazardivide} is from the 2LAC \cite{2LAC}.
In the analysis of the {\it Fermi}-LAT data, the $\gamma$-ray spectral index $\Gamma_\gamma$ is determined
from a single power-law fit to all the data in the 0.1 -- 100 GeV range, though the energy flux is determined
by separate power-law fits in five energy bands (0.1 -– 0.3, 0.3 -– 1, 1 -– 3, 3 -– 10, and 10
-– 100 GeV).

As can be seen, most FSRQ have $\Gamma_\gamma>2.2$ and $L_\gamma> 10^{39}\,{\rm W}$ ($ 10^{46}\,{\rm erg}\,{\rm s^{-1}}$),
while the BL Lac class is mainly in a region where
$\Gamma_\gamma <2.2$ and $L_\gamma\lesssim 10^{39}\,{\rm W}$ ($10^{46}\,{\rm erg}\,{\rm s^{-1}}$).
A rather neat separation between these two classes takes place at
$L_\gamma \sim 10^{39}\,{\rm W}$ ($10^{46}\,{\rm erg}\,{\rm s^{-1}}$),
and is therefore known as the {\it blazar divide}. This divide is interpreted \cite{Ghisellini2009} as having a
physical origin, possibly reflecting a transition to an accretion regime where the radiatively inefficient and
low Eddington accretors have low $\gamma$-ray luminosities \cite{ghi11}. This is consistent with an
interpretation where the inner edge of an optically thick, geometrically thin
accretion disk moves out in radius with declining mass-accretion rate.
Within the regions between the inner edge of the thin disk and the the black hole is
an advective flow, possibly with a hot X-ray emitting corona. The transition
between the weaker, advection-dominated, low-radiative efficiency accretion flows,
to the high-luminosity, high-radiative efficiency inflows at
an Eddington ratio of $\sim 1$\% could explain the divide.

Except for a transition of the spectral index over a rather restricted range of apparent $\gamma$-ray luminosity $L_\gamma$,
$\Gamma_\gamma$ is insensitive to the value of $L_\gamma$. It is apparent from the distribution of FSRQs and
BL Lac objects with  peak synchrotron frequencies $\nu_s$ that a correlation between $\Gamma_\gamma$ and $\nu_s$ follows.
Modeling a large sample of blazars yields parameters that can then be correlated,
such as the comoving electron Lorentz factor at the peak of the SED, making synchrotron photons with frequency $\nu_s$.
Ref.\ \cite{gtf10} argue that this distribution is consistent with a cooling scenario.

\subsection{Spectral-index diagrams}

\begin{figure}[!t]
  \centering
  \includegraphics[width=6.7cm]{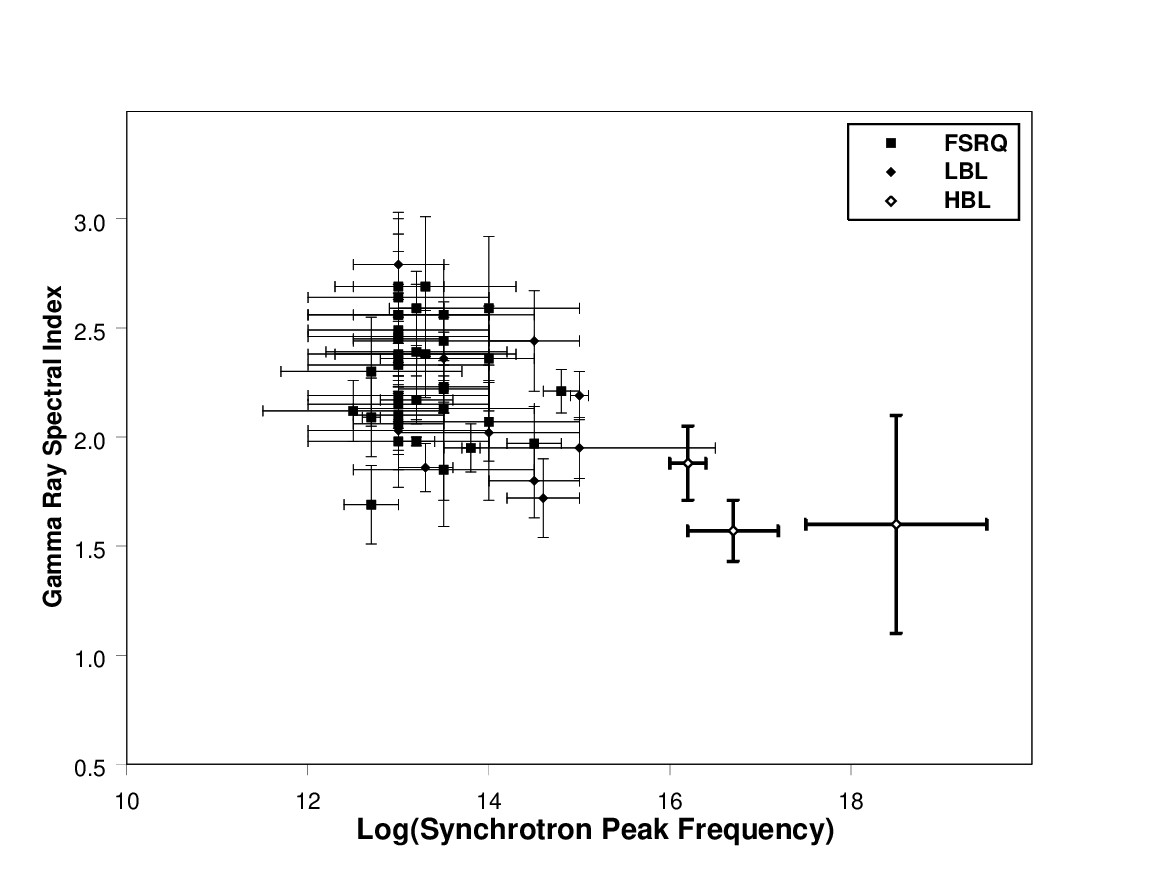}
  \includegraphics[width=7.5cm]{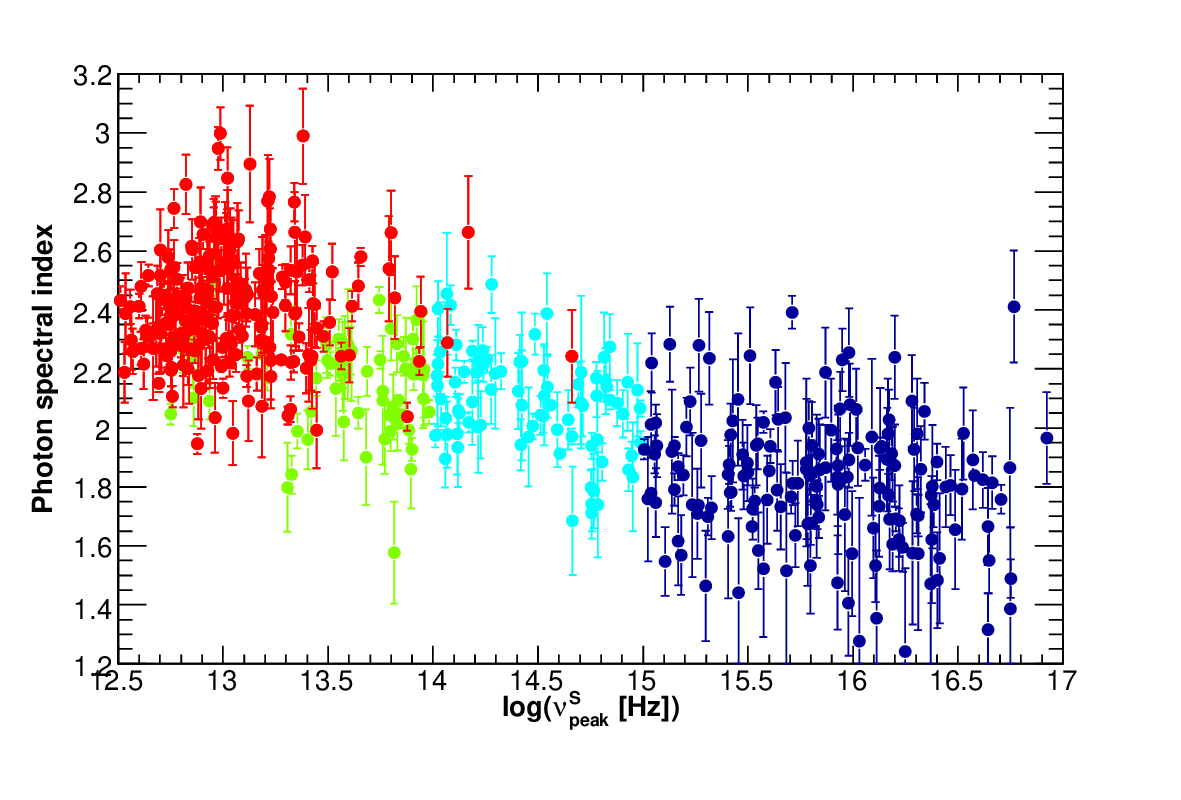}
  \caption{{\bf Spectral index versus peak synchrotron frequency.}  {\it Left:} Spectral-index diagram
based on the EGRET  sample with 50 AGN  \cite{Nandikotkur2007}. {\it Right:} Spectral-index diagram based on the {\it Fermi}-LAT sample of 886 AGN from \cite{Abdo2010} (Red:  FSRQs,  green:LSP-BL
    Lacs, light blue:  ISP-BL Lacs, dark blue:  HSP-BL Lacs). The {\it Fermi}-LAT results  clearly show
    that the HE $\gamma$-ray spectral index $\Gamma_\gamma$ becomes harder when  $\nu^{\rm s}_{\rm peak}$ increases,
    giving the significant correlation apparent in this figure.}
  \label{fig:indexvsvs}
\end{figure}

A further robust correlation in blazar physics are the blazar spectral-index diagrams relating $\Gamma_\gamma$
with either the peak synchrotron frequency $\nu_s$ or
peak Compton frequency $\nu_{\rm C}$. The size of the {\it Fermi}-LAT sample and the improvement of the energy
resolution over previous instruments now shows a clear correlation between the photon spectral index
$\Gamma_\gamma$ of HE AGN and $\nu^{\rm s}_{\rm peak}$ (Fig. \ref{fig:indexvsvs}). On the left is an early
attempt to construct such a diagram from the EGRET data \cite{Nandikotkur2007}, and on the right is {\it Fermi}-LAT
data from the 2LAC  showing the
strong correlation of $\Gamma_\gamma$ with $\nu_s$. The general trend can be described by the
simple relation
\begin{equation}
\Gamma_\gamma \cong d - k\log \nu_s\;,
\label{Gammagammavsnus}
\end{equation}
where $d$ and $k$ are constants.  The corresponding highly correlated $\Gamma_\gamma$ vs. $\nu_{\rm C}$ can
also be found in Fig.\ 29 of \cite{fermibright}.  The ratio of the $\gamma$-ray and synchrotron peak luminosities gives the Compton
dominance $A_{\rm C}$, which can be plotted as a function of $\nu_s$ or $\nu_{\rm C}$.  The $A_{\rm C}$ vs.\ $\nu_s$ plot is known
to be highly correlated \cite{Fossati1998}, and recent data shows that it displays an ``L" shape
\cite{Finke2013} consistent with the blazar sequence diagram of Ref.\ \cite{mey11} shown in
Fig.\ \ref{fig:sequence}.

This correlation can be explained \cite{der15} assuming an equipartition model\footnote{The model assumes equipartition
between non-thermal electron and magnetic-field energy densities.} with a log-parabola electron
energy distribution (i.e. an electron energy density $N(E) \propto E^{a+b\log E}$, \cite{der14}),
noting that the synchrotron, SSC, and external Compton radiations in the
Thomson regime produce a broadened log-parabola SED, the slope of which in the $\gamma$-ray domain is of the form of
eq.\ (\ref{Gammagammavsnus}) \cite{2004A&A...422..103M}. Note that the peak synchrotron frequencies of FSRQs
appear clustered in the range $10^{12.5}~{\rm Hz} \lesssim\nu_s\lesssim 10^{13.5}$ Hz.  Another interesting
feature of Fig.\ \ref{fig:indexvsvs}, right, is that the photon spectral indices do not reach values
significantly smaller than $\Gamma_\gamma=1.5$, which is near the flattest spectral index predicted in shock
acceleration models \cite{Malkov2000}.

%%%%%%%%%%%%%%%%%%%%%%%%%%%%%%%%%%%%%%%%%%%%%%%%%%%%%%%%%
\section{Blazar Models}
\label{sec:unso}
%%%%%%%%%%%%%%%%%%%%%%%%%%%%%%%%%%%%%%%%%%%%%%%%%%%%%%%%%

The blazar paradigm and astrophysics of
blazar spectral modeling have
been recently reviewed  \cite{der14a,dm09,bhk12}.  What may
be worth noting in addition to the information found there is that
the  way in which  the electron energy distribution (EED) is established is
one of the distinctive features of a blazar model, and so should be well understood.
As to theoretical efforts to understand $\gamma$-ray and multiwavelength results,
perhaps most useful models are spectral and, in the best
case,  dynamical.  There has been tremendous theoretical and modeling progress
throughout the CGRO and {\it Fermi-}LAT eras, in
parallel with advances in ground-based VHE astronomy,
to measure quasi-simultaneous blazar spectra across a broad wavelength range.
These data test the models, and a model that cannot be rejected is not very useful.

\begin{figure}[!t]
  \centering
  \includegraphics[width=7.0cm,height=5.0cm]{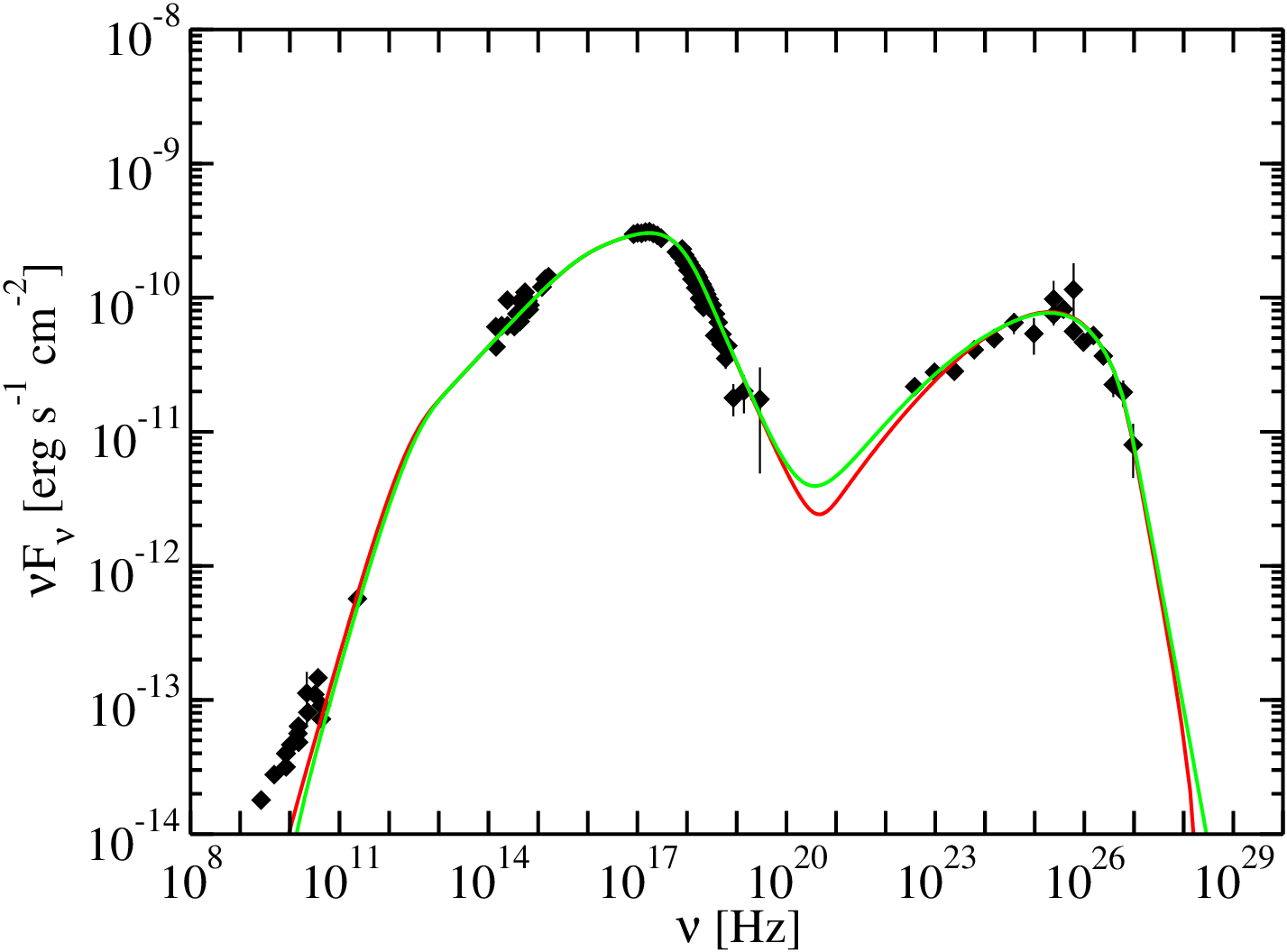}\hskip0.5in
 \includegraphics[width=7.5cm,height=4.9cm]{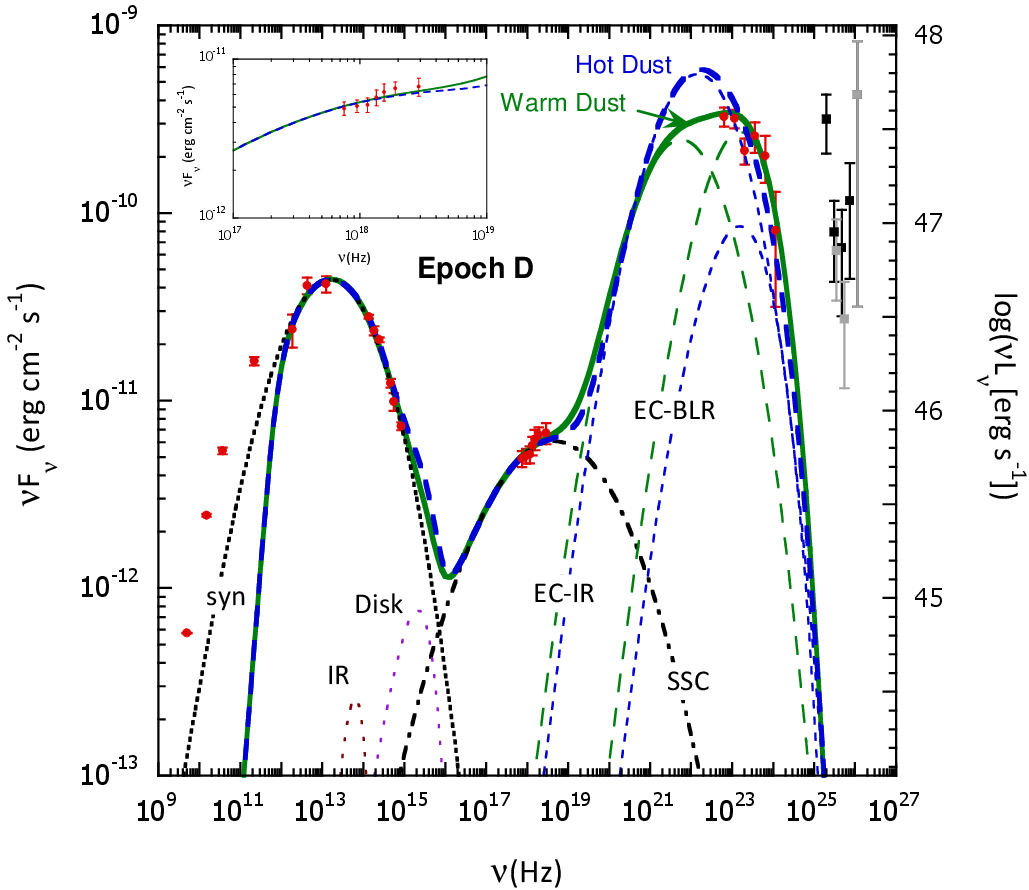}
  \caption{{\bf Blazar model SEDs.}
Left: SED of Mrk 421 measured in multiwavelength campaigns including {\it Fermi}
and MAGIC $\gamma$-ray telescopes \cite{abd11b}.  Two one-zone model
fits are shown, with $t_{var} = 1$ hr and 1 day for the green and red curves, respectively.
Right~: Data shows Epoch D SED of 3C 279
\cite{hay12}.   Blazar model fits \cite{der14}  show separate
spectral components in the fit, including  the IR torus and accretion-disk
emission, beamed synchrotron (syn), SSC, Compton-scattered IR (EC-IR)
radiaion for warm and hot dust, and Compton-scattered
BLR radiation (EC-BLR). Non-simultaneous VHE MAGIC data for 3C 279 \cite{2011A&A...530A...4A}
are shown for comparison.  Total emission spectra include
synchrotron self-absorption. Inset shows detail of fits at X-ray energies.}
  \label{fig:3C279sed}
\end{figure}

Fig.\ \ref{fig:3C279sed}, left, shows synchrotron/synchrotron self-Compton (SSC) spectral modeling
\cite{abd11b} applied to Mrk 421, the prototypical HSP BL Lac.  Adequate fits can be obtained from
conventional one-zone SSC models with an EED described by a broken power law with exponential cutoff, or with
a log-parabola EED.  In comparison, Fig.\ \ref{fig:3C279sed}, right, shows data for an epoch D (corresponding
to the brightest $\gamma$-ray flare) SED of 3C 279
\cite{hay12}. An approach \cite{der14} assuming a log-parabola EED and near-equipartition conditions between
nonthermal electron and magnetic-field energy density gives the spectral models for epoch D as shown.

%%%%%%%%%%%%%%%%%%%%%%%%%%%%%%%%%%%%%%%%%%%%%%%%%%%%%%%%%
\subsection{Leptonic Models}
%%%%%%%%%%%%%%%%%%%%%%%%%%%%%%%%%%%%%%%%%%%%%%%%%%%%%%%%%

The models shown in Fig.\ \ref{fig:3C279sed} are leptonic, and leptonic models are in most cases able to give
good fits to multiwavelength blazar SED data. But even in the case of 3C 379, as shown by the
non-contemporaneous MAGIC data \cite{2011A&A...530A...4A} shown there, the observation of a VHE flare can strain a leptonic
model, or at least lead to predictions regarding the properties of the synchrotron component, because the
hardened electron distribution from a VHE flare should show up in the synchrotron SED.

A wide range of leptonic blazar spectral models have emerged in the
last 20 years covering the most important
aspects of jet physics. Any leptonic blazar model must comprise,
in rough order of decreasing importance, the following features:
\begin{itemize}
\item relativistic boosting;
\item electron energy distribution;
\item magnetic field;
\item synchrotron radiations;
\item synchrotron self-Compton radiations;
\item external Compton radiations;
\item synchrotron self-absorption;
\item accretion-disk and IR torus emissions;
\item internal $\gamma\gamma$ absorption;
\item direct accretion-disk photons (important when the $\gamma$-ray emission region is within 100 -- $1000 R_{\rm S}$ ;
\item external extragalactic background light (EBL) absorption \cite{Horns2015};
\item EBL-induced cascade radiations (or assumptions about the minimum intergalactic magnetic field) \cite{Horns2015};
\item external source $\gamma\gamma$ absorption (e.g., in the inner jet environment);
\item cascade radiation spectrum from pairs formed by internal absorption;
\item higher-order radiative processes, e.g., bremsstrahlung, ionization losses, etc.;
\end{itemize}
leaving aside the details of the particle acceleration mechanism,
whether dynamic or snapshot modeling is being simulated, and the
statistical method to fit data. And this
leaves out a whole plethora of additional
interesting complexities when hadrons are in the mix,
as described next.
The issues associated with leptonic blazar modeling of $\gamma$-ray sources
 have kept astrophysicists busy since 1992.

The short timescale puzzle mentioned above, in Section 5, has motivated
work into magnetic reconnection and jets-within-jets models.
In magnetic reconnection models \cite{gub10}, the short variability time is realized by an electron beam driven by reconnection taking place in a sub-volume of a larger region whose size scale is related to the Schwarzschild radius, $R_{\rm S}$. To compensate for the small comoving size $R^\prime = f \Gamma R_{\rm S}$, with $f < 1$ implied by observations, reconnection is assumed to drive relativistic outflows or beams of particles with sufficiently large Lorentz factor that Doppler boosting compensates for the smaller available energy content in the small blob.  The observed luminosity  $L_{obs} \cong L_{iso} f^2 \left( {\delta_p/\delta_j}\right)^4$ so, provided that the plasma Doppler factor $\delta_p$ and jet Doppler factor $\delta_j$ are suitably chosen, the large apparent power is preserved even from the smaller size scale of the region. Other approaches include jet-within-jet or turbulent cell models \cite{mj10,np12,mar12}, and Poynting-dominated jet models \cite{nal12}. Plasma instability-induced short variability behavior has also been considered \cite{ssb12}.

%%%%%%%%%%%%%%%%%%%%%%%%%%%%%%%%%%%%%%%%%%%%%%%%%%%%%%%%%
\subsection{ Hadronic Models}
%%%%%%%%%%%%%%%%%%%%%%%%%%%%%%%%%%%%%%%%%%%%%%%%%%%%%%%%%

An additional hadronic component could ``rescue" a one-zone model by providing a $\gamma$-ray hardening from
cosmic-ray proton- or ion-induced radiations in the blazar jet.  So-called ``orphan flares," as in the Whipple
flare observed from 1ES 1959+650 \cite{2004ApJ...601..151K}, that show up in the $\gamma$-ray component, but
lack corresponding counterpart flares expected from leptonic processes, have been used to implicate hadronic
processes \cite{2005ApJ...621..176B}.  To the opposite of leptonic processes, hadronic components  would leave
no corresponding feature in the synchrotron part of the SED, because the $\gamma$ rays made by proton and ion synchrotron
processes would be in the GeV -- TeV range, and the photon-lepton cascades induced by photopion production
$p,A + \gamma \rightarrow p,A+ \pi$ would cascade down to $\gamma$ rays, with no direct effect on the
synchrotron component.  Ultra-relativistic protons accelerated by the blazar that reach intergalactic space
can induce photopion secondaries within a few hundred Mpc by interactions with cosmic microwave background
photons, and photopair secondaries over one or two Gpc through interactions with photons of the EBL
\cite{Horns2015,2010APh....33...81E,2010PhRvL.104n1102E}. These emissions can produce an anomalous $\gamma$-ray cascade
component that could confuse attempts to measure the extragalactic optical and infrared background using
$\gamma\gamma$ attenuation with the EBL. Such weakly variable VHE components may produce the SED of weakly
variable BL Lac objects, like 1ES 0347-121 shown in Fig.\ \ref{fig:HESSlcs}.

Proton/ion synchrotron radiation provides one
hadronic mechanism that could be important in blazars, but
essentially requires
a highly magnetized jet. The more favored hadronic
mechanisms in blazars are
photohadronic processes.  One line of
evidence that these processes
operates in blazars depend on the detection
of high-energy neutrinos, and whether they could account for
some of the PeV or lower energy (30 TeV -- 500
TeV) Ice~Cube neutrinos \cite{aar14}. Tentative evidence
linking radio/$\gamma$-ray blazars
to Ice~Cube neutrinos is given in Ref.\ \cite{2014A&A...566L...7K}.

\subsubsection{Proton/Ion Synchrotron Radiation}

We derive (or rederive)  magnetic-field and particle-energy requirements
for proton- and ion-synchrotron radiation.
The particle energy-loss rate by synchrotron
radiation, averaged over pitch angle, is (with primed parameters are in the frame comoving with the plasma)
\begin{equation}
-({dE^\prime\over dt^\prime})_{syn} = {4\over 3}({Q^2\over mc^2})^2 \,c\,
\frac{B^{\prime 2}}{2 \mu_0} \,\gamma^{\prime 2}
= {Z^4\over A^2}\mu^2\,{4\over 3} \, c \, \sigma_{\rm T} \, U^\prime_{B^\prime}\,\gamma^{\prime 2}\;,
\label{dEpoverdtp}
\end{equation}
where $E^\prime = m c^2\gamma^\prime$, $\mu \equiv m_e/m_p$, $m = Am_p$ is the mass of an ion with charge $Z$,
$\sigma_{\rm T}$ is the Thomson cross section, and $U^\prime_{B^\prime}$ is the magnetic energy density. The
comoving-frame time $t^\prime_{var} = \Gamma t_{var}$, where $t_{var}$ is the variability time in the source
frame.  The efficiency for a particle to lose its energy through synchrotron losses is $\eta \equiv (
{t_{var}^\prime/ E^\prime})|({dE^\prime/dt^\prime})_{syn}|\large\;.$ Combining the requirement that $\eta >
1$, that is, efficient dissipation of energy via synchrotron losses, and the Hillas
condition\footnote{Expressing that the size of the accelerator region should not be smaller than the
  accelerated particle's Larmor radius.} $E^\prime <Q
B^\prime R^\prime$, gives a range of allowed energies for which efficient synchrotron radiation is possible,
contingent on the strength of the comoving magnetic field $B^\prime$ expressed in Tesla being $\gtrsim 37 \times 10^{-4} \,A^{4/3}
Z^{-5/3} (\Gamma/10)^{-2/3} t_4^{-2/3}$, where $t_4=t_{var}/(10^4 \, {\rm s})$. The corresponding energies of escaping
cosmic-ray particles (should they be able to escape before losing most of their energy) is $E \cong \Gamma
E^\prime \approx 3\times 10^{20} A^{4/3} Z^{-2/3} (\Gamma/10)^{4/3} t_4^{1/3}$ eV.  This is the reason proton-
and ion-synchrotron models of blazars typically require $B^\prime \gtrsim 10^{-2}$ T, which makes a large demand
on jet power, given that equipartition fields (between nonthermal electron and magnetic field) are a few
$10^{-4}$~T. This does not even take into account that the efficient dissipation only applies to the very
highest-energy particles. The proton/ion synchrotron radiation is emitted at $E_\gamma^{syn} \cong 2.0
(A^3/Z^2)(\Gamma/10)$ TeV, which for protons is a factor $m_p/m_e$ larger than the well-known leptonic value
at $\approx 100 \Gamma$ MeV.  What is problematic for highly magnetized jets is that for the same synchrotron
SED, a larger $B^\prime$ means that the EED peaks at lower comoving Lorentz factors. This not only makes it
harder to scatter SSC radiation to TeV energies, but requires that an increasingly large flux of the blazar
SED has to have an hadronic origin.  See, e.g., \cite{aha02,mue03} for proton/ion synchrotron blazar
models. Minimizing magnetic power by considering smaller sizes of the emission zone makes that zone more
opaque to $\gamma\gamma\rightarrow$ e$^+ +$ e$^-$ attenuation, which makes it difficult to explain luminous
VHE/TeV radiation from blazars with a proton/ion-synchrotron model.

\subsubsection{Photohadronic radiation}

Meson production from photohadronic processes in blazar jets is an attractive idea
not only because it operates in systems with relativistic particles and large radiation
field energy densities, like the blazar environment, but also because it can
be established with high confidence by detecting high-energy neutrinos coincident in direction and time with
a flaring blazar.  The internal synchrotron and direct
accretion-disk radiation in blazars is unlikely to be as
effective for neutrino production as the external BLR and IR radiation fields \cite{ad01}.
We can simply estimate the photomeson efficiency for jet protons
passing through the BLR. Note first that
the photohadronic production efficiency of cosmic-ray protons in a blazar jet is hardly distinguishable
from the corresponding losses for rectilinear propagation of a cosmic-ray proton through the BLR \cite{dmi14}.
Therefore
the photohadronic energy-loss rate for a particle of energy $E>E_{thr}$
is $-({dE/ dt})_{p\gamma\rightarrow \pi} \cong n_{ph}\hat\sigma E c$,
where $\hat\sigma \cong 70\mu$b is roughly the product of the photopion cross section and inelasticity above
threshold, and $n_{ph}$ is the density of photons. The threshold condition is
$(E/m_pc^2)(E_\gamma/m_ec^2) \gtrsim 2m\mu/m_e \cong 300$, so that for Ly $\alpha$ photons dominating
the BLR radiation, $E_\gamma/m_ec^2 = 2\times 10^{-5}$, and $E_p \gtrsim 20$ PeV.

The efficiency to extract proton energy via photomeson processes is $\eta_{p\gamma\rightarrow \pi} =
(R/c)|({dE/ dt})_{p\gamma\rightarrow \pi}|/E$, where $R = 0.1R_{0.1pc}$ pc is the characteristic BLR radius.
Using the BLR scaling relation $R_{0.1pc}\sim L_{46}^{1/2}$ (e.g., \cite{gt08}), where
$L_{46}$ is the accretion-disk luminosity in units of $10^{46}$ erg~s$^{-1}=10^{39}$~W, then the
photon energy density $u_{ph}$ through which the cosmic rays pass is $u_{ph} \cong L\tau/4\pi R^2 c \cong 3\times10^{-3}
(\tau/0.1) $ J~m$^{-3}$, where $\tau$ is an average covering factor through the BLR. Hence we find that $\eta
\cong 3R_{0.1pc}(\tau/0.1)$\%.  If ultra-high energy cosmic-ray acceleration takes place in the inner jets of
blazars, we can expect $\sim 1$ -- 10\% efficiency \cite{mid14,dmi14}, and a prediction for the neutrino
fluxes of blazars given that the baryon loading is normalized by the density of blazars and average power.
Note also that Bethe-Heitler photopair losses can provide an injection source of high-energy electrons and
positrons, though it is usually less important than photopion production.

Regarding recent progress in hadronic and lepto-hadronic modeling (given that the synchrotron component is
widely considered to be nonthermal lepton synchrotron radiation), note that \cite{bsk13} introduces a new
lepto-hadronic model. In most cases, leptonic model fits work well, but in flaring sources, implementation of a hadronic
component can improve the fits, though at the expense of jet power. In the work of \cite{cer15}, a
lepto-hadronic model that includes both lepton and hadron synchrotron and photohadronic processes,
operates in the unusual class of HSP BL Lac objects, explaining the weakly VHE variability
of the HSP BL Lac objects 1ES 0229+200, 1ES 0347-121 (cf.\ Fig.\ \ref{fig:HESSlcs}, upper right),
and 1ES 1101-232 as due to the long timescale
for hadronic cooling processes.
Recent modeling \cite{pet15} connects the production of PeV neutrinos with the spectral properties of
particular blazars that are candidate ultra-high-energy cosmic ray (UHECR) sources.

%\%\%\%\%\%\%\%\%\%\%\%\%\%\%\%\%\%\%\%\%\%\%\%\%\%\%\%\%

\subsection{AGN contribution to the extragalactic $\gamma$-ray background}

The background $\gamma$-ray glow in the intergalactic medium far outside the Milky Way would, according to
present understanding, be resolved into scores of blazars and myriad faint radio galaxies and star-forming
galaxies that are below present sensitivity. There could be an additional contribution from dark-matter
annihilation \cite{pbjct20016} that would be enhanced in the relaxed environments of old and dwarf elliptical
galaxies. Determination of the extragalactic $\gamma$-ray background from the {\it Fermi}-LAT data requires
subtraction of Milky Way point sources and diffuse emission, including the Fermi bubble radiation.  What's
left over is the extragalactic $\gamma$-ray background, including resolved and unresolved extragalactic point
sources of radiation, leaving only a weak model dependence from the uncertain quasi-isotropic $\gamma$-ray
halo formed by cosmic-ray electrons in our Galaxy.

The  diffuse isotropic $\gamma$-ray background (DIGB), which is the
%EGB including  most recent Fermi-LAT report of the EGB finds
the total EGB spectrum including any residual, approximately
isotropic, Galactic foregrounds \cite{ack15},
is reported by the {\it Fermi}-LAT collaboration to be of the form
\begin{equation}
%\epsilon I_\epsilon^{DIGB} \cong 1.4 ({E_\gamma\over 100{\rm ~MeV}})^{-0.32\pm0.02}\;\exp({-E_\gamma\over 279\pm 52{\rm~GeV}})\;{\rm~keV~cm}^{-2}{{\rm s}}^{-1}{\rm sr}^{-1}\;,\; E_\gamma> 100{\rm ~MeV}\;,
\epsilon I_\epsilon^{DIGB} \cong 10^{-2} ({E_\gamma\over 100{\rm ~MeV}})^{-0.32}\;\exp({-E_\gamma\over 280{\rm~GeV}})\;
{{\rm~GeV}\over {\rm m}^{2}{\rm ~s~}{\rm sr}}\;,\; E_\gamma> 0.1{\rm GeV}\;.
\label{eIe}
\end{equation}
The high-energy cutoff involves, undoubtedly, EBL effects, which
is the subject of another paper in this {\it Comptes Rendus} issue \cite{Horns2015}.

One might imagine that the superposition of countless star-forming galaxies, or radio galaxies, could make up
most of the EGB, but in fact they each probably make up $\lesssim 20$\% each (and their relative contribution
is frequency-dependent).  A typical $L^*$ (the characteristic luminosity scale of the galaxy luminosity
function; see, e.g., \cite{2005ApJ...627L..89C}) spiral like the Milky Way, with a $\gamma$-ray luminosity
$L_\gamma\cong 10^{32} {\rm W} \times L_{39}$  or $10^{39}$ erg s$^{-1} \times L_{39}$ \cite{2010ApJ...722L..58S},
has a space density of $\approx 1$ per
300 Mpc$^3$, giving an approximate star-forming galaxy $\gamma$-ray emissivity $\dot\varepsilon_{L^*}\approx
10^{39}L_{39}/300$ Mpc$^{-3}$, implying a $\gamma$-ray intensity from $L^*$ galaxies $ I_{L^*}\approx R_{\rm
  H}\dot\varepsilon_{L^*}/4\pi \cong 8\times 10^{-8} \zeta_z L_{39}$ GeV cm$^{-2}$ s$^{-1}$ sr$^{-1}$).  The factor $\zeta_z
\approx$ 2 -- 3 accounts for increased star formation rate at $z \gtrsim 1$.  Taking into account a bolometric
factor implies that the $\gamma$-ray intensity from star-forming galaxies is at the 10 -- 20\% level of the
total EGB or DIGB intensity in eq.\ (\ref{eIe}). The cores of radio galaxies and radio galaxy emissions
\cite{2011ApJ...733...66I} probably make up another sub-dominant but non-negligble fraction of the EGB.

\begin{figure}[!t]
  \centering
  \includegraphics[width=6.6cm]{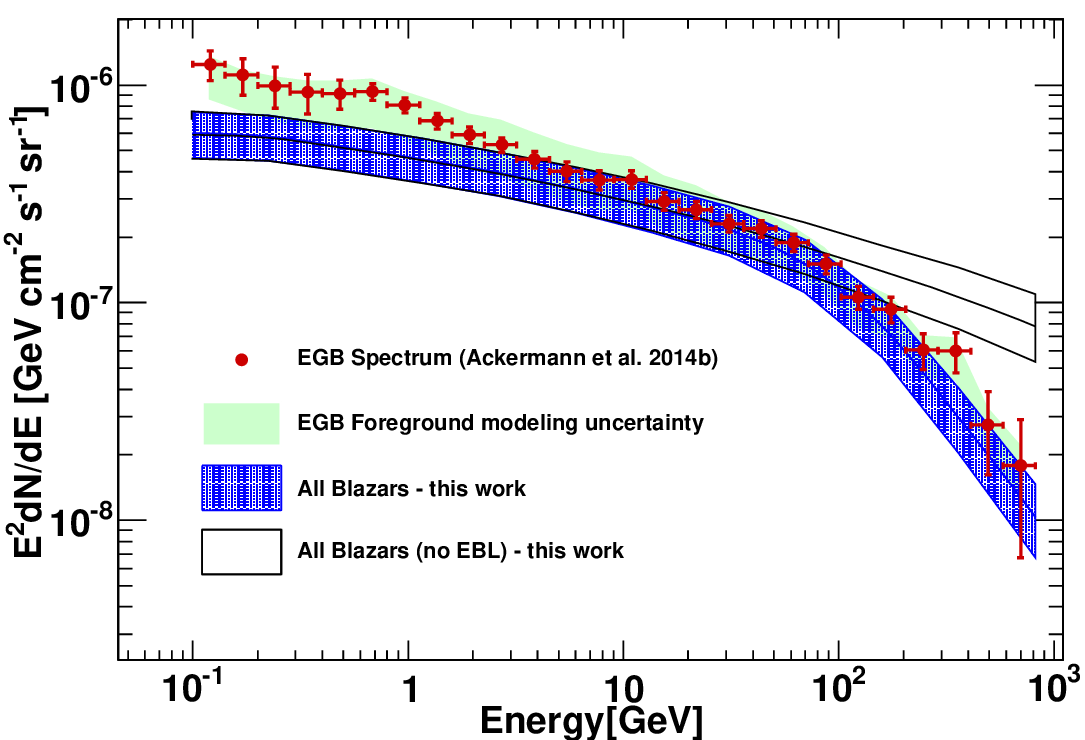}\hskip0.2in
  \includegraphics[width=7.0cm]{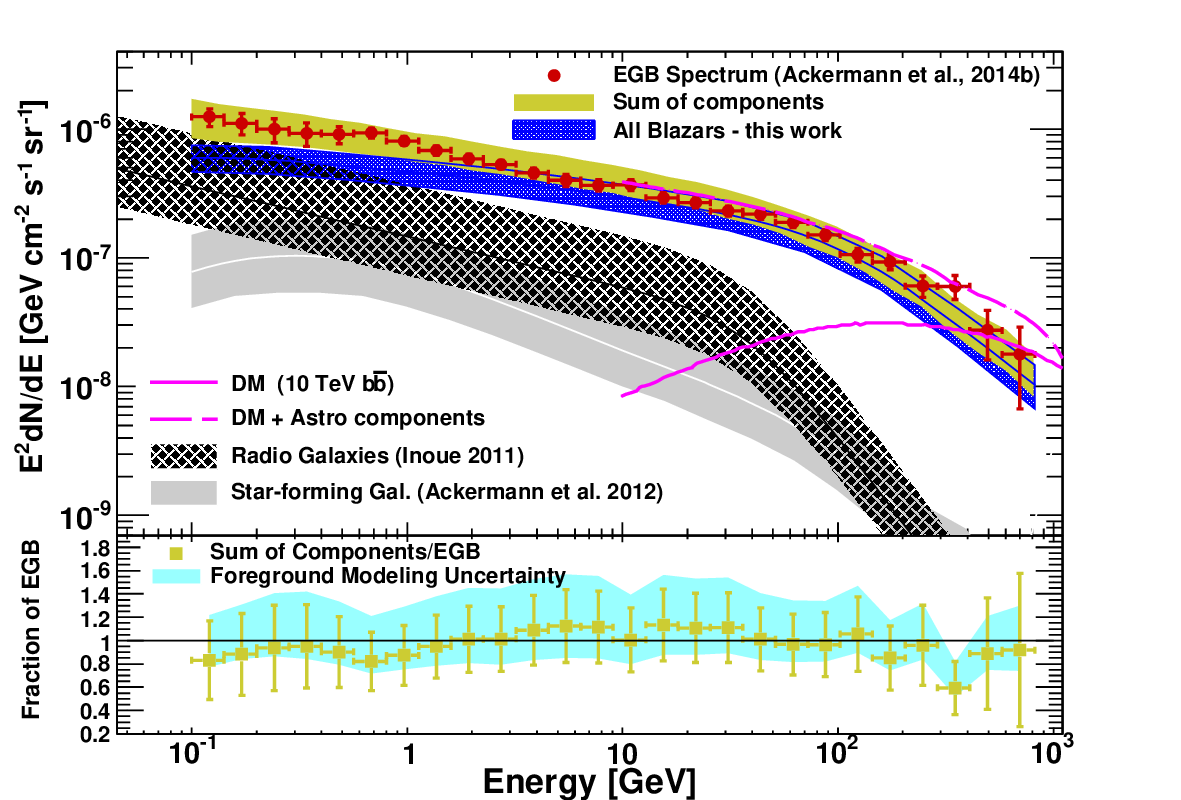}
  \caption{{\bf Contributions of various source classes to the EGB in the analysis of
 Ajello et al. (2015) \cite{2015ApJ...800L..27A}}.
{\it Left:} Integrated emission of
blazars are shown with and without redshift corrections, in comparison
with the EGB intensity \cite{ack15}. {\it Right:} Decomposition of the EGB intensity into
contributions from blazars, radio galaxies, and star-forming galaxies, leaving
only a residual intensity from dark matter.   }
  \label{fig:egb}
\end{figure}

The contribution of BL Lac objects \cite{Ajello2014} and FSRQs \cite{2012ApJ...751..108A} to the EGB is
achieved quantitatively by constructing the evolving luminosity distribution of the two source populations.  A
recent decomposition of the EGB based upon luminosity function of bright {\it Fermi}-LAT $\gamma$-ray blazars
\cite{2015ApJ...800L..27A} finds that $\approx 50$\% of the total EGB is resolved into blazars, but only
$\lesssim 20$\% of the remainder can originate from blazars (Fig. \ref{fig:egb}). Using models for the radio galaxy and
star-forming populations leaves the maximum contribution from other sources, including dark matter.  This
technique provides one of the strongest constraints on dark matter cross sections. See also
Ref.\ \cite{2011ApJ...736...40S}.

%%%%%%%%%%%%%%%%%%%%%%%%%%%%%%%%%%%%%%%%%%%%%%%%%%%%%%%%%

\section{Conclusions: $\gamma$-ray AGN}
\label{sec:conc}

The HE and VHE $\gamma$-ray sky has undergone a sea change in the last decade, thanks primarily
to the {\it Fermi} Gamma ray Space Telescope in space and new arrays of IACTs on the ground,
including the $4\times 13$ m VERITAS array, the $2\times 17$ m MAGIC telescopes, and the $4\times 13$ m H.E.S.S.-I
array which, with the addition of the big 28~m telescope, makes the H.E.S.S.-II.
The High Altitude Water Cherenkov
(HAWC) observatory, the successor to the Milagro all-sky TeV telescope, is now inaugurated
and taking data with its full complement of 300 tanks.

At the same time, only limited advances have been made on AGN science in the MeV regime.
Above 1 MeV, INTEGRAL has not been more sensitive than COMPTEL on CGRO, and only a small
fraction of its time was spent off the Galactic plane looking at extragalactic AGN.
The ``MeV blindness" ends below $\sim 100$ keV with the Swift-BAT surveys,
hard X-ray spectroscopy of sources with NuSTAR, and the upcoming ASTRO-H, while the {\it Fermi}-LAT pair-production
tracker design becomes most sensitive well above $\sim 100$ MeV. At the present
moment, there is no consensus
and few designs for a successor to {\it Fermi}-LAT in the $\approx 0.1$ -- 100 GeV range.

\begin{figure}[!t]
  \centering
  \includegraphics[width=10.0cm]{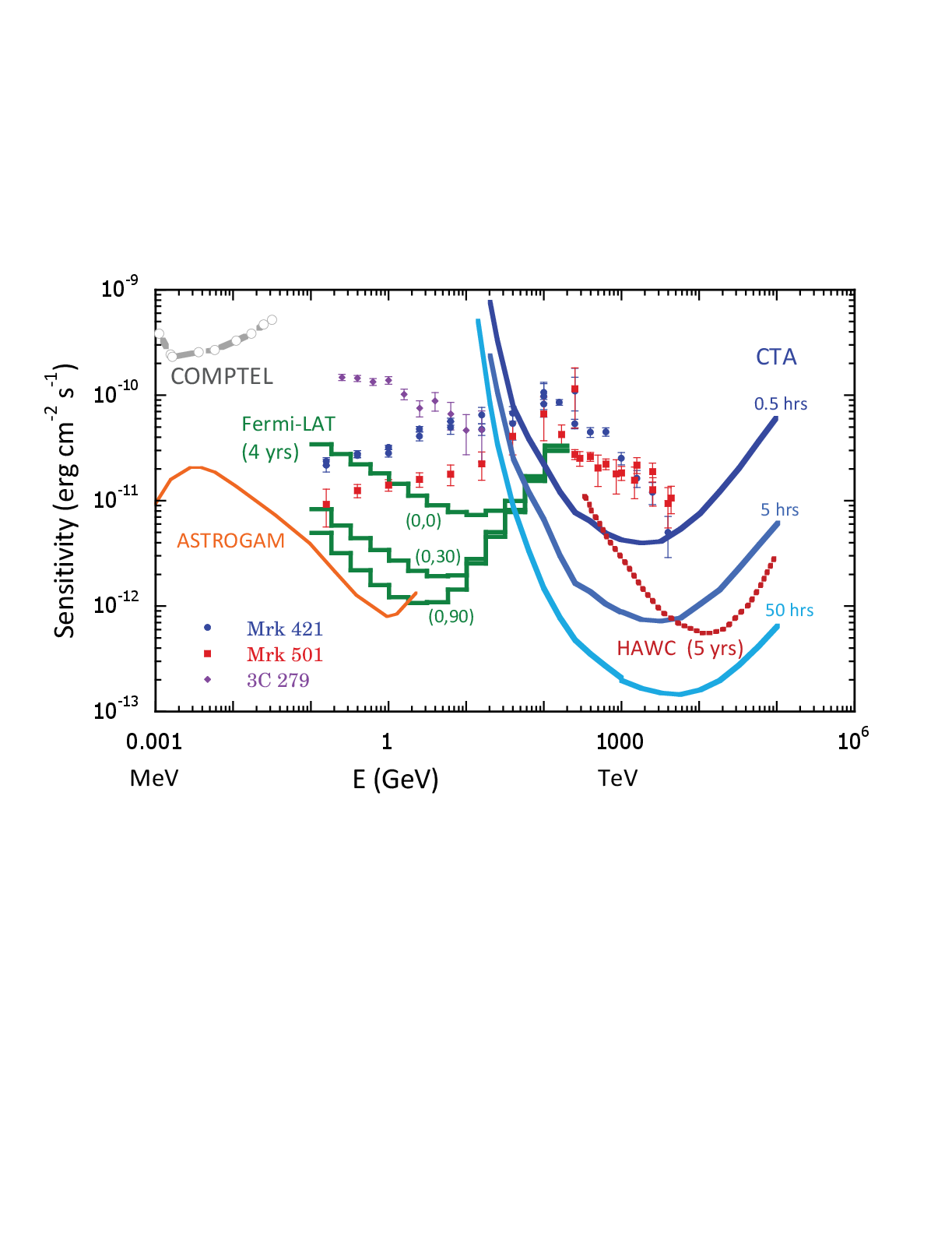}
  \caption{{\bf Sensitivity of $\gamma$-ray observatories.} Nominal CTA sensitivity requirements for exposures of
0.5 hrs, 5 hrs, and 50 hrs, as labeled \cite{2011ExA....32..193A}. The field of view of CTA is a few degrees
wide. The HAWC sensitivity is for 5 yrs of exposure \cite{jk}. The {\it Fermi}-LAT differential sensitivity
plot \cite{FermiLAT} is for four years, in 4 bins per energy decade, at Galactic longitude $\ell$ and latitude
$b$ of $(\ell,b) = (0,0), (0, 30),$ and $(0,90)$, as labeled, that is, at the Galactic center, intermediate
latitudes, and at the north Galactic pole, respectively. The sensitivity uses Pass 7 {\it Fermi}-LAT instrument response to detect
 a point source of $\gamma$ rays with a $-2$ power-law photon number spectrum, and $>10$ photons
per energy band. The Pass 8 instrument responses made available recently might change these estimations.
%The histograms are for 3 different locations (shown in Galactic coordinates).
The COMPTEL sensitivity is for the entire CGRO mission from 1991 -- 2000 \cite{2004NewAR..48..193S}. The ASTROGAM sensitivity is for three years in the scanning mode at high Galactic latitude  \cite{jk}.  }
  \label{fig:sensitivity}
\end{figure}

The next big activity in $\gamma$-ray astronomy is, of course, the Cherenkov Telescope Array (CTA), with its
siting in both the Northern and Southern Hemispheres recently announced.  CTA, and other proposed $\gamma$-ray
activities, are discussed elsewhere in this volume \cite{jk}. With regard to
$\gamma$-ray AGN specifically, Fig.\ \ref{fig:sensitivity} shows {\it Fermi}-LAT, CTA, and HAWC sensitivity
curves, in comparison with $\nu F_\nu$ SEDs of Mrk 421, Mrk 501, and 3C 279 (data from the same epochs as in
Fig.\ \ref{fig:sequence}). The sensitivity curves are relative to SEDs, thus to differential photon fluxes multiplied by $E^2$, so the
data comparison should be made with care, not to mention the differences in exposure and field of view in the
different curves, as described in the caption to Fig.\ \ref{fig:sensitivity}.  The sensitivity graph of
ASTROGAM \cite{jk}, recently proposed for an ESA M4 opportunity, is also shown in Fig.\ \ref{fig:sensitivity}.

One thing that will surely hamper blazar studies in the future is the lack of spectral definition in the $\sim 0.2$ -- 100 MeV
range, where separate IR radiation fields deep in the black hole are exposed through their Compton signatures.
This is the regime of the hypothetical ``extreme" blazar, whose $\nu F_\nu$ peak synchrotron frequency is at
MeV energies. This is where the $\nu F_\nu$ $\gamma$-ray Compton peaks of LSP blazars are found, and where the SSC emission component
gives way to the EC component when there is an external radiation field.
This is where, if blazars share any similarity with $\gamma$-ray bursts \cite{fpGRB}, a quasi-thermal
photosphere from the blazar engine could be detected.

\subsection{GeV-TeV connection}

The combination of {\it Fermi}-LAT and the current generation of IACTs allows for unprecedented $\gamma$-ray
coverage, with the high-energy component of blazar SEDs sampled with hitherto unmatched resolution, permitting a much
improved characterization of the peak emission and $\gamma$-ray spectrum \cite{Senturk2013}. Given the slight overlap in
energy with ground-based detectors, {\it Fermi}-LAT, with its wide field of view, has increased
the efficiency of catching VHE emission of flaring blazars with IACTs. The higher accuracy of spectral measurements
in both HE and VHE has improved the search for direct evidence of EBL-induced spectral steepening in GeV -- TeV blazars,
since the HE and VHE spectral photon indices for a
sample of GeV-TeV blazars show a clear correlation with redshift \cite{Sanchez2013}. The size of the spectral
break is in all cases compatible or larger than the one expected from HE -- VHE $\gamma$ rays interacting with
the EBL, which can sometimes mean that the {\it intrinsic} SED is not peaking where the {\it observed} SED
peaks.

{\it Fermi}-LAT continues to work hand in glove with the IACTs and water Cherenkov detectors.
The population explosion of VHE sources shown in Fig.\ \ref{fig:tevcat} right was, after 2008,
in significant part due to the {\it Fermi}-LAT detection
of hard, $\gtrsim 100$ GeV photons from high-latitude sources. The sky at $\gtrsim 10$ GeV, as reported in
the First {\it Fermi}-LAT Hard Source Catalog (1FHL) \cite{2013ApJS..209...34A}, contains 514 objects, of which $\sim 75$\% are
BL Lac objects. The Second Catalog of hard {\it Fermi}-LAT sources (2FHL)
at $> 50$ GeV \cite{2015arXiv150302664A}
based on Pass 8 analysis comprises $\sim 350$ sources over the full sky,
of which $\sim 145$ are found in the VHE catalog.
Above 10 GeV, the Fermi bubbles stand out above the background in the {\it Fermi}-LAT sky maps,
but are difficult to see with IACTs because of their extension. With a systematic VHE survey,
statistical studies comparing
VHE $\gamma$-ray spectral indices with luminosity and peak frequencies,
such as those found in with {\it Fermi}-LAT.
HAWC will provide the best large field of view
$\gamma$-ray sky maps above $\approx 1$ TeV, and will be sensitive
to day-scale flaring of the brightest BL Lac objects.

%\vskip1.0in

\subsection{Whither $\gamma$-ray AGN astronomy?}

Now that the {\it Fermi} Gamma ray Space Telescope has begun its 8$^{th}$ year of operations, and the
VERITAS, MAGIC, and H.E.S.S.\ IACT arrays are, within $\approx 5$ yrs, to be overshadowed by CTA, we may ask
``whither $\gamma$-ray astronomy of AGN?" that is, where are we headed? The important thing is that technology
is evolving, even when the observatories go dark. The extraordinarily dynamic SEDs of $\gamma$-ray AGN now
measured with ground-based and space-based facilities reveal the nature of the extreme black-hole jet
environment, and tell us how black holes accumulate mass and dissipate energy. In statistical samples, they
tell us how blazars and radio galaxies evolve over cosmic time. Both observationally and theoretically, major
advances have been made in measuring and deciphering the SEDs of $\gamma$-ray emitting blazars in the quarter
century since the $\gamma$-ray blazar class was discovered.  Some of the big puzzles that $\gamma$-ray AGN
astronomy can address and solve are: What powers the jets of supermassive black holes? What causes the
variability time scale to be so short during intense flaring periods?  Do shocks or turbulence in blazar jets
accelerate the ultra-high energy cosmic rays \cite{cras2014}, or do UHECRs have a different origin?  What acceleration and
radiation physics explains the peak synchrotron and Compton frequencies of blazars of different types?  How
much can we limit the dark-matter cross section by modeling the $\gamma$-ray background light?

In the last decade, the number of $\gamma$-ray emitting AGN has increased by more than an order of
magnitude, and new classes of extragalactic $\gamma$-ray emitters have emerged in both the HE and VHE regimes. The better
sensitivity of these observatories has also improved our knowledge of the time-dependent behavior of AGN,
further pushing the need for dynamical radiative modeling, which are more complex than static models but allow
 details about the acceleration mechanisms to be modeled. The AGN paradigm receives new challenges,
for example, the inability of simple kinematic shell models to explain rapid variability, which may be achieved
through magnetic reconnection; breakdown of one-zone models;  extremely hard spectra of flaring events in FSRQs;
identification of robust correlations in statistical samples of blazars; etc. The advent of the next generation of
ground-based $\gamma$-ray telescopes, namely HAWC and CTA, will open up the $\gamma$-ray sky and provide
data to answer important questions in high-energy astronomy and particle astrophysics, but the lack of a sensitive
MeV-regime space telescope will impede a deeper understanding of $\gamma$-ray AGN and, indeed, the high-energy sky.

%%%%%%%%%%%%%%%%%%%%%%%%%%%%%%%%%%%%%%%%%%%%%%%%%%%%%%%%%
\section*{Acknowledgements}
%%%%%%%%%%%%%%%%%%%%%%%%%%%%%%%%%%%%%%%%%%%%%%%%%%%%%%%%%

We thank
 Dr. B.\ Lott,
 Dr.\ J.\ Finke,  Dr.\ C.~C.\ Cheung, and Dr.\ M.\ Ajello for discussions and correspondence,
 Dr. D.~J.\ Thompson and Dr.\ F.\ Krauss for suggestions and corrections,
 Dr.\ A.\ Marscher for permission to reproduce Fig.\ \ref{fig:radiogamma}, left,
 Dr.\ L.\ Fuhrmann for permission to reproduce Fig.\ \ref{fig:radiogamma}, right,
 Dr.\ G.\ Fossati for permission to reproduce Fig.\ \ref{fig:sequence}, left, and
 Dr.\ E.\ Meyer for permission to reproduce Fig.\ \ref{fig:sequence}, center.
The work of C.D.D.\ is supported by the Chief of Naval Research.

%%%%%%%%%%%%%%%%%%%%%%%%%%%%%%%%%%%%%%%%%%%%%%%%%%%%%%%%%


\begin{thebibliography}{00}
%%%%%%%%%%%%%%%%%%%%%%%%%%%%%%%%%%%%%%%%%%%%%%%%%%%%%%%%%
\bibitem{ohm} Ohm, S., 2015, this volume.
\bibitem{ack15} Ackermann, M., Ajello, M., Albert, A., et al.\ 2015, Astrophys.\ Jour.\ 799, 86.
\bibitem{dg95} Dermer, C.~D., \& Gehrels, N.\ 1995, Astrophys.\ Jour.\ 447, 103.
\bibitem{Urry1995} C.\ M.\ Urry and P.\ Padovani, PASP, 107 (1995) 803.
\bibitem{fr74} Fanaroff, B.~L., \& Riley, J.~M.\ 1974, MNRAS 167, 31P.
\bibitem{gio12} Giommi, P., Padovani, P., Polenta, G., et al.\ 2012, MNRAS 420, 2899.
\bibitem{intro} B. Degrange \& G. Fontaine, C. R. Physique, 16 (2015) 587.
\bibitem{Swanenburg1978} B. N. Swanenburg \textit{et al}, Nature 275 (1978) 298.
\bibitem{djt} D. J. Thomson, 2015, C. R. Physique, 16 (2015) 600.
\bibitem{bd83}  Bassani, L., \& Dean, A.~J.\ 1983, Space Sci.\ Rev.\ 35, 367
\bibitem{dsm92} Dermer, C.~D., Schlickeiser, R., \& Mastichiadis, A.\ 1992, Astron.\ Astrophys.\ 256, L27
%\bibitem{har99} Hartman, R.~C., Bertsch, D.~L., Bloom, S.~D., et al.\ 1999, Astrophys.\ Jour.\ Supp.\ 123, 79
\bibitem{Hartman1999}    R. C. Hartman \textit{et al}, Astrophys.\ Jour.\ Supp.\ 123 (1999) 79
\bibitem{nama} M. de Naurois \& D. Mazin, 2015, C. R. Physique, 16 (2015) 610.
\bibitem{wee89} Weekes, T.~C., Cawley, M.~F., Fegan, D.~J., et al.\ 1989, Astrophys.\ Jour.\ 342, 379
\bibitem{pun92} Punch, M., Akerlof, C.~W., Cawley, M.~F., et al.\ 1992, Nat.\ 358, 477
\bibitem{Biteau2013} J. Biteau, PhD thesis, 2013, pastel-00822242, version 1.
\bibitem{dm09} C.D.\ Dermer \& G.\ Menon, High Energy Radiation from Black Holes (Princeton) (2009).
\bibitem{1985ApJ...289..109U} Unwin, S.~C., Cohen, M.~H., Biretta, J.~A., et al.\ 1985, Astrophys.\ Jour.\ 289, 109.
\bibitem{kon81} K\"onigl, A.\ 1981, Astrophys.\ Jour.\ 243, 700
\bibitem{bbr84} Begelman, M.~C., Blandford, R.~D., \& Rees, M.~J.\ 1984, Rev.\ Mod.\ Phys.\ 56, 255.
\bibitem{fic94} Fichtel, C.~E., Bertsch, D.~L., Chiang, J., et al.\ 1994, Astrophys.\ Jour.\ Supp.\ 94, 551.
\bibitem{tho95} Thompson, D.~J., Bertsch, D.~L., Dingus, B.~L., et al.\ 1995, Astrophys.\ Jour.\ Supp.\ 101, 259.
\bibitem{LBAS} Abdo, A.~A., Ackermann, M., Ajello, M., et al.\ 2009, Astrophys.\ Jour.\ 700, 597.
\bibitem{1LAC} Abdo, A.~A., Ackermann, M., Ajello, M., et al.\ 2010, Astrophys.\ Jour.\ 715, 429.
\bibitem{2LAC}  M. Ackermann \textit{et al}, Astrophys.\ Jour.\ 743 (2011) 171; (e)  Astrophys.\ Jour.\ 806 (2015) 144.
\bibitem{3LAC}  Ackermann, M., Ajello, M., Atwood, W., et al.\ 2015, Astrophys.\ Jour.\ 810, 14.
\bibitem{sre92} Sreekumar, P., Bertsch, D.~L., Dingus, B.~L., et al.\ 1992, Astrophys.\ Jour.\ Lett.\ 400, L67.
\bibitem{abd10SMC} Abdo, A.~A., Ackermann, M., Ajello, M., et al.\ 2010, Astron.\ Astrophys.\ 523, AA46.
\bibitem{abd10sb} Abdo, A.~A., Ackermann, M., Ajello, M., et al.\ 2010, Astrophys.\ Jour.\ Lett.\ 709, L152.
\bibitem{ace09} Acero, F., Aharonian, F., Akhperjanian, A.~G., et al.\ 2009, Science 326, 1080.
\bibitem{VER09} VERITAS Collaboration, Acciari, V.~A., Aliu, E., et al.\ 2009, Nature 462, 770.
\bibitem{suel} M. Su \& C. van Eldik, 2015, C. R. Physique, 16 (2015) 686.
\bibitem{san15} D.A.\ Sanchez, F.\ Brun, C.\ Couturier, A.\ Jacholkowska, J.-P.\ Lenain, in the Fourth {\it Fermi} Symposium.
\bibitem{sha13} M.~S.\ Shaw, R.~W.\ Romani,  G.~Cotter, et al. 2013, Astrophys.\ Jour.\ 764, 135.
\bibitem{Ajello2014}  M. Ajello \textit{et al}, Astrophys. Jour.\ 780 (2014) 73.
\bibitem{abd11} Abdo, A.~A., Ackermann, M., Ajello, M., et al.\ 2011, Astrophys.\ Jour.\ Lett.\ 733, LL26.
\bibitem{Abdo2010}  A. A. Abdo {\it et al}, Astrophys.\ Jour.\ 720 (2010) 912.
\bibitem{mg85} Marscher, A.~P., \& Gear, W.~K.\ 1985, Astrophys.\ Jour.\ 298, 114.
\bibitem{fuh14} L.\ Fuhrmann, S.\ Larsson, J.\ Chiang, et al.\ 2014, MNRAS 441, 1899.
\bibitem{MAGN} Abdo, A.~A., Ackermann, M., Ajello, M., et al.\ 2010, Astrophys.\ Jour.\ 720, 912.
\bibitem{lh05} Lister, M.~L., \& Homan, D.~C.\ 2005, Astron.\ Jour.\ 130, 1389.
\bibitem{2012A&A...538L...1K} Kadler, M., Eisenacher, D., Ros, E., et al.\ 2012, Astron.\ \& Astrophys.\ 538, L1.
\bibitem{ale10} Aleksi{\'c}, J., Antonelli, L.~A., Antoranz, P., et al.\ 2010, Astrophys.\ Jour.\ Lett.\ 723, L207.
\bibitem{ner10} Neronov, A., Semikoz, D., \& Vovk, I.\ 2010, Astron.\ Astrophys.\ 519, LL6.
\bibitem{geo08} Georganopoulos, M., Sambruna, R.~M., Kazanas, D., et al.\ 2008, Astrophys.\ Jour.\ Lett.\ 686, L5.
\bibitem{abd10cena} Abdo, A.~A., Ackermann, M., Ajello, M., et al.\ 2010, Science, 328, 725.
\bibitem{mig11} Migliori, G., Grandi, P., Torresi, E., et al.\ 2011, Astron.\ Astrophys.\ 533, AA72.
\bibitem{abd09c} Abdo, A.~A., et al.\ 2009, Astrophys.\ Jour.\ Lett.\ 707, L142.
%(RL-NLS1s)
\bibitem{pog00} Pogge, R.~W.\ 2000, New Astron.\ Rev.\ 44, 381.
\bibitem{Bottcher2002}  M. B\"ottcher and C. Dermer, ApJ, 564 (2002) 86.
\bibitem{Aharonian2003} F. Aharonian {\it et al}, Astron.\ Astrophys.\ 403 (2003) 1.
%\bibitem{Abdo2010}  A. A. Abdo {\it et al}, Astrophys.\ Jour.\ 720 (2010) 912.
\bibitem{Acero2015} F. Acero \textit{et al}, ApJS 218 (2015) 23.
\bibitem{Abramowski2012} Abramowski, A., {\it et al}, Astrophys. Jour. 2012, 746, 2.
\bibitem{Aleksic2014} Aleksi\'{c}, J., {\it et al}, Science, 346, 1080.
\bibitem{2001ASPC..250..294W} Werner, P.~N., Worrall, D.~M., \& Birkinshaw, M.\ 2001, Particles and Fields in Radio Galaxies Conference, 250, 294.
\bibitem{werner}http://www.werner.lu/pn/phd/ngc6251.html.
\bibitem{2008ApJ...680..911S} Stawarz, {\L}., Ostorero, L., Begelman, M.~C., et al.\ 2008, Astrophys.\ J.\ 680, 911.
\bibitem{2014A&A...562A...4M} M{\"u}ller, C., Kadler, M., Ojha, R., et al.\ 2014, Astron.\ \& Astrophys.\ 562, A4.
\bibitem{ack12} Ackermann, M., Ajello, M., Allafort, A., et al.\ 2012, Astrophys.\ Jour.\ 747, 104.
\bibitem{bla90} Blandford, R.~D.,
Netzer, H., Woltjer, L., eds.\ Courvoisier, T.~J.-L.,
\& Mayor, M.\ Saas-Fee, 1990, Active Galactic Nuclei.
\bibitem{Horns2015} Horns, D. \& Jacholkowska, A., 2015, this volume.
\bibitem{abd103C279} Abdo, A.~A., Ackermann, M., Ajello, M., et al.\ 2010, Nature 463, 919.
\bibitem{abd11mrk501} Abdo, A.~A., et al.\ 2011, Astrophys.\ Jour., 727, 129.
\bibitem{abd09M87} Abdo, A.~A., Ackermann, M., Ajello, M., et al.\ 2009, Astrophys.\ Jour.\ 707, 55.
\bibitem{abd09NGC1275} Abdo, A.~A., Ackermann, M., Ajello, M., et al.\ 2009, Astrophys.\ Jour.\ 699, 31.
\bibitem{henri06} Henri, G. \& Saug\'e, L., \ 2006, Astrophys.\ Jour.\ 640, 185.
\bibitem{gtc05} Ghisellini, G., Tavecchio, F., \& Chiaberge, M. 2005, Astron.\ Astrophys.\ 432, 401.
\bibitem{gk03} Georganopoulos, M., \& Kazanas, D.\ 2003, Astrophys.\ Jour.\ Lett.\ 594, L27.
\bibitem{1966ApJ...146..634P} Pauliny-Toth, I.~I.~K., \& Kellermann, K.~I.\ 1966,  Astrophys.\ Jour.\ 146, 634.
\bibitem{1973ApJ...186..791J} Jones, T.~W., \& Burbidge, G.~R.\ 1973, Astrophys.\ Jour.\ 186, 791.
\bibitem{1993ApJ...411..133K} Kniffen, D.~A., Bertsch, D.~L., Fichtel, C.~E., et al.\ 1993, Astrophys.\ Jour.\ 411, 133.
\bibitem{2011ApJ...733L..26A} Abdo, A.~A., Ackermann, M., Ajello, M., et al.\ 2011, Astrophys.\ J.\ 733, LL26.
\bibitem{2011ApJ...733...19T} Tanaka, Y.~T., Stawarz, {\L}., Thompson, D.~J., et al.\ 2011, Astrophys.\ J.\ 733, 19.
\bibitem{2013ApJ...766L..11S} Saito, S., Stawarz, {\L}., Tanaka, Y.~T., et al.\ 2013, Astrophys.\ J.\ 766, LL11.
\bibitem{2013MNRAS.431..824B} Brown, A.~M.\ 2013, MNRAS 431, 824.
\bibitem{2011A&A...530A..77F} Foschini, L., Ghisellini, G., Tavecchio, F., Bonnoli, G., \& Stamerra, A.\ 2011, Astron.\ \& Astrophys.\ 530, AA77.
\bibitem{2013MNRAS.430.1324N} Nalewajko, K.\ 2013, MNRAS 430, 1324.

\bibitem{Lott14} Lott, B., ``Assessing the short-timescale variabilty in LAT blazars," {\it Fermi} Symposium, Nagoya, Japan (2014).
\bibitem{2000ApJ...539L...9F} Ferrarese, L., \& Merritt, D., \ 2000, Astrophys. \ Jour. \ 539, 9.
\bibitem{urr00} Urry, C.~M., Scarpa, R., O'Dowd, M., et al.\ 2000, Astrophys.\ J.\ 532, 816.
\bibitem{1995ApJ...449L..99M} Macomb, D.~J., Akerlof, C.~W., Aller, H.~D., et al.\ 1995, Astrophys.\ J.\ 449, L99.
\bibitem{Aharonian2007b} Aharonian, F.~A., {\it et al}, 2007, Astrophys. Jour., 664, L71.
\bibitem{2012A&A...539A.149H} H.E.S.S.~Collaboration, Abramowski, A., Acero, F., et al.\ 2012, Astron.\ Astrophys.\ 539, A149.

\bibitem{2007ApJ...669..862A} Albert, J., Aliu, E., Anderhub, H., et al.\ 2007, Astrophys.\ J.\ 669, 862.
\bibitem{2012AIPC.1505..514F} Fortson, L., VERITAS Collaboration,
\& {\it Fermi}-LAT Collaborators 2012, American Institute of Physics Conference Series, 1505, 514.
\bibitem{2008MNRAS.384L..19B} Begelman, M.~C., Fabian, A.~C., \& Rees, M.~J.\ 2008, MNRAS 384, L19.
\bibitem{Aharonian2007} Aharonian, F.~A., {\it et al}, 2007, Astron.\ Astrophys.\ 473, 25.
\bibitem{2013arXiv1307.8091C} Cerruti, M., \& for the VERITAS Collaboration 2013, arXiv:1307.8091.
\bibitem{2011ApJ...730L...8A} Aleksi{\'c}, J., Antonelli, L.~A., Antoranz, P., et al.\ 2011, Astrophys.\ J.\ 730, LL8.
\bibitem{2008Sci...320.1752M} MAGIC Collaboration, Albert, J., Aliu, E., et al.\ 2008, Science, 320, 1752.
\bibitem{2011A&A...530A...4A} Aleksi{\'c}, J., Antonelli, L.~A., Antoranz, P., et al.\ 2011, Astron.\ \& Astrophys.\ 530, AA4.
\bibitem{2010ApJ...710L.126M} Marscher, A.~P., Jorstad, S.~G., Larionov, V.~M., et al.\ 2010, Astrophys.\ Jour.\ 710, L126.
\bibitem{2008ApJ...676L..13V} Vercellone, S., Chen, A.~W., Giuliani, A., et al.\ 2008, Astrophys.\ J.\ 676, L13.
\bibitem{2009ApJ...699..817A} Abdo, A.~A., Ackermann, M., Ajello, M., et al.\ 2009, Astrophys.\ J.\  699, 817.
\bibitem{2010ApJ...710.1271A} Abdo, A.~A., Ackermann, M., Ajello, M., et al.\ 2010, Astrophys.\ J.\ 710, 1271.
\bibitem{2010ApJ...721.1383A} Ackermann, M., Ajello, M., Baldini, L., et al.\ 2010, Astrophys.\ J.\  721, 1383.
%\bibitem{2011ApJ...733L..26A} Abdo, A.~A., Ackermann, M., Ajello, M., et al.\ 2011, Astrophys.\ J.\ 733, LL26.
\bibitem{1997ApJ...476..692M} Mattox, J.~R., Wagner, S.~J., Malkan, M., et al.\ 1997, Astrophys.\ J.\ 476, 692.
\bibitem{2010ApJ...722..520A} Abdo, A.~A., Ackermann, M., Ajello, M., et al.\ 2010, Astrophys.\ Jour.\ 722, 520.
\bibitem{2010ApJ...717L.118P} Poutanen, J., \& Stern, B.\ 2010, Astrophys.\ J.\ 717, L118.
\bibitem{2011MNRAS.417L..11S} Stern, B.~E., \& Poutanen, J.\ 2011, MNRAS, 417, L11.
\bibitem{Cerruti2013} Cerruti, M., Dermer, C.~D., Lott, B., Boisson, C., \& Zech, A.\ 2013, Astrophys.\ Jour.\ 771, LL4.
\bibitem{der14} Dermer, C.~D., Cerruti, M., Lott, B., Boisson, C., \& Zech, A.\ 2014, Astrophys.\ Jour.\ 782, 82.
\bibitem{2000ApJ...545..107B}  B{\l}a{\.z}ejowski, M., Sikora, M., Moderski, R., \& Madejski, G.~M.\ 2000, Astrophys.\ J.\ 545, 107.
\bibitem{2008Natur.452..966M} Marscher, A.~P., Jorstad, S.~G., D'Arcangelo, F.~D., et al.\ 2008, Nature 452, 966.
\bibitem{2011ApJ...726L..13A} Agudo, I., Jorstad, S.~G., Marscher, A.~P., et al.\ 2011, Astrophys.\ Jour.\ 726, LL13.
\bibitem{2013arXiv1303.2039A} Agudo, I., Marscher, A.~P., Jorstad, S.~G., et al.\ 2013, arXiv:1303.2039.
\bibitem{2010ApJ...722L...7P} Pushkarev, A.~B., Kovalev, Y.~Y., \& Lister, M.~L.\ 2010, Astrophys.\ Jour.\ 722, L7.
\bibitem{2011ApJS..194...29R} Richards, J.~L., Max-Moerbeck, W., Pavlidou, V., et al.\ 2011, Astrophys.\ Jour.\ Supp.\ 194, 29.
\bibitem{2010ApJ...714L.303F} Finke, J.~D., \& Dermer, C.~D.\ 2010, Astrophys.\ Jour.\ 714, L303.
\bibitem{2014ApJ...790...45P} Pacciani, L., Tavecchio, F., Donnarumma, I., et al.\ 2014, Astrophys.\ Jour.\ 790, 45.
\bibitem{2015arXiv150204699H} Hayashida, M., Nalewajko, K., Madejski, G.~M., et al.\ 2015, arXiv:1502.04699.
\bibitem{2012MNRAS.425.2519N} Nalewajko, K., Begelman, M.~C., Cerutti, B., Uzdensky, D.~A., \& Sikora, M.\ 2012, MNRAS 425, 2519.
\bibitem{2013ApJ...771...57A} Ackermann, M., Ajello, M., Albert, A., et al.\ 2013, Astrophys.\ Jour.\  771, 57.
\bibitem{2013JInst...8P6008A} Anderhub, H., Backes, M., Biland, A., et al.\ 2013, Journal of Instrumentation, 8, P06008.
\bibitem{Fossati1998}    G. Fossati {\it et al}, MNRAS, 299 (1998) 433.
\bibitem{mey11} Meyer, E.~T., Fossati,
G., Georganopoulos, M., \& Lister, M.~L.\ 2011, Astrophys.\ Jour.\ 740, 98.
\bibitem{Lind1985} Lind, K. R., \& Blandford , R. D.\ 1985, Astrophys.\ Jour.\ 295, 358.
\bibitem{Aharonian2009} Aharonian, F., et al.\ 2009, Astrophys.\ Jour.\ 696, 150.
\bibitem{hay12} Hayashida, M., Madejski, G.~M., Nalewajko, K., et al.\ 2012, Astrophys.\ Jour.\ 754, 114.
\bibitem{Ghisellini1998}     G. Ghisellini {\it et al}, MNRAS, 301 (1998) 451.
\bibitem{Finke2013}    J. Finke, ApJ, 763 (3013) 134.
\bibitem{Ghisellini2009}      G. Ghisellini, L. Maraschi, and F. Tavecchio, MNRAS, 396 (2009) 10.
\bibitem{ghi11} Ghisellini, G.,
Tavecchio, F., Foschini, L., \& Ghirlanda, G.\ 2011, MNRAS 414, 2674.
\bibitem{gtf10} Ghisellini, G., Tavecchio, F., Foschini, L., et al.\ 2010, MNRAS 402, 497.
\bibitem{Nandikotkur2007} G. Nandikotkur {\it et al}, Astrophys.\ Jour.\ 657 (2007) 706.
\bibitem{fermibright} Abdo, A.~A., Ackermann, M., Agudo, I., et al.\ 2010, Astrophys.\ Jour.\ 716, 30.
\bibitem{der15}  Dermer, C.~D., Yan, D.,
Zhang, L., Finke, J.~D., \& Lott, B.\ 2015, Astrophys.\ Jour.\ 809, 174.
\bibitem{2004A&A...422..103M} Massaro, E., Perri, M., Giommi, P., Nesci, R., \& Verrecchia, F.\ 2004, Astron.\ Astrophys.\ 422, 103.
\bibitem{Malkov2000} M. Malkov and L. O'C Drury, Rep. Prog. Phys. 64 (2000) 429.
\bibitem{der14a} Dermer, C.~D.\ 2014, Memorie della Societa Astronomica Italiana 86 (2015) 13.
\bibitem{bhk12} B\"ottcher, M., Harris, D.~E., \& Krawczynski, H.\ 2012, Relativistic Jets from Active Galactic Nuclei, Berlin: Wiley.
\bibitem{abd11b} Abdo, A.~A.,  et al.\ 2011, Astrophys.\ Jour.\ 736, 131.
\bibitem{gub10} Giannios, D., Uzdensky, D.~A., \& Begelman, M.~C.\ 2010, MNRAS, 402, 1649.
\bibitem{mj10} Marscher, A.~P., \& Jorstad, S.~G.\ 2010, arXiv:1005.5551.
\bibitem{np12} Narayan, R., \& Piran, T.\ 2012, MNRAS, 420, 604.
\bibitem{mar12} Marscher, A.~P., Jorstad, S.~G., Agudo, I., MacDonald, N.~R., \& Scott, T.~L.\ 2012, arXiv:1204.6707.
\bibitem{nal12} Nalewajko, K., Begelman, M.~C., Cerutti, B., Uzdensky, D.~A., \& Sikora, M.\ 2012, MNRAS 425, 2519.
\bibitem{ssb12} Subramanian, P., Shukla, A., \& Becker, P.~A.\ 2012, MNRAS 423, 1707.
\bibitem{2004ApJ...601..151K} Krawczynski, H., Hughes, S.~B., Horan, D., et al.\ 2004, Astrophys.\ Jour.\ 601, 151.
\bibitem{2005ApJ...621..176B} B{\"o}ttcher, M.\ 2005, Astrophys.\ Jour.\ 621, 176.
\bibitem{2010APh....33...81E} Essey, W., \& Kusenko, A.\ 2010, Astroparticle Physics, 33, 81.
\bibitem{2010PhRvL.104n1102E} Essey, W., Kalashev, O.~E., Kusenko, A.,  \& Beacom, J.~F.\ 2010, Physical Review Letters, 104, 141102.
\bibitem{aar14} Aartsen, M.~G., Ackermann, M., Adams, J., et al.\ 2014, Physical Review Letters, 113, 101101.
\bibitem{2014A&A...566L...7K} Krau{\ss}, F., Kadler, M., Mannheim, K., et al.\ 2014, Astron.\ \& Astrophys.\ 566, L7.
\bibitem{aha02} Aharonian, F.~A.\ 2002, MNRAS 332, 215.
\bibitem{mue03} M{\"u}cke, A., Protheroe, R.~J., Engel, R., Rachen, J.~P., \& Stanev, T.\ 2003, Astroparticle Physics, 18, 593.
\bibitem{ad01} Atoyan, A., \& Dermer, C.~D.\ 2001, Physical Review Letters, 87, 221102.
\bibitem{dmi14} Dermer, C.~D., Murase, K., \& Inoue, Y.\ 2014, Journal of High Energy Astrophysics, 3, 29.
\bibitem{gt08} Ghisellini, G., \& Tavecchio, F.\ 2008, MNRAS 387, 1669
\bibitem{mid14} Murase, K., Inoue, Y.,
\& Dermer, C.~D.\ 2014, Phys.\ Rev.\ D 90, 023007.
\bibitem{bsk13} B{\"o}ttcher, M., Reimer, A., Sweeney, K., \& Prakash, A.\ 2013, Astroph.\ J.\ 768, 54.
\bibitem{cer15} Cerruti, M., Zech, A., Boisson, C., \& Inoue, S.\ 2015, MNRAS 448, 910.
\bibitem{pet15} Petropoulou, M., Dimitrakoudis, S., Padovani, P., Mastichiadis, A., \& Resconi, E.\ 2015, MNRAS 448 (2015) 2412.
\bibitem{pbjct20016} P. Brun \& J. Cohen-Tanugi, 2016, this volume.
\bibitem{2005ApJ...627L..89C} Cooray, A., \& Milosavljevi\'{c}, M. \ 2005, Astrophys. \ Jour. \ 627, 89.
\bibitem{2010ApJ...722L..58S} Strong, A.~W., Porter, T.~A., Digel, S.~W., et al.\ 2010, Astrophys.\ J.\ 722, L58.
\bibitem{2015ApJ...800L..27A} Ajello, M., Gasparrini, D., S{\'a}nchez-Conde, M., et al.\ 2015, Astrophys.\ J.\ 800, LL27.
\bibitem{2011ApJ...733...66I} Inoue, Y.\ 2011, Astrophys.\ J.\ 733, 66.
\bibitem{2012ApJ...751..108A} Ajello, M., Shaw, M.~S., Romani, R.~W., et al.\ 2012, Astrophys.\ J.\ 751, 108.
\bibitem{2011ApJ...736...40S} Stecker, F.~W., \& Venters, T.~M.\ 2011, Astrophys.\ J.\ 736, 40.
\bibitem{jk} Kn\"odlseder, J., 2016, this volume.
\bibitem{2011ExA....32..193A} Actis, M., Agnetta, G., Akhperjanian, A., et al.\ 2011, Experimental Astronomy, 32, 193.
\bibitem{FermiLAT} {\it Fermi}-LAT performance at http://www.slac.stanford.edu/exp/glast/groups/canda/lat\_Performance.htm.
\bibitem{2004NewAR..48..193S} Sch{\"o}nfelder, V.\ 2004, New Astron.\ Rev.\ 48, 193.
\bibitem{fpGRB} F. Piron, 2016, this volume.
\bibitem{Senturk2013}    G. D. Sent\"urk {\it et al}, Astrophys.\ Jour.\ 764 (2013) 119.
\bibitem{Sanchez2013}      D. Sanchez, S. Fegan, and B. Giebels, A\&A, 554 (2013) 75.
\bibitem{2013ApJS..209...34A} Ackermann, M., Ajello, M., Allafort, A., et al.\ 2013, Astrophys.\ Jour.\ Supp.\ 209, 34.
\bibitem{2015arXiv150302664A} Ajello, M., Dom{\'{\i}}nguez, A., Gasparrini, D., Cutini, S., \& for the {\it Fermi}-LAT Collaboration 2015, arXiv:1503.02664.
\bibitem{cras2014} C. R. Physique 15, issue 4 (2014) 297-384, edited by A. Letessier-Selvon.

\end{thebibliography}
\end{document}